\documentclass[11pt]{article}
\pdfoutput=1

\usepackage{euscript}
\usepackage{amssymb}
\usepackage{amsfonts}
\usepackage{amsbsy}
\usepackage{amsmath}
\usepackage{epsfig}
\usepackage{amsthm}
\usepackage{amscd}
\usepackage{amstext}

\usepackage[pdftex]{hyperref}

\def\ben{\begin{equation}}
\def\een{\end{equation}}
\def\half{{\textstyle{\frac{1}{2}}}}
\def\qtr{{\textstyle{\frac{1}{4}}}}
\let\a=\alpha \let\b=\beta \let\g=\gamma \let\d=\delta \let\e=\varepsilon
   \let\k=\kappa
 \let\m=\mu    
 \let\t=\tau

\let\w=\omega

\let\pa=\partial
\def\be{\begin{equation}}
\def\ee{\end{equation}}
\def\ba{\begin{array}}
\def\ea{\end{array}}

\def\dalemb#1#2{{\vbox{\hrule height .#2pt
       \hbox{\vrule width.#2pt height#1pt \kern#1pt
               \vrule width.#2pt}
       \hrule height.#2pt}}}

\newcommand{\bea}{\begin{eqnarray}}
\newcommand{\eea}{\end{eqnarray}}

\newcommand{\Tr}{{\rm Tr} }

\def\R{{{\mathbb R}}}

\def\Z{{{\mathbb Z}}}

\def\Lag{{\mathcal{L}}}

\def\ocal{{\mathcal{O}}}

\begin{document}

\begin{flushright}
arXiv:0903.3246 [hep-th]
\end{flushright}

\begin{center}

\vspace{1cm} { \LARGE {\bf Lectures on holographic methods for \\ condensed matter physics}}

\vspace{1cm}

Sean A. Hartnoll

\vspace{0.8cm}

{\it Jefferson Physical Laboratory, Harvard University,
\\
Cambridge, MA 02138, USA \\}

\vspace{0.6cm}

{\tt hartnoll@physics.harvard.edu} \\

\vspace{2cm}

\end{center}

\begin{abstract}

These notes are loosely based on lectures given at the CERN Winter School on Supergravity, Strings and Gauge theories, February 2009 and at the IPM String School in Tehran,
April 2009. I have focused on a few concrete topics and also on addressing questions that
have arisen repeatedly. Background condensed matter physics material is included as motivation and easy reference for the high energy physics community. The discussion of holographic techniques progresses from equilibrium, to transport and to superconductivity.

\end{abstract}

\newpage
\setcounter{page}{1}

\enlargethispage{\baselineskip}

\tableofcontents

\newpage

\section{Why holographic methods for condensed matter?}

\subsection{Why condensed matter?}

Why, on the eve of the LHC, should
high energy and gravitational theorists be thinking about phenomena that
occur at energy scales many orders of magnitude below
their usual bandwidth? Three types of answer come to mind.

Firstly, the AdS/CFT correspondence \cite{Maldacena:1997re} is a unique approach to
strongly coupled field theories in which certain questions
become computationally tractable and conceptually more
transparent. In condensed
matter physics there are many strongly coupled systems
that can be engineered and studied in detail in laboratories.
Some of these systems are of significant technological interest.
Observations in materials involving strongly correlated electrons
are challenging traditional condensed matter paradigms that were based around
weakly interacting quasiparticles and the theory of
symmetry breaking \cite{anderson}. It seems reasonable to hope,
therefore, that the AdS/CFT correspondence may
be able to offer insight into some of these nonconventional
materials.

Secondly, condensed matter systems may offer an arena
in which many of the fascinating concepts of high energy
theory can be experimentally realised. The standard model
Lagrangian and its presumptive completion are unique in our
universe. There will or will not be supersymmetry.
There will or will not be a conformal sector.
And so on. In condensed matter physics there are many
effective Hamiltonians. Furthermore, an increasing number of Hamiltonians
may be engineered using, for instance, optical lattices \cite{Greiner}.
As well as novel realisations of theoretical ideas,
ultimately one might hope to engineer
an emergent field theory with a known AdS dual, thus leading to
experimental AdS/CFT (and reversing the usual relationship between string theory and the
standard model).

Thirdly, and more philosophically, the AdS/CFT correspondence
allows a somewhat rearranged view of nature in which the traditional
classification of fields of physics by energy scale is less important.
If a quantum gravity theory can be dual to a theory with many
features in common with quantum critical electrons, the
question of which is more `fundamental'
is not a meaningful question. Instead, the emphasis is on concepts
that have meaning on both sides of the duality. This view has
practical consequences. For instance, seeking a dual
description of superconductivity one realises that there might be
loopholes in black hole `no-hair' theorems and one is led
to new types of black hole solutions.

These lectures will be about the first type of answer.  We shall
explore the extent to which the AdS/CFT correspondence
can model condensed matter phenomena.

\subsection{Quantum criticality}
\label{sec:qcp}

Although quantum critical systems are certainly not the only condensed matter
systems to which holographic techniques might usefully be applied,
they are a promising and natural place to start.
Quantum critical points have a spacetime scale invariance
that provides a strong kinematic connection to the simplest
versions of the AdS/CFT correspondence. Furthermore,
a lack of weakly coupled quasiparticles often makes
quantum critical theories difficult to study using traditional
methods. Outside of AdS/CFT there are no models
of strongly coupled quantum criticality in 2+1 dimensions in which
analytic results for processes
such as transport can be obtained.\footnote{An
example of a solvable 1+1 dimensional model with a quantum
critical point is the Ising model in a transverse magnetic field,
see e.g. chapter 4 of \cite{sachdev}.}

Quantum critical theories arise at continuous phase transitions
at zero temperature. A zero temperature phase transition
is a nonanalyticity in the ground state of an (infinite) system as
a function of some parameter such as pressure of applied magnetic field.
The quantum critical point may or may not be the zero
temperature limit of a finite temperature phase transition.
Note in particular that the Coleman-Mermin-Wagner-Hohenberg
theorem \cite{sym} prevents spontaneous breaking of a continuous
symmetry in 2+1 dimensions
at finite temperature, but allows a zero temperature phase transition.
In such cases the quantum phase transition becomes a crossover
at finite temperature.\footnote{
In certain 2+1 dimensional systems the quantum critical point can
connect onto a Berezinksy-Kosterlitz-Thouless transition
at finite temperature. Also, strictly infinite $N$ evades the theorem
as fluctuations are suppressed.}

Typically, as the continuous quantum critical point is approached,
the energy of fluctuations about the ground state (the `mass gap') vanishes
and the coherence length (or other characteristic lengthscale)
diverges with specific scaling properties.
In a generic nonrelativistic theory, these two scalings (energy and
distance) need not be inversely related, as we will discuss in detail
below. The quantum critical theory itself is scale invariant.

Quantum critical points can dominate regions of the phase diagram
away from the point at which the energy gap vanishes. For instance,
in regions where the deformation away from criticality due to an
energy scale $\Delta$ is less important than the deformation due to
a finite temperature $T$, i.e. $\Delta < T$, then the system should
be described by the finite temperature quantum critical theory.
This observation leads to the counterintuitive fact that the imprint of the
zero temperature critical point grows as temperature is increased.
This phenomenon is illustrated in figure 1.

\begin{figure}[h]
\begin{center}
\includegraphics[height=6cm]{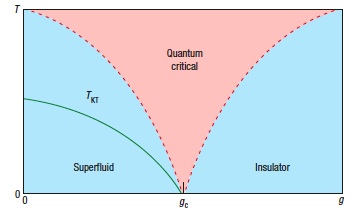}
\end{center}
\caption{Typical temperature and coupling phase diagram near a quantum critical point. The two low temperature phases are separated by a region described by a scale-invariant theory at finite temperature. The solid line denotes a possible Kosterlitz-Thouless transition. Figure taken from reference \cite{sachdev2}.}
\end{figure}

To get a feel for quantum critical physics and its relevance,
we now discuss several examples of systems that display quantum
criticality. These will include both lattice models and experimental setups.
Our discussion will be little more than an overview -- the reader is encouraged
to follow up the references for details.
We shall focus on 2+1 dimensions, as we often will throughout these lectures.
In several cases we will explicitly write down an action for the quantum
critical theory. Typically the critical theory is strongly coupled and so
the action is not directly useful for the analytic computation of many quantities
of interest. Even in a large $N$ or (for instance) $d=4-\epsilon$
expansion, which effectively make the fixed point perturbatively accessible,
time dependent processes, such as charge transport, are not easy
to compute. This will be one important motivation for turning to the AdS/CFT
correspondence. The correspondence will give model theories that
share feature of the quantum critical theories of physical interest, but
which are amenable to analytic computations while remaining strongly
coupled.

\subsubsection{Example: The Wilson-Fisher fixed point}
\label{sec:WF}

Let $\Phi$ be an $N$ dimensional vector. The theory described by the
action
\be\label{eq:WF}
S[{\Phi}] = \int d^3x \left( \left( \pa \Phi \right)^2 + r \Phi^2 + u \left(\Phi^2 \right)^2 \right) \,,
\ee
becomes quantum critical as $r \to r_c$ (in mean field theory
$r_c=0$ but the value gets renormalised) and is known as the
Wilson-Fisher fixed point. At finite $N$ the relevant coupling $u$ flows to
large values and the critical theory is strongly
coupled. The derivative in (\ref{eq:WF})
is the Lorentzian 3-derivative (i.e. signature (--,+,+)) and we have set a velocity
$v=1$. This will generally not be the speed of light.

We now briefly summarise two lattice models in which the theory (\ref{eq:WF})
describes the vicinity of a quantum critical point, as reviewed in \cite{sachdev2, sachdev}.

The first model is an insulating quantum magnet.
Consider spin half degrees of freedom $S_i$ living on a square lattice
with the action
\be\label{eq:AF}
H_\text{AF} = \sum_{\langle ij \rangle} J_{ij} S_i \cdot S_j \,,
\ee
where $\langle ij \rangle$ denotes nearest neighbour interactions
and we will consider antiferromagnets, i.e. $J_{ij} > 0$. Now choose
the couplings $J_{ij}$ to take one of two values, $J$ or $J/g$ as shown in figure 2.
The parameter $g$ takes values in the range $[1,\infty)$.

\begin{figure}[h]
\begin{center}
\includegraphics[height=6cm]{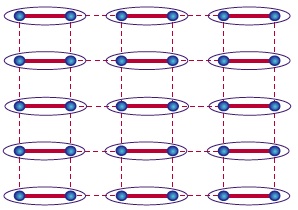}
\end{center}
\caption{At large $g$, the dashed couplings are weaker ($J/g$) than the solid ones ($J$). This favours pairing into spin singlet dimers as shown. Figure taken from reference \cite{sachdev2}.}
\end{figure}

The ground state of the model (\ref{eq:AF}) is very different in the two limits
 $g \to 1$ and $g \to \infty$. At $g = 1$ all couplings between spins are equal,
 this is the isotropic antiferromagnetic Heisenberg model,
 and the ground state has N\'eel order characterised by
\be\label{eq:neel}
\langle S_i \rangle = (-1)^i \Phi \,,
\ee
where $(-1)^i$ alternates in value between adjacent lattice sites.
We can na\"ively picture this state as the classical ground state
in which neighbouring spins are anti-alligned. Here $\Phi$
is a three component vector. The low energy excitations about
this ordered state are spin waves described by the action
(\ref{eq:WF}) with $N=3$ and $\Phi^2$ fixed to a finite value.
Spin rotation symmetry is broken in this phase.

In the limit of large $g$, in contrast, the ground state is given by
decoupled dimers. That is, each pair of neighbouring spins with a
coupling $J$ (rather than $J/g$) between them forms a spin singlet.
At finite but large $g$, the low energy excitations are triplons. These
are modes in which one of the spin singlet pairs is excited to a triplet
state. The triplons have three polarisations and are again described by
the action (\ref{eq:WF}) with $N=3$ but about a vacuum with $\Phi=0$.

These two limits suggest that the low energy dynamics of the
coupled-dimer antiferromagnet (\ref{eq:AF}) is captured by the
action (\ref{eq:WF}) across its phase diagram and that there will be
a quantum critical point at an intermediate value of $g$ described
by the $N=3$ Wilson-Fisher fixed point theory. This indeed appears
to be the case, with $g_c \approx 1.91$ found numerically.
For further details and references, including applications of this model to the compound
TlCuCl$_3$, see \cite{sachdev2}.

A second lattice model realising the Wilson-Fisher fixed point is the
boson Hubbard model with filling fraction one (i.e. with the same
number of bosons as lattice sites). This model describes bosons
$b_i$ on a square lattice with Hamiltonian
\be\label{eq:BH}
H_{BH} = - t \sum_{\langle ij \rangle} \left(b_i^\dagger b_j + b_j^\dagger b_i \right)
+ U \sum_i n_i \left( n_i - 1\right) \,.
\ee
Here $n_i = b_i^\dagger b_i$ is the site occupation number. The total boson
number commutes with the Hamiltonian and we are considering the
case in which the number of boson is constrained to equal the number of
lattice sites (canonical ensemble).
The first term in (\ref{eq:BH}) allows hopping between adjacent sites
while the second is an on-site repulsive interaction between bosons. The
$-1$ in the last term ensures there is no penalty for single occupancy. The Hamiltonian has
a $U(1) = SO(2)$ symmetry: $b_i \to e^{i \phi} b_i$.

Once again, the ground state is very different in two limits. When $U/t \to \infty$
the sites decouple, there is one boson per site
and no fluctuations. A mean field analysis \cite{sachdev} is then sufficient to determine that the
ground state has $\langle b_i \rangle = 0$. In the opposite limit, $U/t \to 0$
the model becomes quadratic in the $b_i$ and can be solved exactly.
Passing to a grand canonical ensemble, the (necessarily nonzero) chemical potential is easily
seen to imply that the ground state now has a condensate $\langle b_i \rangle \neq 0$.

The two limits considered suggest that there is a superfluid-insulator transition at
an intermediate value of $U/t$. This is indeed the case. The $U(1)$ symmetry is spontaneously
broken on one side of the transition and the critical theory is given in terms
of the Ginzburg-Landau order parameter $\langle b_i \rangle$ which we can rewrite as a two-component vector $\Phi$. The symmetries of the problem are then almost enough to conclude that
the theory at the transition is the Wilson-Fisher fixed point (\ref{eq:WF}) with $N=2$.
In fact, having integer filling is necessary to nontrivially eliminate a term in the quantum critical
action that is first order
in time derivatives: more details and references can be found in \cite{sachdev}, chapter 10,
the original work is \cite{fisher}. This model arises in the description of bosonic atoms in optical
lattices, see e.g. \cite{bloch} for a review.

\subsubsection{Example: Spinons and emergent photons}
\label{sec:zA}

Let $A$ be an (emergent 2+1 dimensional) photon
and $z$ a complex spinor described by
\be\label{eq:zA}
S[{z,A}] =  \int d^3x \left( \left| \left( \pa - i A \right) z \right|^2 + r |z|^2 +
u  \left( |z|^2 \right)^2  + \frac{1}{2 e_0^2} F^2 \right) \,,
\ee
where $F=dA$. As before, derivatives are Lorentzian and we have
set a velocity to one. This theory becomes quantum critical as
$r \to r_c$. In fact, in the phase $r < r_c$, this theory is equivalent to
our previous example (\ref{eq:WF}), with $N=3$, via the mapping
\be\label{eq:Phiz}
\Phi = \bar z_\a \sigma_{\a\b} z_\b \,,
\ee
where $\sigma$ are the Pauli matrices. The photon must be introduced
to gauge the phase redundancy of the $z$ parametrisation and the Maxwell term is
generated upon renormalisation. However, the change to
spinon and photon variables leads to an inequivalent path integral. The new formulation
turns out to allow the theory to capture different quantum
critical points that mediate transitions between two distinct ordered phases (i.e.
a different symmetry is broken on each side of the transition).
Generically second order phase transitions
separate an ordered and a disordered phase and the critical theory describes
fluctuations of the order parameter. This was the case for the examples we discussed
in the previous subsection. At `deconfined' quantum critical points, this
`Landau-Ginzburg-Wilson paradigm' does not hold and instead the quantum
critical theory is described in terms of degrees of freedom that are
not present in either of the ordered phases \cite{deconfined, deconfined2, sachdev2}.

We will now briefly discuss two distinct lattice models in which the spinon-photon action
(\ref{eq:zA}) describes the system at quantum criticality, closely following \cite{sachdev2}.

The first model is again an insulating quantum magnet. Similarly to the previous
subsection, we wish to induce a phase transition from a N\'eel ordered antiferromagnetic
phase to a phase in which the spin rotation symmetry is unbroken.
The difference will be that we will start with a model which is invariant under
the $\Z_4$ rotational symmetry of a square lattice. This symmetry will
be spontaneously broken in the spin singlet phase, leading to a
`valence-bond solid' (VBS) state.

Consider the square lattice spin half model
\be\label{eq:VBS}
H_\text{VBS} = J \sum_{\langle ij \rangle} S_i \cdot S_j -
Q \sum_{\langle ijkl \rangle} \left(S_i \cdot S_j - \qtr \right) 
\left(S_k \cdot S_l - \qtr \right)\,,
\ee
with $J,Q > 0$ and $\langle ijkl \rangle$ denotes the sites
on a plaquette. Recall that $S_i \cdot S_j = - 3/4$ in a spin
singlet state and $S_i \cdot S_j = 1/4$ in a spin one state.
The second term in the Hamiltonian (\ref{eq:VBS}) therefore favours
states in which the four spins on a plaquette pair up into
two spin singlets.
The Hamiltonian does not pick out any particular
pair of adjacent spins in the plaquette, however, and is therefore $\Z_4$
invariant.

The ground states in differing limits are characterised as follows.
In the limit $Q/J \to 0$, we are once again back at the isotropic Heisenberg
antiferromagnet and the ground state will have N\'eel order
(\ref{eq:neel}). Spin wave fluctuations of $\Phi$ can be expressed in terms
of the spinon variables via (\ref{eq:Phiz}), leading to the action (\ref{eq:zA})
in a phase with $\langle z \rangle \neq 0$. The manifest $U(1)$ symmetry of the
action (\ref{eq:zA}) is Higgsed in this phase.

As $Q/J \to \infty$ the model develops an expectation value for the operator
\be\label{eq:PsiO}
\Psi = (-1)^{j_x} S_j \cdot S_{j+\hat x} + i (-1)^{j_y} S_j \cdot S_{j + \hat y} \,.
\ee
This is called VBS order.
The operator $\Psi$ can be thought of as measuring the tendency of neighbouring
spins to pair into singlets. It is not obvious that $\Psi$ will condense, although
it is plausible given the second term in (\ref{eq:VBS}).
The precise form of (\ref{eq:PsiO}) is chosen so that $\Psi$
transforms by a phase under $\Z_4$ lattice rotations. Therefore a condensate
$\langle \Psi \rangle \neq 0$ spontaneously breaks lattice rotation symmetry while
preserving spin rotation symmetry. An obvious excited state above the VBS ground state
is when a singlet breaks into a pair of independent spin half modes that can now move freely,
these are the spinons $z$ in (\ref{eq:zA}). Note that spinons did not exist in the coupled
dimer model of the previous subsection because the locations of the dimer singlets
were fixed; thus the spin halves could not move freely and instead one had triplon
excitations. The gauge boson $A$ can be thought of in the following way.\footnote{The gauge
boson can be directly related to the relative phases of the different spin pairings in the
ground state wavefunction. This connection reveals that the spinons are charged.
The gauge field is a spin singlet excitation.}
Firstly recall
that in 2+1 dimensions we can dualise $A$ to a scalar $\zeta$
\be
\star_3 F = d \zeta \,.
\ee
This operation reveals a `dual' global $U(1)$ symmetry $\zeta \to \zeta + \delta \zeta$.
This symmetry is to be identified with the space rotational symmetry that is spontaneously
broken in the VBS phase. The operator $\Psi$ can then be identified with the
`monopole operator'
\be
\Psi \sim e^{i 2 \pi \zeta/e_0^2} \,,
\ee
where squiggle denotes the presence of non-universal factors that depend on the microscopic model.
This identification is determined by matching the transformation under the $U(1)$ shift symmetry.

Now, the rotation symmetry of the lattice model (\ref{eq:VBS}) was
$\Z_4$, not $U(1)$. The breaking $U(1) \to \Z_4$ can be achieved by adding
terms to the action (\ref{eq:zA}), such as
\be\label{eq:breaking}
\Delta S = \int d^3x \lambda \cos \frac{8 \pi \zeta}{e_0^2} \,.
\ee
This interaction, describing the insertion of monopole defects, will manifestly give the
Goldstone boson $\zeta$ a mass. The claim of \cite{deconfined, deconfined2}
is that as the critical point is approached from the VBS phase, terms like (\ref{eq:breaking}) become
irrelevant, the $\Z_4$ symmetry is enhanced to $U(1)$ and the Goldstone boson
becomes a massless, quantum critical, excitation. This is the emergent gauge field at the quantum
critical point.

The upshot of the preceding (necessarily superficial)
discussions is that the action (\ref{eq:zA}) describes the
quantum critical point separating phases with N\'eel and VBS order, respectively. The description
is in terms of gapless spinons and a photon which are not the order parameters of either
phase. See \cite{sachdev2} for
further details and references on this model, and to references for
experimental systems showing VBS order, including the underdoped cuprates. We should
also note that the `particle-vortex duality' underlying the deconfined criticality we are describing
is formally very similar to the mirror symmetry of supersymmetric 2+1 dimensional gauge theories
\cite{Sachdev:2008wv}.

A second lattice model realising the critical theory (\ref{eq:zA}) is a boson Hubbard-like
model with filling fraction $\half$, i.e. with half as many bosons as lattice sites.
We will be very brief, for details and references see \cite{sachdev2}.
With a non-integer filling fraction it is more difficult to obtain insulating phases, that is, to suppress
fluctuations into empty neighbouring sites, and so one needs to supplement the boson Hubbard Hamiltonian
(\ref{eq:BH}) with off-site repulsive interactions such as
$\sum_{\langle ij \rangle} \Lambda_{ij} n_i n_j$, with $\Lambda_{ij} > 0$.
In the hard-core boson limit, in which
double occupancy of sites is forbidden, there is a direct map between the Hilbert space
of the half filled boson model and the antiferromagnetic models we have considered: One
simply maps vacant sites to spin up and filled sites to spin down. This shows
that hard-core boson models admit (under this map) states with both N\'eel and VBS order.
The superfluid phase maps onto a N\'eel ordered phase.
Tuning the relative strength of hopping and repulsive interactions, these models can
exhibit superfluid/N\'eel to VBS transitions in the same universality class as the antiferromagnetic
lattice models we have just discussed.

\subsubsection{Connection to nonconventional superconductors}
\label{sec:SC}

The phenomenology of conventional superconductors is extremely well explained
by BCS theory and its extensions, see e.g. \cite{Parks}. In these theories a charged
fermion bilinear operator condenses due to an attractive interaction between fermionic
quasiparticles (dressed electrons) that becomes
strong at low energies \cite{Polchinski:1992ed}. These bilinears are called Cooper pairs
and in BCS theory the attractive interaction is mediated by phonons (lattice vibrations).

It is now clear that there exist superconductors in nature
that are not described by BCS theory. There are several meanings that `non-BCS' might
have. One is that the attractive interaction between the fermionic quasiparticles
is not due to phonons. An example is a spin-spin interaction mediated by, for instance,
`paramagnons'. Another, more radical, departure from BCS theory
would be if the normal state of the system,
at temperatures just above the onset of superconductivity, were inherently strongly
interacting and did not admit a weakly interacting quasiparticle description at all.
Natural circumstances under which this latter possibility might arise are if the onset
of superconductivity occurs in the vicinity of a quantum critical point.
We will now discuss two classes of systems in which superconductivity and quantum
criticality may occur in close proximity: the `heavy fermion' metals and (more speculatively)
the cuprate `high-T$_c$' superconductors.

The heavy fermion metals are compounds in which the effective
charge carrying quasiparticle has a mass of the order of hundreds of
times the bare (`standard model') electron mass. The large mass
arises due to the Kondo effect: hybridisation between conducting
electrons on the one hand and fixed strongly correlated electrons
that behave as a lattice of magnetic moments (the Kondo lattice)
on the other. This setup can be described by the model (e.g. \cite{heavyNat})
\be\label{eq:kondolattice}
H_\text{K.L.} = \sum_{ij,\a} t_{ij} c^\dagger{}_{i\a} c_{j\a} +
J_K \sum_iS_i \cdot s_i + \sum_{ij} I_{ij} S_i \cdot S_j \,.
\ee
Here $S_i$ is the fixed magnetic moment at the site $i$,
$c^\dagger_{i\a}$ creates a conduction electron at the site $i$ with
spin $\a$ (i.e. spin up or down) and $s_i$ is the spin
operator for the conduction electrons:
$s_i = c^\dagger{}_{i\a} \sigma_{\a \a'} c_{i\a'} $.
The first term allows the conduction electrons to hop from site
to site, the second term is the Kondo exchange interaction between conduction
and magnetic electrons and the third describes the magnetic interactions
between the fixed electrons.

The last, magnetic interaction, term in the model (\ref{eq:kondolattice}) is of the form
that we reviewed for insulating systems in the previous sections.
We might expect therefore that heavy fermion metals will exhibit
quantum critical behaviour as the various couplings are tuned.
This is indeed the case: quantum critical points in heavy fermion metals
have been extensively studied, with detailed measurements of the
onset of magnetic order and the associated quantum critical scaling
of various observables. The zero temperature phase transitions
are realised by tuning the pressure, external magnetic field or chemical
composition (doping). For a review see \cite{heavyNatP}.
The dynamics of the quantum critical points is very rich in these
systems. As well as magnetic ordering one must take into
account a crossover in which the local moments become screened
by the conduction electrons. See for instance \cite{heavyNat, heavyNatP}.

Usually magnetic order competes with superconductivity because there is a free energy
cost in screening the resulting magnetic field inside a superconductor.
It might appear surprising, therefore, that several heavy fermion metals, such as CeIn$_3$ and CePd$_2$Si$_2$ \cite{heavyAF, heavySC}, are found to have the following phase diagram as a
function of temperature and pressure.

\begin{figure}[h]
\begin{center}
\includegraphics[height=8cm]{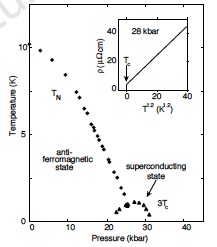}
\end{center}
\caption{Phase diagram of CePd$_2$Si$_2$ as a function
of temperature and pressure. The bottom left phase is antiferromagnetically ordered whereas the bottom middle phase is superconducting. Figure taken from reference \cite{heavyAF}.}
\end{figure}

In these materials superconductivity develops right at the edge of
antiferromagnetic order. Various properties of the materials
are found to be consistent with a simple picture of magnetically
mediated superconductivity (see e.g. \cite{heavyAF, heavySC}) in which
there is an effective interaction between two electronic quasiparticles
\be\label{eq:ssinteraction}
V(r,t) = - g^2 \chi_m(r,t) s \cdot s' \, . 
\ee
Here $s,s'$ are spins, $g$ is a coupling and $\chi_m$ is the
magnetic susceptibility. The susceptibility becomes large at
the onset of antiferromagnetism. When $s$ and $s'$ form a singlet
it turns out that (\ref{eq:ssinteraction}) is repulsive near the origin
but then oscillates in sign with a period of order the lattice
spacing. Thus there is an attractive interaction between the quasiparticles
when a finite distance apart.
This forces the resulting `Cooper pair' operator
to have a nonzero angular momentum ($\ell=2$), leading to a
d-wave superconductor, as is observed.

The interaction (\ref{eq:ssinteraction}) only makes sense if there
are weakly interacting quasiparticles. This picture seems
to work at some level for the materials at hand. However, given
the nearby quantum critical point and the associated non-Fermi liquid
behaviour, observed in many heavy fermion compounds, it might be
instructive to have a more nonperturbative approach \cite{heavySC}.

The cuprate high-T$_c$ superconductors typically have the following phase
diagram as a function of temperature and hole doping (that is, reducing
the number of conducting electrons per Cu atom in the copper oxide planes by
chemical substitution, e.g. La$_{2-x}$Sr$_x$CuO$_4$)

\begin{figure}[h]
\begin{center}
\includegraphics[height=7cm]{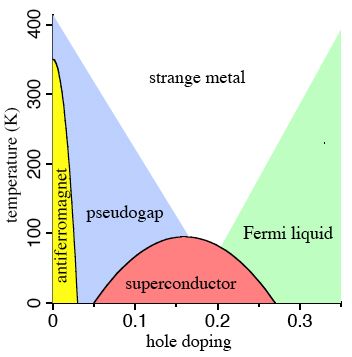}
\end{center}
\caption{Schematic temperature and hole doping phase diagram for a high-$T_c$ cuprate. There are antiferromagnetic and a superconducting ordered phases.
Figure taken from \cite{kitpblog}.}
\end{figure}

This phase diagram is obviously similar to that of the heavy fermion compounds in
figure 3. One important difference is that the antiferromagnetic
phase is separated from the superconducting dome by the still mysterious
`pseudogap' region. The precise role of critical magnetic fluctuations in mediating
superconductivity in these materials is contested, although at the very least
it correctly anticipated the d-wave nature of the Cooper pairs \cite{scalapino}.
The similarity of phase diagrams along with the many observed non-Fermi liquid properties
of the pseudogap region may suggest that, as with the heavy fermion metals,
there is a quantum critical point beneath the superconducting dome in the high-T$_c$ compounds.
Some, as yet inconclusive, evidence for such a quantum critical point is reviewed in
\cite{broun}. Even if there is a quantum phase transition under the dome, one then has to confirm
that it is continuous and important for the dynamics of the pseudogap region and the
superconducting instability more generally.

It is also possible that there are distinct quantum phase
transitions located in an extended phase diagram (in which one adds
a new axis to figure 4 that may or may not be accessible experimentally) that
nevertheless influence pseudogap physics. See figure 10 in \cite{sachdev2}.
Evidence for this picture includes experimental indications of
a `stripped' insulating phase at hole doping $x=\frac{1}{8}$ in the cuprates. Stripes
arise in the insulating VBS phases that we discussed in previous sections,
in which the charge density is spatially inhomogeneous and the lattice
rotation symmetry is broken. If one imagines the holes pairing to form
bosons, then, as we briefly mentioned above, such a VBS state could
emerge from a superfluid-insulator
transition in a boson Hubbard-type model with filling fraction $f = \frac{1}{16}$.
For further experimental and theoretical references see \cite{sachdev2}.

The upshot of the above discussions is that quantum critical points
are certainly present in heavy fermion superconductors and may be
present in the high temperature cuprate superconductors. It is possible
that in these materials quantum critical physics has a role to play
in understanding the onset of superconductivity.

\section{Applied AdS/CFT methodology}

\subsection{Geometries for scale invariant theories}

There many different paths to the AdS/CFT correspondence. Rather than motivate
or support the correspondence at this point (see e.g. \cite{Horowitz:2006ct}) let us take the
correspondence as given and ask what it achieves. 

Firstly, the AdS/CFT correspondence makes manifest the semiclassical nature
of the large $N$ limit in certain gauge theories. It allows us to compute field theory
quantities at large $N$ using saddle point methods.\footnote{
This has been a long term goal for gauge field theories. For vector field theories,
large $N$ limits are often soluble directly within field theory. Gauge theories, in
contrast, have more complicated interactions, as manifested in the necessity of the
't Hooft double line notation to count the $N$ dependence of Feynman diagrams, see e.g.
\cite{coleman}.}
The remarkable
fact is that the action functional that we have to use, and the corresponding
classical saddles, appear completely unrelated to the gauge theory degrees
of freedom. We are instructed to expand around classical solutions to a `dual' gravitational
theory with at least one more spatial dimension than the original gauge theory.
The first principles emergence of spacetime geometry from field theory is far from understood
(preliminary ideas can be found in, for instance, \cite{Berenstein:2005aa}). In the applied AdS/CFT business we shall take this, extremely useful, phenomenon for granted.

Secondly, the AdS/CFT correspondence geometrises the field theory energy scale.
The fact that the renormalisation group is expressed in terms of differential
equations hints at a notion of locality in energy scale. In the dual gravitational
description arising in the AdS/CFT framework, the energy scale is treated geometrically
on an equal footing to the spatial directions of the field theory. This is the `extra' spatial
dimension of the gravitational theory, and allows scale dependent phenomena such as
confinement and temperature to be conceptualised in new ways.

The absolute minimal structure that we need for an applicable AdS/CFT duality
will therefore be the following correspondence:
\be
\begin{array}{c}
\text{Large $N$ gauge theory} \\
d \; \text{spacetime dimensions}
\end{array}
\quad \leftrightsquigarrow \quad
\begin{array}{c}
\text{Classical gravitational theory} \\
d+1 \; \text{spacetime dimensions.}
\end{array}
\ee
There is much more structure than this in fully fledged examples
of the AdS/CFT correspondence. These lectures will concentrate instead
on the minimal properties necessary for the whole construction to be
possible at all. The logic and hope is that these will be the most robust
aspects of the correspondence and therefore a good starting point for
comparing with real world systems.

In a Wilsonian approach to field theory, the theory is (most commonly) defined
either at an ultraviolet cutoff or via an ultraviolet fixed point which renders
the theory valid at all scales. A fixed point is the simplest place to start
for the AdS/CFT correspondence. At the fixed point itself, the theory is
scale invariant. The scaling symmetry need not act the same way on
space (momentum) and time (energy). Assuming spatial isotropy, in general
we can have the scaling action
\be
t \to \lambda^z t \,, \qquad \vec x \to \lambda \vec x \,. 
\ee
Here $z$ is called the dynamical critical exponent, first introduced
as an anisotropic space and time scaling of the renormalisation
group in \cite{hertz}. A priori $z$ can take any positive value.

At the fixed point the symmetry algebra of the field theory will contain
the generators of rotations $\{M_{ij}\}$, translations $\{P_i\}$,
time translations $H$ and dilatations $D$. These generators satisfy the
standard commutation relations for $\{M_{ij}, P_k, H\}$ together with
the action of dilatations
\be\label{eq:D1}
\left[ D, M_{ij}\right] = 0 \,, \qquad \left[D, P_i\right] = i P_i \,, \qquad
\left[D, H\right] = i z H \,.
\ee
This symmetry algebra is sometimes called a Lifshitz algebra,
as it generalises the symmetry of Lifshitz fixed points, which
have $z=2$ and describe tricritical points in which
one nearby phase is inhomogeneous (see e.g. \cite{chaikin}), to general $z$.

The AdS/CFT logic suggests that we look for a spacetime metric
in one higher dimension than the field theory in which these
symmetries are realised geometrically. One is lead to the following
metric
\be\label{eq:lifshitz}
ds^2 = L^2 \left( - \frac{dt^2}{r^{2z}} + \frac{dx^i dx^i}{r^2} + \frac{dr^2}{r^2} \right) \,.
\ee
The Killing vectors generating the algebra are
\be\label{eq:lifkilling}
M_{ij} = - i (x^i \pa_j - x^j \pa_i) \,, \quad P_i = - i \pa_i \,, \quad H = - i \pa_t \,, \quad
D = -i (z \, t \, \pa_t + x^i \pa_i + r \, \pa_r ) \,.
\ee
The metric (\ref{eq:lifshitz}) was first written down in \cite{Kachru:2008yh}.
The claim of the AdS/CFT correspondence is that the physics of
a strongly coupled scale invariant theory is captured in the large $N$ limit
by classical dynamics about the background metric (\ref{eq:lifshitz}).
At this point we have said nothing about the two dual theories in question,
beyond the fact that the `bulk' gravitational dynamics includes a metric
$g_{\mu\nu}$. The metric (\ref{eq:lifshitz}) is robust: independently
of the dynamics of the theory, the only effect of bulk quantum corrections
(i.e. $1/N$ corrections) are to renormalise the values of the radius $L$
and the dynamical critical exponent $z$ \cite{Adams:2008zk}.

Various comments should be made about the background (\ref{eq:lifshitz}) .
All curvature invariants are constant, with a curvature scale $1/L$.
However, the metrics all have pp curvature singularities at the `horizon' at
$r \to \infty$, unless $z=1$. Recall that a pp singularity means that the tidal
forces diverge in a parallel propagated orthonormal frame.\footnote{Specifically,
take the radially ingoing null geodesic with tangent $T = r^{2z} \pa_t + r^{1+z} \pa_r$.
A parallely propagated null-orthonormal frame is completed with the
vectors $N = \pa_t - r^{1-z} \pa_r$ and $X_i = \pa_{x^i}$.  The tidal force
$R_{\mu \nu \rho \sigma} T^\mu X_i^\nu T^\rho X_i^\sigma = (z-1) r^{2z}$
diverges as $r \to \infty$ unless $z=1$.} These are genuine singularities, as
infalling observers are not able to continue through the $r=\infty$ surface
(which is reached in a finite proper time).
They are rather mild null singularities, however, and likely to be resolved at finite temperature
and to be acceptable within a string theory framework, cf. \cite{Horowitz:1997uc}. Indeed they may
be indicative of interesting physics. However, it does mean that
one should be careful in using these spaces for $z \neq 1$ and that a global
geodesic completion of the space does not exist.

The case $z=1$ is nothing other than Anti-de Sitter space. In this case the symmetry
of the spacetime, and hence of the dual scale-invariant theory is substantially
enhanced. Besides rotations, spacetime translations and dilatations, the theory
enjoys Lorentz boost symmetry and special conformal symmetries. As well as being
regular, Anti-de Sitter spacetime has the virtue of being a solution to a simple $d+1$
dimensional theory of gravity, namely general relativity with a negative cosmological constant
\be\label{eq:einstein}
S = \frac{1}{2 \kappa^2} \int d^{d+1}x \sqrt{-g} \left(R + \frac{d (d-1)}{L^2} \right) \,.
\ee
Partly for this reason, much of these lectures will use examples with $z=1$.
Lifshitz invariant spacetimes with $z \neq 1$ can also be obtained from more complicated
actions (see e.g. \cite{Kachru:2008yh, Adams:2008zk,Taylor:2008tg}).

For $z>1$ these spaces are candidate duals to nonrelativistic field theories. Besides the absence
of Lorentz boost symmetry, this fact is reflected in their causal structure.
As we move towards the `boundary' $r=0$ (the boundary is in the
direction in which $g_{x_i x_i}$ diverges and should more properly be
thought of as a conformal boundary, as the spacetime itself is infinite in extent),
the metric component $g_{tt}$ diverges faster
than $g_{x_i x_i}$. This means that the lightcones are flattening out and so the effective
speed of light is diverging, as one would expect for a nonrelativistic theory.
The technical expression of this fact is that arbitrarily near to the boundary, the spacetime is not
causally distinguishing. That is to say, at $r \to 0$ distinct spatial points $x \neq y$ at some time
$t=t_0$ have identical causal futures and pasts.

We have seen how the Poincar\'e group can emerge geometrically at the special value of $z=1$
(together with its additional conformal symmetries).
A different important structure that can be added to the basic algebra of rotations
and space and time translations are Galilean boosts. The Galilean boosts are vectors
$K_i$ that in classical mechanics generate the transformation $\{x_i \to x_i + v_i t, t \to t\}$
and satisfy the algebra
\be
[M_{ij}, K_k] = i (\delta_{ik} K_j - \delta_{jk} K_i) \,, \quad [P_j, K_i] = 0 \,, \quad [H, K_i] = - i P_i \,. 
\ee
In quantum mechanics, however, it has been argued that physically relevant systems require
a central extension of this algebra \cite{Inonu}\footnote{More precisely,
 \cite{Inonu} show that irreducible representations of the Galilean algebra in which translations and boosts commute do not admit states with definite position
 or velocity. We will see shortly that a dual geometric realisation of the Galilean algebra
 automatically includes a number operator symmetry. The conclusion of \cite{Inonu} however may be too strong, see e.g. \cite{Bagchi:2009my} for a `massless' (non-extended) Galilean algebra that may be of physical interest.}
\be
[P_j, K_i] = 0 \quad \rightsquigarrow \quad [P_j, K_i]  = - i \delta_{ij} N \,.
\ee
For instance, in the action of the Galilean group on the Schr\"odinger equation for a free
particle, $N=m$ is the mass. In general $N$ can be interpreted as a number operator that
counts particles with a certain fixed mass. The operator $N$ commutes with the whole
Galilean algebra (in particular $[H,N]=0$), and therefore the system has a conserved particle number.

The Galilean algebra can be extended to include an action of dilatations. As well
as the operations of equation (\ref{eq:D1}) above, dilatations act on the Galilean boosts
and particle number as
\be\label{eq:moredilaton}
[D, K_i] = i (1-z) K_i \,, \quad [D,N] = i (2-z) N \,.
\ee
These relations are determined by all the previous commutation relations
considered together with the Jacobi identity. The final symmetry algebra,
involving $\{M_{ij}, P_i, H, D, K_i, N\}$ is an algebra for a Galilean and scale
invariant field theory. It is often called the Schr\"odinger algebra, as it generalises
the symmetry of the Schr\"odinger equation for a free particle, which has $z=2$
\cite{Hagen:1972pd, Niederer:1972zz}.
We should note that the case $z=2$ allows for an extra `special conformal' generator
to be added to the algebra. A modern field theoretic discussion of the $z=2$ Schr\"odinger
algebra can be found in \cite{Nishida:2007pj} .

If we wish to study strongly coupled Galilean-invariant conformal field theories using the
AdS/CFT correspondence we need to realise the Schr\"odinger algebra geometrically.
The extra generators $N$ and $K_i$, on top of the Lifshitz algebra, are problematic.
It is not possible to arrange for the whole algebra of a $d$ dimensional Schr\"odinger invariant
field theory to act as the isometries of
a $d+1$ dimensional spacetime. This lead \cite{Son:2008ye, Balasubramanian:2008dm}
to push the rules of AdS/CFT slightly and consider a candidate dual
spacetime in $d+2$ dimensions, namely:
\be\label{eq:schrodinger}
ds^2 = L^2 \left( - \frac{dt^2}{r^{2z}} - \frac{2 dt d \xi}{r^2}+ \frac{dx^i dx^i}{r^2} + \frac{dr^2}{r^2} \right) \,.
\ee
As with the Lifshitz symmetry metric (\ref{eq:lifshitz}), it was shown in
\cite{Adams:2008zk} that $1/N$ corrections can only renormalise the values of $z$ and $L$.
The generators of the Schr\"odinger algebra are geometrically given by
\be\label{eq:schaction}
K_i = -i (-t \pa_i + x^i \pa_\xi) \,, \quad N = -i \pa_\xi \,, \quad D = -i (zt \pa_t + x^i \pa_i +
(2-z) \xi \pa_\xi + r \pa_r) \,,
\ee
while $M_{ij}$, $P_i$ and $H$ are the same as in equation (\ref{eq:lifkilling}).
Note that while the Lifshitz metric (\ref{eq:lifshitz}) was time reversal invariant
($t \to -t$), the Schr\"odinger metric (\ref{eq:schrodinger}) is not. The Schr\"odinger metric
for $z=2$ was first embedded into string theory in \cite{Herzog:2008wg, Adams:2008wt, Maldacena:2008wh}.

We see in (\ref{eq:schrodinger}) and (\ref{eq:schaction}) that the extra extra dimension of the spacetime 
(i.e. beyond the holographic direction of scale) is directly related to the particle number $N$. The particle number is given by the momentum
in the $\xi$ direction. It is common in the AdS/CFT correspondence for global symmetries in
field theory to appear in this way as extra dimensions in the gravitational dual. What is unusual in the present case is firstly that the $\xi$ direction is null, $|| \pa_\xi||^2=0$, and secondly that the
$N$ generator arises in the commutator of two spacetime symmetries.

It is clear from the commutator of dilatations with the number operator (\ref{eq:moredilaton})
that $z=2$ is special.
At this value of the dynamical critical exponent, the dilatations commute with the number operator.
Hence operators in the algebra can be simultaneously labelled by a scaling dimension and a particle
number. The commutator (\ref{eq:moredilaton}) is a mathematical expression
of the fact that mass is dimensionless at $z=2$ (having already set $\hbar =1$), which is why the free Schr\"odinger equation can be scale invariant with this particular time and space scaling.

In many systems of physical interest the spectrum of the particle number operator (the spectrum of masses) is quantised. To implement this fact in the bulk geometry we need to periodically identify the
$\xi$ direction
\be
\xi \sim \xi + 2 \pi L_\xi \,.
\ee
This identification introduces a mass scale $1/L_\xi$. We can see from (\ref{eq:schaction})
that dilatations do not preserve the length $L_\xi$ unless $z=2$ and hence are no longer isometries of the background. This is a reflection of exactly the same
phenomenon we noted in the previous paragraph: mass is only dimensionless if $z=2$. The conclusion is that we cannot have a scale invariant Galilean theory with a nontrivial discrete mass/particle number spectrum unless $z=2$. One must either dispense of scale invariance or discreteness of particle number.

A further complication with these backgrounds, relevant for all $z$, is that periodically identifying a null circle is a potentially dangerous thing to do. This has been emphasised in \cite{Maldacena:2008wh}. For instance, in a string theory embedding, strings winding the circle become light. This concern may be less
acute if one is thinking of the spacetime as primarily providing a geometric realisation of scale invariant kinematics, as opposed to a precise string theory dual of a particular field theory (i.e. if one is taking a more `phenomenological' approach to the AdS/CFT correspondence). The problem is also ameliorated by considering the theory at a finite number density \cite{Maldacena:2008wh}, which is indeed a physically sensible thing to do.

The metric (\ref{eq:schrodinger}) has Lorentzian signature for either sign of the $dt^2$ term. The sign of this term has at least three physical consequences, however. Firstly, it seems that if the sign is reversed with the orientation of $\xi$ kept fixed (i.e. with a certain direction corresponding to positive mass), then the space can become unstable to modes with a large particle number \cite{Hartnoll:2008rs}. Secondly, if $g_{tt}$ is taken to have a positive sign, opposite to (\ref{eq:schrodinger}), then the dual theory may no longer be causally non-relativistic. This is because the $dt d\xi$ term in (\ref{eq:schrodinger}) grows at the same rate as the $dx^i dx^i$ term as we go towards the boundary $r \to 0$; we need the $dt^2$ term to be negative and large in order for the lightcones to flatten.\footnote{In fact, the `nonrelativistic' (nondistinguishing) causality properties of (\ref{eq:schrodinger}) are more dramatic than in the Lifshitz spacetime (\ref{eq:lifshitz}). With the sign of $g_{tt}$ as in (\ref{eq:schrodinger}) all points at some fixed $t=t_0$ share the same causal future and past! \cite{Herzog:2008wg, Hubeny:2003sj}}
Finally, if $z \neq 2$ and $z \neq 1$, changing the sign of this term can lead to geodesic incompleteness at the boundary $r \to 0$ and pp singularities in the spacetime \cite{Hartnoll:2008rs}.

Despite this lengthly introduction to various possible gravity duals for general dynamical critical
exponents, in the rest of these lectures we will focus on the `relativistic' case $z=1$.
We will do this firstly because the holographic description of the other cases is still under development at the time of writing; conceptual uncertainties remain and basic computations remain to be done. Secondly, the $z=1$ case admits a rather universal  and minimal gravitational description, in terms of $d+1$ dimensional Einstein gravity (\ref{eq:einstein}).

\subsubsection{Aside: So what is {\it z} in the real world?}

Before listing the values of the dynamical critical exponent $z$
in some example quantum critical systems, we should emphasise
one physical consequence of the value of $z$: It determines
the critical dimension of interactions. For simplicity we can consider
an $N$ component vector $\Phi$ with a $(\Phi^2)^2$ interaction.
See e.g. \cite{Horava:2008jf} for a discussion of
gauge theories with gauge group of rank $N$. In this subsection we will work
in Euclidean time $\tau$, as the real time description of actions which
are non-analytic in frequencies is subtle. All frequencies that appear
in this subsection are similarly Euclidean frequencies.

Recall that the renormalisation group maps field theories to field theories
by a two step process. Start with a field theory with UV cutoff $\Lambda$ on
both momenta and energies. Suppose for instance we had a free field theory of the form
\be\label{eq:free}
S = \int^{\{\Lambda,\Lambda\}} \frac{d^{d-1}kd\w}{(2\pi)^d} \left(r + k^2 + |\w|^{2/z} \right)
|\Phi(\w,k)|^2 \,.
\ee
For $z=2$ one can also have $-i \w$ instead of $|\w|$.
Firstly one integrates out modes with momenta and
energies between some lower cutoff $\Lambda'$ and the original cutoff $\Lambda$.
In this free theory the integration does not generate any new interactions. With the
benefit of hindsight, we will lower the position and momentum cutoffs by
different amounts:
\be
\Lambda'_k = e^{-l} \Lambda \,, \qquad \Lambda'_\w = e^{-z l} \Lambda \,,
\ee
for some $l>0$. The action becomes
\be
S = \int^{\{\Lambda'_k, \Lambda'_\w \}} \frac{d^{d-1}kd\w}{(2\pi)^d} \left(r + k^2 + |\w|^{2/z} \right)
|\Phi(\w,k)|^2 + \text{constant}\,,
\ee
The second step is to rescale the momenta, energies and field $\Phi$ in order to restore the
action to its original form with a rescaled value of the `coupling' $r$. If we let
\be\label{eq:scaling}
k' = e^l k \,, \qquad \w' = e^{z l} \w \,, \qquad \Phi'(\w',k') = e^{-(z + d+1) l/2}\Phi(\w,k) \,,
\ee
then the action becomes
\be
S = \int^{\{\Lambda,\Lambda\}} \frac{d^{d-1}k'd\w'}{(2\pi)^d} \left(r e^{2l} + k'^2 + |\w|'^{2/z} \right)
|\Phi'(\w',k')|^2 + \text{constant}\,.
\ee
What this exercise shows us is that the theory can be renormalised to lower energies and momenta
by the scalings (\ref{eq:scaling}). Now suppose we add a quartic interaction
\be
S_{\text{int.}} = \int^{1/\Lambda} d^{d-1}x d\tau \, u \, (\Phi^2)^2 \,. 
\ee
We would like to know whether this interaction becomes stronger or weaker as we flow to lower energies.
From (\ref{eq:scaling}), noting that a Fourier transform implies that
$\Phi'(\tau',x') = e^{(z + d-3) l/2}\Phi(\tau,x)$, we have that $u \to u'$ with
\be
u' = e^{(5-z-d) l} u \,.
\ee
It follows that the coupling $u$ is irrelevant (becomes weaker at low energies) if
\be
d > d_c = 5 - z \,.
\ee
Setting $z=1$ we recover the well known result that the critical spacetime dimension of
relativistic $\Phi^4$ theory is $d=4$. However if $z>1$, we see that the critical dimension is
lowered. Thus for instance if we are interested in $d=2+1$ dimensional theories we see
that the interaction is irrelevant if $z=3$ and marginal if $z=2$. This fact was first noted in
\cite{hertz} and implies that `nonrelativistic' ($z > 1$) quantum critical points are increasingly
amenable to a perturbative treatment. This observation 
provides another motivation for concentrating on $z=1$ critical points in these
lectures; in 2+1 dimensions and in this class of models, at least, they are the most difficult
to study by other means!

We now give a sampling of values of $z$ that arise in models of physical interest:

${\bf \star \, z=1}$: All of the explicit examples we gave in sections \ref{sec:WF} and \ref{sec:zA} had $z=1$. Recall that these included N\'eel to VBS order transitions in insulating quantum magnets and superfluid to insulating transitions in boson Hubbard models with integer filling. We noted that these transitions were of interest for physical systems including the cuprate superconductors and atoms in optical lattices.\footnote{The theories in sections \ref{sec:WF} and \ref{sec:zA} were furthermore Lorentz invariant. A priori, $z=1$ does not imply Lorentz invariance as one could imagine having several modes propagating with linear dispersion but different velocities. In all known strongly coupled IR quantum critical points the renormalisation flow drives the velocities to be equal and this believed to be a general phenomenon. Similarly, setting $z=1$ in the metric (\ref{eq:lifshitz}) automatically lead to Lorentz symmetry.} For some more theoretical and experimental examples, see \cite{subirrecent}.

${\bf \star \, z=2}$: An example of a transition with $z=2$ is the onset of antiferromagnetism in clean `itinerant' (as opposed to localised) fermion systems. For details see chapter 12 of \cite{sachdev}, as we will be brief. These transitions are relevant for the quantum critical physics of the heavy fermion metals that we discussed in section \ref{sec:SC} above. One starts with a model for magnetically interacting itinerant spin half fermions $c^\dagger_\a$:
\be
H_0 = \int \frac{d^{d-1}k}{(2\pi)^{d-1}} \left( \epsilon(k) - \mu \right) c^\dagger_\a(k) c_\a(k) +
\frac{1}{2} \int d^{d-1}x  d^{d-1}x' J(x-x') s(x) \cdot s(x') \,, 
\ee
where $\mu$ is the chemical potential, $\epsilon(k)$ the free quasiparticle energy, $J(x-x')$ an effective exchange interaction and $s(x) = c^\dagger_\a(x) \sigma_{\a\b} c_\b(x)$ is the spin of the quasiparticle. Near the onset of antiferromagnetism one derives an effective action for a spin density wave condensate
$\Phi$ (with $N=3$ components):
\be\label{eq:K}
\langle s(x) \rangle = \Phi \cos (K \cdot x) \,,
\ee
where $K$ is the ordering wavevector. In the limit of long wavelengths and small fluctuations one finds the one loop effective action for fluctuations of $\Phi$:
\be\label{eq:SAF}
S_{iA.F.} = 
\int\frac{d\w d^{d-1}k}{(2\pi)^d} \gamma |\w| |\Phi(\w,k)|^2 +  \int d\tau d^{d-1}x \left[(\pa_x \Phi)^2 + r \Phi^2  + u \left(\Phi^2\right)^2 \right] \,.
\ee
Here $\gamma$ is a damping timescale due to interactions with gapless quasiparticles at the Fermi surface that are coupled to $\Phi$. The necessity of $z=2$ scaling is manifest in this action.
Following our discussion at the start of this section, as one approaches the critical point $r \to r_c$, the quartic coupling is irrelevant if $d=3$ and marginal if $d=2$.

A Fourier transform of the kinetic term in the action (\ref{eq:SAF}) leads to the long range in time interaction $- \int d\tau_1 d\tau_2 \Phi(\tau_1) \Phi(\tau_2)/(\tau_1 - \tau_2)^2$. Note that although this theory has $z=2$ classically, it is not Galilean invariant. Thus there is no symmetry preventing $z$ from getting corrections due to higher order loops in the quartic coupling $u$.

${\bf \star \, z=3}$: This case arises in the onset of ferromagnetism in clean itinerant fermion systems.
Ferromagnetism also occurs in, for instance, heavy fermions metals \cite{heavySC}.
For some closely related $z=3$ quantum critical points see e.g. \cite{z3a,z3b}.
The setup is similar to the antiferromagnetic case just discussed except that now the order parameter is simply
$\Phi = \langle s(x) \rangle$. One again computes a one loop effective action for $\Phi$
to obtain, see e.g. \cite{hertz} or \cite{simons} problem 6.7.8,
\be
S_{i F.} =  \int \frac{d\w \, d^{d-1}k}{(2\pi)^d} \frac{|\w|}{v |k|} |\Phi(\w,k)|^2 +
\int d\tau d^{d-1}x \left[(\pa_x \Phi)^2 + r \Phi^2  + u \left(\Phi^2\right)^2 \right] \,.
\ee
Noting that the first term has an inverse wavevector appearing in the time derivative term,
we see that this theory requires $z=3$. The quartic interaction is thus irrelevant in both $d=2$
and $d=3$. There is a nice physical interpretation of the factor of the inverse wavevector that
appears. Ferromagnetic order, unlike antiferromagnetism, carries a net spin and so is a conserved quantity in this model. Therefore
the relaxation timescale must diverge in the homogeneous limit $k \to 0$.

${\bf \star \, z=\text{{\bf nonuniversal}}}$: There is no restriction that $z$ be integer.
As well as the $z=2$ itinerant antiferromagnetic transitions we have just discussed, the
Kondo lattice model (\ref{eq:kondolattice}) for the heavy fermion metals also
admits `locally critical' quantum phase transitions. At these transitions Kondo screening
of the localised impurities vanishes simultaneously with the onset of N\'eel ordering.
This transition is not understood at the level of the previous cases we have considered. An uncontrolled self-consistent computation of the Green's function in \cite{heavyNat} obtain that near such critical points the dynamical spin susceptibility satisifies
\be
\chi(\w,k) = \frac{1}{A (K- k)^2 + B |\w|^\a} \,. 
\ee
Here $K$ is the ordering wavevector (as in (\ref{eq:K})) and $\a$ depends on microscopic lattice properties. These critical points have
$z = 2/\a$ that is nonuniversal and has to be measured in experiment. For CeCu$_{6-x}$Au$_x$
at critical doping one finds $\a \approx 0.75$ and hence $z \approx 2.6$ \cite{heavyNat}.

\subsection{Finite temperature at equilibrium}

The various spacetimes we have just considered admitted a dilatation symmetry, corresponding
to the scale invariance of the dual field theory. Thinking of the scale invariant theory as describing
the high energy (UV) physics we can consider deforming the theory by relevant operators or
by considering ensembles such as finite temperature or chemical potential.\footnote{Of course, the
scale invariant theory may itself be the low energy limit of a different field theory or lattice system. At energies well below the lattice cutoff, for instance, we are free to take the critical theory as our starting point: this is the phenomenon of universality.} We expect these effects to break the dilatation symmetry of the spacetime. As scale invariance is recovered at energies well above the characteristic scale of the deformation, we expect that the spacetime should also recover scaling invariance as we go towards the `boundary'. In the case of $z=1$ (which we shall restrict to without comment from now on) this is the technical requirement that the spacetime be `asymptotically Anti-de Sitter'. We shall shortly see how this works in practice.

For concreteness, let us write out Anti-de Sitter (AdS) spacetime explicitly in some convenient coordinates
\be\label{eq:ads}
ds^2 = L^2 \left( - \frac{dt^2}{r^2} + \frac{dr^2}{r^2} + \frac{dx^i dx^i}{r^2} \right) \,.
\ee
If we wish to break the scale invariance of this metric, while preserving rotations and spacetime translations, we should consider spacetimes of the form
\be\label{eq:ansatz}
ds^2 = L^2 \left( - \frac{f(r) dt^2}{r^2} + \frac{g(r) dr^2}{r^2} + \frac{h(r) dx^i dx^i}{r^2}  \right) \,.
\ee
We have introduced three nontrivial functions of the radial coordinate: $f(r), g(r)$ and $h(r)$.
There is clearly a certain gauge freedom in parameterising this metric; $g(r)$ can be chosen
freely by changing variables $r \to \hat r (r)$. If $f \neq h$ then this metric also breaks Lorentz invariance, as would be expected for finite temperature or finite chemical potential physics. If we were describing a renormalisation group flow triggered by
a Lorentz scalar operator, then we should set $f=h$. To recover scale invariance at high energies we should impose, for instance, that $f,g,h \to \text{const.}$ (sufficiently quickly) as $r \to 0$.

To find specific solutions for $f,g$ and $h$, we need some equations of motion.
Let's see what follows from the simplest theory that has the AdS metric (\ref{eq:ads})
as a solution, namely the Einstein gravity action (\ref{eq:einstein}). The equations of motion are
\be
R_{\mu \nu} = - \frac{d}{L^2} g_{\mu \nu} \,.
\ee
Plugging the metric ansatz (\ref{eq:ansatz}) into these equations, one finds the
Schwarzschild-AdS solution
\be\label{eq:schwarzads}
ds^2 = \frac{L^2}{r^2} \left(- f(r) dt^2 + \frac{dr^2}{f(r)} + dx^i dx^i \right) \,,
\ee
where
\be\label{eq:fsads}
f(r) = 1 - \left(\frac{r}{r_+} \right)^{d} \,.
\ee
We see that this solution introduces one dimensionless parameter $r_+/L$,
which we now need to interpret in field theory. We can see that $f \to 1$
as $r \to 0$ and hence this spacetime is asymptotically AdS as required. However,
as we go into the spacetime, to the infrared (IR) region of large $r$, we find that
there is a horizon at $r=r_+$. That is, $g_{tt}$ vanishes and hence the surface
at $r=r_+$ is infinitely redshifted with respect to an asymptotic observer.
The appearance of a black hole (with a planar horizon, the horizon is $\R^2$)
immediately suggests that the IR physics we have just found corresponds
to placing the scale invariant theory at a finite temperature
\cite{Hawking:1974rv}.

An elegant and robust argument showing that horizons correspond
to thermally mixed states can be found, for instance, in chapter 3 of 
\cite{Susskind:2005js}. We shall follow a closely related
Euclidean argument that is sufficient for our purposes \cite{Gibbons:1976ue}. 
Within a semiclassical regime we can think of the partition function
of the bulk theory (which, together with asymptotically AdS boundary
conditions, is to be equivalent to the partition function of the large $N$
field theory) as a path integral over metrics. Mimicking instanton
computations in field theory, one looks for Euclidean saddle points of the
bulk theory. Given the dominant saddle $g_\star$, the partition function is
\be\label{eq:partition}
Z = e^{-S_E[g_\star]} \,,
\ee
where $S_E[g_\star]$ is the Euclidean action evaluated on the saddle.
In the path integral, the action must include the Gibbons-Hawking
boundary term to give the correct (Dirichlet) variational problem and
furthermore a constant boundary counterterm in order to render the
action finite  (e.g. \cite{Henningson:1998gx, Balasubramanian:1999re})
\be\label{eq:fullaction}
S_E =  - \frac{1}{2 \kappa^2} \int d^{d+1}x \sqrt{g} \left(R + \frac{d (d-1)}{L^2} \right) 
 + \frac{1}{2 \kappa^2}  \int_{r \to 0} d^dx \sqrt{\gamma} \left(-2 K + \frac{2 (d-1)}{L} \right) \,,
\ee
where $\gamma$ is the induced metric on the boundary $r \to 0$,
$n^\mu$ is an outward pointing unit normal vector to the boundary
and $K =  \gamma^{\mu\nu} \nabla_\mu n_\nu$ is the trace of the
extrinsic curvature. We have omitted intrinsic curvature terms in the
boundary action in (\ref{eq:fullaction}). See e.g. \cite{de Haro:2000xn}.

One such saddle is obtained by analytic continuation of the
Schwarzschild-AdS metric; setting $\tau = i t$. That is
\be\label{eq:schwarzE}
ds^2_\star = \frac{L^2}{r^2} \left( f(r) d\tau^2 + \frac{dr^2}{f(r)} + dx^i dx^i \right) \,.
\ee
The fact that $f$ vanishes at $r=r_+$ now places a constraint on this
Euclidean signature spacetime. In order for the space to be regular
at $r=r_+$ (and hence, to be a genuine stationary point of the action)
we must periodically identify $\tau$ with periodicity
\be\label{eq:periodic}
\tau \sim \tau + \frac{4 \pi}{| f'(r_+)|} = \tau + \frac{4 \pi r_+}{d} \,.
\ee
This condition is most easily obtained by introducing coordinates
$\rho^2 = \alpha (r-r_+)$ and $\phi = \beta \tau$ and choosing
the constants $\alpha$ and $\beta$ such that as $r \to r_+$ the
$\{\rho,\phi\}$ part of the metric looks like $d\rho^2 + \rho^2 d\phi^2$.
Absence of a conical singularity requires $\phi$ to have period $2\pi$.

We can now deduce the consequence of the identification of imaginary time $\tau$
for the dual field theory. The basic object in our bulk theory is the metric $g_{\mu \nu}$. This metric
tends to a certain value $g_{(0) \mu \nu}$ on the boundary
\be\label{eq:asymptoticmetric}
g_{\mu \nu}(r) = \frac{L^2}{r^2} g_{(0) \mu \nu}  + \cdots \quad \text{as} \quad r \to 0 \,.
\ee
It is very natural to interpret $g_{(0) \mu \nu}$ as the background metric of the field
theory (pulling back to the boundary, of course, so that there is no $g_{rr}$ component).
The metric is not dynamical in field theory, but the field theory can be defined in
any fixed background metric. In (\ref{eq:asymptoticmetric})
we have factored out the overall scaling with the holographic direction $r$ and furthermore
factored out an $L^2$ so that the boundary metric is dimensionful. There is an
ambiguity in this definition of the boundary metric, $g_{(0) \mu \nu}$, which corresponds
to the fact that a conformally invariant theory is only sensitive to the conformal class of its background
metric (in particular, it is not sensitive to the overall scale of the metric). 

From (\ref{eq:schwarzE}) and (\ref{eq:asymptoticmetric}) we see that the background metric for the field theory is $ds^2 = d\t^2 + dx^i dx^i$, with $\t$ periodically identified by (\ref{eq:periodic}).
It is a well known fact\footnote{Recall: $\langle \ocal \rangle_T = \int {\mathcal{D}}\phi(x) \langle \phi(x),t \mid \ocal e^{-H/T} \mid \phi(x),t \rangle =  \int {\mathcal{D}}\phi(x) \langle \phi(x),t
\mid \ocal \mid \phi(x),t + i/T \rangle$.} that studying field theory with a periodically
identified Euclidean time corresponds to considering the theory in equilibrium at a finite
temperature. The temperature is the inverse of the periodicity. Thus
we find that the physics of the Schwarzschild-AdS black hole is dual
to field theory at a finite temperature give by
\be\label{eq:T}
T = \frac{d}{4 \pi r_+} \,.
\ee
In a scale invariant theory at finite temperature and in equilibrium there
is no other scale with which to compare the temperature. Therefore, all
nonzero temperatures should be equivalent. We can see this in the
Schwarzschild-AdS metric (\ref{eq:schwarzads}): the scaling
$(r,t,x^i) \to r_+ (r,t,x^i)$ eliminates $r_+$ from the metric. In a scale invariant
theory there are only two inequivalent temperatures: zero and nonzero.

Given the temperature, we can obtain other thermodynamic quantities
by evaluating the partition function (\ref{eq:partition}).
The action (\ref{eq:fullaction}) evaluated on the Euclidean Schwarzschild-AdS metric is found to be
\be\label{eq:onshell}
S_E = - \frac{L^{d-1}}{2 \k^2 r_+^d} \frac{V_{d-1} }{T} = - \frac{(4 \pi)^{d} L^{d-1}}{2 \k^2 d^d} V_{d-1} T^{d-1} \,,
\ee
where $V_{d-1}$ is the spatial volume in field theory units (i.e. with no factors of $L$).
From (\ref{eq:onshell}) we see that in order to be in the semiclassical gravity regime
we need that the spacetime is weakly curved in Plank units, namely $L^{d-1}/\k^2 \gg 1$.
Given that we expect the semiclassical gravity regime to be tied to a large $N$ limit
in field theory, we can anticipate that $L^{d-1}/\k^2 \sim N^\#$, where $\#$ is some
positive power. The AdS radius $L$ is not a lengthscale in the dual field theory, which is
scale invariant. For this reason $L$ is often set to 1, although we shall not do so. In
expressions that have a field theory meaning, $L$ will always appear divided by Planck
lengths, giving a dimensionless constant that is proportional to $N$ to some power.

From the value of the action (\ref{eq:onshell}) we obtain the free energy
\be
F = - T \log Z = T S_E[g_\star] = - \frac{(4\pi)^d L^{d-1}}{2 \k^2 d^d} V_{d-1} T^d \,,
\ee
and the entropy
\be
S = - \frac{\pa F}{\pa T} = \frac{(4\pi)^{d} L^{d-1}}{2 \kappa^2 d^{d-1}} V_{d-1} T^{d-1}\,.
\ee
As a check of our computation we can note that this expression for the entropy
is equal to the area of event horizon divided by $4 G_N$, where in our conventions
Newton's constant is $G_N = \kappa^2/8\pi$. This area-entropy relation is universally expected
to be true for event horizons.

To summarise the story so far: we have argued that AdS space provides
a geometric dual for scale invariant theories with $z=1$. The most universal
deformation away from pure AdS is the Schwarzschild-AdS black hole.
The black hole is dual to a finite temperature. The free energy and other
thermodynamic quantities are computed in terms of the temperature and
radius of curvature of AdS in Planck units (equivalently $N^\#$).
In terms of static and isotropic backgrounds there is not much more to be done
with pure Einstein gravity. In order to describe more features of the dual field
theory, we need to add structure to the bulk theory.

\subsection{Finite chemical potential and magnetic field at equilibrium}
\label{sec:mu}

A common additional structure that arises in condensed matter systems
(and elsewhere) is a $U(1)$ symmetry.
This could be, for instance but not necessarily, the electromagnetic $U(1)$ symmetry.
In this section we will consider the gravitational dual of theories with
a global $U(1)$ symmetry. The electromagnetic $U(1)$ symmetry
in nature is of course gauged. However, there are at least two reasons
why photons can be correctly neglected in many condensed matter processes.
Firstly the electromagnetic
coupling is observed to be small.\footnote{This statement is not always true. For
instance, 3+1 dimensional photons can mediate an effectively strong interaction in a 2+1
theory. See e.g. \cite{Son:2007ja}. However, the higher dimensional
Coulomb interaction is marginally irrelevant in the 2+1 dimensional
theory and so becomes weak at low temperatures \cite{markus}.} Secondly, the electromagnetic interaction is screened
in a charged medium. Of course, almost all of condensed matter physics
is ultimately due to electromagnetism (and the Pauli exclusion principle). When talking about neglecting photons we mean
that there is an effective field theory description of the dynamics involving effective
degrees of freedom and that in this description there are charged fields but no
gauge bosons for the $U(1)$ symmetry (i.e. no photons).
In such processes in which `virtual photons' are not important, 
the electromagnetic symmetry can be treated as a global symmetry.
If we wish to consider the response of the theory to an electromagnetic source
it is sufficient to consider a background electromagnetic field.
Indeed, this is a standard procedure throughout condensed matter theory;
for example one computes the conductivity by considering electrons, or particular collective
modes thereof, in background fields.

So what is the dual to a global $U(1)$ symmetry in field theory? We can
take our cue from the symmetries we have already discussed in previous
sections. Another global symmetry the field theory possesses (in a fixed Minkowski
background metric, say) is $SO(d-1)$ rotational invariance. In the bulk this symmetry symmetry also appears, but it is gauged. Namely, it is part of the diffeomorphism invariance of general
relativity: we can act on our AdS spacetime with a local $SO(d-1)$ rotation and we simply
obtain AdS again in a different coordinate system. This observation suggests the general correspondence
\be
\begin{array}{c}
\text{Global symmetry (field theory)} \\
d \; \text{spacetime dimensions}
\end{array}
\quad \leftrightsquigarrow \quad
\begin{array}{c}
\text{Gauged symmetry (gravity)} \\
d+1 \; \text{spacetime dimensions.}
\end{array}
\ee
Another fact that makes the above correspondence natural is that gauge symmetries
include the subgroup of `large' gauge symmetries, that is, symmetries which
act nontrivially as global symmetries on the boundary of spacetime. In an AdS/CFT
framework we can precisely identify this global subgroup of the bulk gauged symmetry
as the global symmetry group of the dual field theory.

To describe the physics of the global $U(1)$ symmetry we should therefore add
a Maxwell field to our bulk spacetime. The minimal bulk action is thus
Einstein-Maxwell theory\footnote{There is an interesting very simple
extension of Einstein-Maxwell theory, which is to include a coupling between
the field strength and the Weyl tensor, e.g. \cite{Ritz:2008kh}.}
\be\label{eq:einsteinmaxwell}
S = \int d^{d+1}x \sqrt{-g} \left[\frac{1}{2 \kappa^2} \left(R + \frac{d (d-1)}{L^2}
\right)  - \frac{1}{4 g^2} F^2 \right] \,.
\ee
Here $F=dA$ is the electromagnetic field strength.
At this point, without any charged matter in the bulk, the Maxwell coupling $g^2$
could be absorbed in the Maxwell field. We introduce the coupling
now for future convenience.

In thermal equilibrium there are two new background scales we can now introduce
in the field theory in a way that preserves rotational symmetry.
One is a chemical potential $\mu = A_{(0)t}$
and the other, which only preserves rotational symmetry in 2+1 dimensions,
is a background magnetic field $B = F_{(0)xy}$.
As we saw previously with the temperature, $T$, these new scales
must cause deformations away from a pure AdS spacetime as we move away from
the boundary and into
the IR region. Also as with the metric in (\ref{eq:asymptoticmetric}), the background
Maxwell potential of the field theory is read off from the boundary value
of the bulk Maxwell potential
\be\label{eq:asymptoticgauge}
A_\mu(r) = A_{(0)\mu} + \cdots \quad \text{as} \quad r \to 0 \,.
\ee
As with the metric, we are pulling back to the boundary so that there is no $A_r$ component

We now need to search for solutions to Einstein-Maxwell theory of the form
(\ref{eq:ansatz}) together with a nonvanishing Maxwell field
\be\label{eq:Aansatz}
A = A_t(r) dt + B(r) x \, dy \,.
\ee
The second term in this expression will break the isotropy of the field theory unless
there are only two spatial dimensions (i.e. $d=3$). We will firstly consider the case
of no magnetic field in arbitrary dimensions and then will consider the $d=3$ case
separately. The Einstein equations of motion are
\be\label{eq:EMeom}
R_{\mu\nu} - \frac{R}{2} g_{\mu\nu} - \frac{d(d-1)}{2 L^2} g_{\mu\nu} = \frac{\kappa^2}{2 g^2}
\left(2 F_{\mu \sigma} F_{\nu}{}^{\sigma} - \frac{1}{2} g_{\mu\nu} F_{\sigma \rho} F^{\sigma \rho} \right) \,,
\ee
while the Maxwell equation is
\be
\nabla_\mu F^{\mu \nu} = 0 \,.
\ee
Looking for solutions to these equations of the form (\ref{eq:ansatz}) and (\ref{eq:Aansatz})
one finds the Reissner-Nordstrom-AdS black hole
\be\label{eq:RNads}
ds^2 = \frac{L^2}{r^2} \left(- f(r) dt^2 + \frac{dr^2}{f(r)} + dx^i dx^i \right) \,,
\ee
where
\be\label{eq:RNf}
f(r) = 1 - \left(1 + \frac{r_+^2 \mu^2}{\gamma^2} \right) \left(\frac{r}{r_+}\right)^d +
\frac{r_+^2 \mu^2}{\gamma^2} \left(\frac{r}{r_+}\right)^{2(d-1)} \,.
\ee
In this expression we defined
\be\label{eq:gamma}
\gamma^2 = \frac{(d-1) g^2 L^2}{(d-2) \k^2} \,,
\ee
which is a dimensionless measure of the relative strengths of the gravitational
and Maxwell forces.
The scalar potential is
\be\label{eq:At}
A_t = \mu \left[1 - \left(\frac{r}{r_+}\right)^{d-2} \right] \,.
\ee
The constant term in $A_t$ cannot be chosen arbitrarily. This is because there is a bifurcate Killing horizon of the Killing vector $\pa/\pa_t$ at $r=r_+$, and so the one form $A$ will not be well defined there unless $A_t$ vanishes \cite{Kobayashi:2006sb}.

By comparing with (\ref{eq:asymptoticgauge}) we see that the chemical potential is $\mu$.
The temperature can be found as previously by analytic continuation to a Euclidean signature solution
(note that $A_t$ becomes pure imaginary under this process). The periodicity of imaginary time
is again given by the first relation in (\ref{eq:periodic}) and hence we obtain the temperature
\be\label{eq:RNT}
T = \frac{1}{4 \pi r_+} \left(d -  \frac{(d-2) r_+^2 \mu^2}{\gamma^2} \right) \,.
\ee
Periodically identifying the time coordinate gives another reason to enforce
that $A_t$ vanish at $r=r_+$. If $A_t(r_+)$ were finite one could obtain a finite Wilson loop $\oint A$
around the vanishing Euclidean time circle, indicating that the gauge connection is singular.

An important feature of (\ref{eq:RNT}) relative to the zero chemical potential case (\ref{eq:T}) is that
the temperature can become zero continuously. Recall that with no chemical potential we could scale out $r_+$ and hence all nonzero temperatures were equivalent. Here we can again scale out $r_+$, but we are left with the scale set by $\mu$ and therefore with the dimensionless ratio $T/\mu$, which can be continuously taken to zero. In a scale invariant theory all dimensionless equilibrium quantities can only depend on temperature and chemical potential through this ratio -- there are no other scales.

The thermodynamic potential is obtained from evaluating the Euclidean action on the analytically continued solution, just as in the previous section.
The action is again (\ref{eq:fullaction}) together with the Maxwell $F^2$ term (which appears with a $+$ sign in the Euclidean action $S_E$). No additional counterterms
are necessary because the Maxwell field falls off sufficiently quickly near the boundary in the dimensions of interest ($d \geq 3$). We are working in the grand canonical ensemble\footnote{To work instead in the canonical ensemble, fixed charge density $\rho$, we should add a boundary term to the Euclidean action: $\Delta S_E = \frac{1}{g^2} \int_{r\to 0} d^dx \sqrt{\gamma} n^a F_{ab} A^b$. This term changes the variational problem so that one must keep the field strength $n^a F_{ab}$ fixed at the boundary rather than the potential $A_a$. It can be seen to imply the standard thermodynamic relation $F = \Omega + \mu Q$. Here $Q = \rho V_2$ is the total charge.}, with $\mu$ fixed, and thus use the notation $\Omega = - T \log Z$ where $Z$ is the partition function defined by the gravitational path integral (\ref{eq:partition}).
One finds
\be\label{eq:Free2}
\Omega = - \frac{L^{d-1}}{2 \kappa^2 r_+^d} \left(1 +  \frac{ r_+^2 \mu^2}{\gamma^2}  \right) V_{d-1} = {\mathcal{F}}\left(\frac{T}{\mu}\right) V_{d-1} T^d\,,
\ee
where the function ${\mathcal{F}}$ is easily obtained by solving (\ref{eq:RNT}) for $r_+$.
This function is a nontrivial output from AdS/CFT. At low temperatures we have
$\Omega \sim a \mu^d + b \mu^{d-1} T + c \mu^{d-2} T^2 +  \cdots$. In particular, the leading nontrivial temperature dependence of the thermodynamic potential is linear\footnote{
It follows that the entropy remains finite at zero temperature. This is disturbing from a field
theory perspective. We have made no assumptions about the theory being supersymmetric and furthermore these black holes would not be supersymmetric at extremality in any case. It has been argued that the entropy of Reissner-Nordstrom black holes is not continuous in the zero temperature limit and that strictly extremal black holes have zero entropy, contradicting the area-entropy relation \cite{Hawking:1994ii,Teitelboim:1994az}. Discussion of this tension can be found in \cite{Horowitz:1996qd, Carroll:2009ma}. A (putative) discontinuity as $T \to 0$ does not evade the field theory discomfort, as arbitrarily low temperatures would still have macroscopic entropy. One possible resolution is that this macroscopic entropy is a consequence of the semiclassical (large $N$) limit. Perhaps tunneling interactions between the degenerate ground states, lifting the degeneracy, are suppressed at large $N$? Alternatively, perhaps extremal Reissner-Nordstrom black holes are never stable ground states of consistent gravity theories \cite{ArkaniHamed:2006dz}.}, as is the leading low temperature dependence of the heat capacity $c =  T \pa S/\pa T$.
At high temperatures one finds $\Omega \sim \mu^{2(d-1)}/T^{d-2} + \cdots$.
Recall that $L^{d-1}/\kappa^2$ is dimensionless (the AdS radius in Planck units) and scales like a positive power of $N$. We can again check that the entropy following from (\ref{eq:Free2}) is $2\pi/\kappa^2$ times the area of the event horizon.

Adding a magnetic field in the case $d=3$ is straightforward. Taking the ansatz (\ref{eq:Aansatz})
one finds the dyonic Reissner-Nordstrom-AdS$_4$ solution. The metric is again of the form (\ref{eq:RNads})
but the function appearing is now
\be\label{eq:fmuB}
f(r) = 1 - \left(1 + \frac{(r_+^2 \mu^2 + r_+^4 B^2)}{\gamma^2} \right) \left(\frac{r}{r_+}\right)^3 +
\frac{(r_+^2 \mu^2 + r_+^4 B^2)}{\gamma^2} \left(\frac{r}{r_+}\right)^{4} \,,
\ee
and the gauge potential is
\be\label{eq:Adyonic}
A = \mu \left[1 - \frac{r}{r_+} \right] dt + B x \, dy \,.
\ee
Note that whereas the chemical potential $\mu$ has mass dimension one in field theory,
the background magnetic field has mass dimension 2.

It is clear from (\ref{eq:fmuB}) that the temperature will simply be given by (\ref{eq:RNT}) with the replacement: $r_+^2 \mu^2 \to r_+^2 \mu^2 + r_+^4 B^2$. That is
\be\label{eq:TBM}
T = \frac{1}{4 \pi r_+} \left(3 -  \frac{r_+^2 \mu^2}{\gamma^2} - \frac{r_+^4 B^2}{\gamma^2} \right) \,.
\ee
The thermodynamic potential on the other
hand is found to depend more asymmetrically on the chemical potential and magnetic field
\be
\Omega = - \frac{L^2}{2 \kappa^2 r_+^3} \left(1 + \frac{r_+^2 \mu^2}{\gamma^2} -  \frac{3 r_+^4 B^2}{\gamma^2} \right) \,.
\ee
Two quantities of immediate interest are the charge density
\be\label{eq:rho}
\rho = - \frac{1}{V_2} \frac{\pa \Omega}{\pa \mu} = \frac{2 L^2}{\k^2} \frac{\mu}{r_+ \g^2} \,,
\ee
and the magnetisation density
\be
m  = - \frac{1}{V_2} \frac{\pa \Omega}{\pa B} = - \frac{2 L^2}{\k^2} \frac{r_+ B}{\gamma^2} \,.
\ee
In all three of these expressions, $r_+$ should be thought of as a function of $\mu,B$ and $T$ via (\ref{eq:TBM}). By scale invariance of the dual theory, dimensionless quantities can only depend on the ratios $T/\mu$ and $T^2/B$. As noted in \cite{Hartnoll:2008vx}, the absence of other scales or small couplings in the field theory will imply that the magnetic susceptibility $\chi = \pa^2\Omega/\pa B^2$ will typically be of order $1/T$. This is very large compared to many `standard' systems such as a free electron gas. One can therefore expect quantum critical theories to be strongly magnetic.

We conclude this section with a comment concerning conformal field theories with background magnetic fields. The Ward identity implementing conformal invariance implies a modified version of the usual tracelessness of the energy momentum tensor:
\be\label{eq:trace}
-\epsilon + 2 P = 2 m B \,, 
\ee
where $\epsilon$ is the energy density and the pressure is related to the thermodynamic potential by $P V_2 = - \Omega$. The thermodynamic pressure differs from the expectation value of the energy momentum tensor by $\langle T^{xx} \rangle = P - m B$. One can check that the thermodynamic quantities obtained from the black hole satisfy the required relation
\be
E + PV = ST + \mu Q \,,
\ee
with the total charge $Q = \rho V_2$. 

\subsection{Relevant operators}

The last two sections have discussed deformations away from scale invariance due to temperature, chemical potential and a background magnetic field. This section will consider another way to deform the low energy physics: perturb the critical action by a relevant operator. In fact we will see that within AdS/CFT such deformations are treated in exactly the same way as the cases we have just studied (i.e. temperature, chemical potential,...), providing a conceptual unification that is not immediately apparent in standard field theory treatments.
While there is a very large literature on renormalisation group flows in AdS/CFT, this technology has not at the time of writing been brought to bear on questions of condensed matter interest. We shall therefore make some more formal developments here.

To see how to introduce relevant operators, it is useful to recast our discussion of
finite temperature and chemical potential in terms of sources in the field theory.
Recall that in discussing the temperature, we argued that the boundary value
of the bulk metric gave the (nondynamical) background metric of the field theory, as in
(\ref{eq:asymptoticmetric}).
A background metric can be thought of as a source for the energy-momentum
tensor of the field theory, $T^{\mu\nu}$.
Recall that the (field theory) energy momentum tensor is defined as
\be
T^{\mu\nu} = \frac{\delta S}{\delta g_{(0) \mu\nu}} \,.
\ee
Suppose we perturb the bulk metric so that its boundary
value, as defined in (\ref{eq:asymptoticmetric}),
becomes $g_{(0)} + \d g_{(0)}$. The change in the field theory
action is $\delta S = \int d^dx \sqrt{-g_{(0)}}
\d g_{(0) \mu \nu} T^{\mu \nu}$. Requiring that the equality of 
bulk and field theory partition functions remains valid as the
background metric is shifted therefore leads to
\be\label{eq:deltag}
Z_\text{bulk}[g \to g_{(0)} + \d g_{(0)}] = \langle \exp \left( i \int d^dx \sqrt{-g_{(0)}}
\d g_{(0) \mu \nu} T^{\mu \nu}\right) \rangle_\text{F.T.} \,.
\ee
This equation is a useful re-expression of the statement that the boundary
value of the bulk metric gives the background field theory metric.

The second case we encountered was the statement that the
boundary value of a bulk Maxwell field gave the (nondynamical)
background field for a global $U(1)$ symmetry of the boundary.
Let us call this the electromagnetic $U(1)$ for verbal convenience.
A background electromagnetic field is a source for the current $J^\mu$.
As with the energy momentum tensor, the current is defined by
the change in the action due to a background field
\be
J^\mu = \frac{\delta S}{\delta A_{(0) \mu}} \,.
\ee
Therefore the change in the field theory action upon introducing
a small background field $\delta A_{(0)}$ is
$\delta S = \int d^dx \sqrt{-g_{(0)}} \d A_{(0) \mu} J^{\mu}$. Imposing the
identification of bulk and field theory partition functions, with the
boundary value for the bulk gauge field giving the background
field in the dual field theory, as defined in (\ref{eq:asymptoticgauge}),
requires
\be\label{eq:deltaA}
Z_\text{bulk}[A \to \d A_{(0)}] = \langle \exp \left( i
\int d^dx \sqrt{-g_{(0)}} \d A_{(0) \mu} J^{\mu}\right) \rangle_\text{F.T.} \,.
\ee
Once again, this last equation is simply a re-expression of the
identification of $A_{(0)}$ as a background field in the field theory.

The results (\ref{eq:deltag}) and (\ref{eq:deltaA}) indicate a correspondence
between specific operators in field theory, $\{J^\mu, T^{\mu\nu}\}$, and
certain fields in the bulk spacetime, $\{A_\mu, g_{\mu\nu}\}$.
The correspondence states that the boundary value of the bulk field gives
a background source for the corresponding dual field theory operator. This statement
suggests the following generalisation
\be
\begin{array}{c}
\text{operator ${\mathcal{O}}$} \\
\text{(field theory)}
\end{array}
\quad \leftrightsquigarrow \quad
\begin{array}{c}
\text{dynamical field $\phi$} \\
\text{(bulk)\,,}
\end{array}
\ee
such that
\be
\label{eq:deltaO}
Z_\text{bulk}[\phi \to \d \phi_{(0)}] = \langle \exp \left( i
\int d^dx \sqrt{-g_{(0)}} \d \phi_{(0)} \ocal\right) \rangle_\text{F.T.} \,.
\ee
This relationship is usually taken as the backbone of the AdS/CFT
correspondence \cite{Witten:1998qj, Gubser:1998bc}.
We will define $\delta \phi_{(0)}$ shortly in terms of the boundary behavior
of the bulk field $\phi$. First, we should note that (\ref{eq:deltaO})
describes nothing other than the perturbation of the scale invariant
field theory Lagrangian by the coupling $\delta \phi_{(0)} \ocal$.
If the operator $\ocal$ is relevant than this perturbation will generate
a renormalisation group flow into the IR.

Assume for simplicity that the operator $\ocal$ is a Lorentz scalar.
Then we expect the spacetime description of the renormalisation
group flow to break scale but not Lorentz invariance. This constrains
the geometry more than in the cases of finite temperature and chemical
potential. We are lead to the metric
\be
ds^2 = L^2 \left(\frac{dr^2}{r^2} + \frac{h(r) (-dt^2 + dx^i dx^i)}{r^2} \right) \,,
\ee
together with a profile for the scalar field
\be
\phi = \phi(r) \,.
\ee
We have set $g(r)=1$ and $f(r)=h(r)$ in (\ref{eq:ansatz}), using
Lorentz invariance and a choice of radial coordinate. To find solutions
of this form we need equations
of motion for the scalar field. A minimal Einstein-scalar action is
\be\label{eq:einsteinscalar}
S = \int d^{d+1}x \sqrt{-g} \left[\frac{1}{2 \kappa^2} \left(R + \frac{d (d-1)}{L^2}
\right)  - \frac{1}{2} (\nabla \phi)^2 - V(\phi) \right] \,.
\ee
There is a large ambiguity in this action for a scalar field, namely
the form of the potential.  Unfortunately, the existence and nature of
solutions do depend significantly on the potential
$V(\phi)$. It is also something of a simplification to assume that the flow driven
by the operator $\ocal$ does not generate other relevant operators in the action. These
would appear in the bulk as further (scalar and other) fields coupling to $\phi(r)$. Of course,
renormalisation flow can also be combined with finite temperature and
chemical potential by using the full Einstein-Maxwell-scalar action.

We will not consider explicit examples of $V(\phi)$ and the corresponding solutions
for $\{h(r),\phi(r)\}$ here. A nice discussion with analytic solutions for cases with one scalar
field, in which the renormalisation group flow leads to a new IR fixed point,
can be found in, for example, \cite{Skenderis:1999mm}. It is also possible for
the flow to lead to a singularity in the IR region of the spacetime. Such
singularities are discussed in \cite{Gubser:2000nd}.

If the scalar field falls off sufficiently quickly (i.e. goes to zero or a constant)
near the boundary, then its equation of
motion can be linearised and the backreaction on the metric becomes negligible. The spacetime
is therefore asymptotically AdS,
so that $h(r) \to 1$ as $r \to 0$, and at small $r$ we have
\be\label{eq:larger}
r^2 \pa^2_r \phi - (d-1) r \pa_r \phi = (L m)^2 \phi \,,
\ee
where $m^2$ is the mass squared of the field, $V(\phi) = \half m^2 \phi^2 + \cdots$.
It follows that the near boundary behaviour of the scalar field is
\be\label{eq:asymptoticphi}
\phi(r) = \left(\frac{r}{L}\right)^{d-\Delta} \phi_{(0)} + \cdots \quad \text{as} \quad r \to 0 \,,
\ee 
where $\Delta$ is one of the solutions of
\be\label{eq:2sols}
(L m)^2 = \Delta (\Delta - d) \,.
\ee
In general this equation will have two solutions: $\Delta$ and $d-\Delta$.
We can now show \cite{Witten:1998qj} that the scaling dimension of the dual
operator is
\be\label{eq:dim}
\text{dim}[{\mathcal{O}}] = \Delta \,.
\ee
This follows from the action of scaling on the spacetime: $\{r,t,x^i\} \to \lambda \{r,t,x^i\}$
and $\phi_{(0)} \to \lambda^{\Delta-d} \phi_{(0)}$, together with the basic relation
(\ref{eq:deltaO}). The expression (\ref{eq:dim}) places a constraint on $\Delta$ because
the scaling dimension
must be greater than the CFT unitary bound, $\Delta \geq (d-2)/2$. We can choose either
solution to (\ref{eq:2sols}) to be $\Delta$ subject to this constraint \cite{Klebanov:1999tb}.
Equation (\ref{eq:asymptoticphi}) then defines $\phi_{(0)}$ as appearing in (\ref{eq:deltaO}).
Note that the term shown in (\ref{eq:asymptoticphi}) may not be the leading term as $r \to 0$,
depending on the choice of solution to (\ref{eq:2sols}).

The operator $\ocal$ will be a relevant or marginal deformation of the theory
if its dimension $\Delta \leq d$. From (\ref{eq:dim}) we see that this implies that
$d-\Delta \geq 0$, and therefore the bulk field $\phi$ indeed goes to a constant or zero
at the boundary $r \to 0$. This is precisely the condition we required in order for our linearised
analysis about an asymptotically AdS space to be consistent. To rephrase: relevant operators
can be turned on in the theory without destroying the asymptotically AdS region of the metric.
This is dual to the fact that relevant operators do not destroy the UV fixed point of the field theory.
In contrast, deforming the theory by irrelevant operators will change the UV structure
of the field theory and of the bulk gravity solution, taking us outside the
best understood AdS/CFT framework (one might still attempt to define the theory at a UV cutoff,
which would correspond to ending the bulk spacetime at some finite $r=\epsilon$).

\subsection{Expectation values}
\label{sec:vevs}

A deformation of the critical theory by any of the processes we have just considered
will generically result in expectation values for the operators involved. We can
compute expectation values using the basic relation (\ref{eq:deltaO}). That relation
implies
\be\label{eq:vev}
\langle \ocal \rangle = - i \frac{\delta Z_\text{bulk}[\phi_{(0)}]}{\delta \phi_{(0)}} \,
\stackrel{N\to\infty}{=} \, \frac{\delta S[\phi_{(0)}]}{\delta \phi_{(0)}} \,,
\ee
where we are dropping factors of $Z_\text{bulk}[0]$ and we have taken the semiclassical (large $N$) limit $Z_\text{bulk} = e^{i S}$. It is useful to formally manipulate this expression further. There is a certain flavour of Hamilton-Jacobi theory in (\ref{eq:vev}), with the radial direction of the bulk spacetime playing the role of time (see e.g. \cite{de Boer:1999xf}). Recall that, on shell, the derivative of the action with respect to an endpoint value of a coordinate is the corresponding momentum. We should also take into account the possibility of boundary terms in the action, so that
\be\label{eq:momentum}
\frac{\delta S[\phi_{(0)}]}{\delta \phi_{(0)}} = \lim_{r \to 0} \left(- \frac{\delta S[\phi_{(0)}]}{\delta \pa_r \phi_{(0)}} + \frac{\delta S_\text{bdy.}[\phi_{(0)}]}{\delta \phi_{(0)}} \right) \equiv \lim_{r \to 0} \Pi[\phi_{(0)}] \,.
\ee
With a slight abuse of the notation introduced in (\ref{eq:asymptoticphi}), we mean $\pa_r\phi_{(0)} = \pa_r [(r/L)^{\Delta-d} \phi]$. The minus sign in the first term is because the boundary is at $r=0$,
the lower limit of integration.

For the bulk metric and Maxwell field we have already discussed the boundary terms that are needed to obtain a finite on shell action. For scalars fields dual to operators with conformal dimension $\Delta$
boundary terms must also be added. A clear and more detailed discussion than we will give can be found in \cite{Marolf:2006nd} for the range of masses in which both falloffs are allowed. There are two cases. If the falloff in (\ref{eq:asymptoticphi}) is the slower of the two falloffs near the boundary (i.e. $d-\Delta <  \Delta$), then the scalar boundary action to be added to (\ref{eq:einsteinscalar}) is
\be\label{eq:scalarboundary}
S_\text{bdy.} = \frac{\Delta-d}{2 L} \int_{r \to 0} d^dx \sqrt{\gamma} \phi^2 \,.
\ee
If instead the falloff defining the boundary value of the field is the faster falloff (this is only possible if $d/2 \geq \Delta \geq (d-2)/2$, the second restriction coming from the CFT unitary bound as we mentioned above) then the boundary action must be taken to be
\be\label{eq:scalarboundary2}
S_\text{bdy.} = -\int_{r \to 0} d^dx \sqrt{\gamma} \left(\phi n^\mu \nabla_\mu \phi + \frac{\Delta}{2 L} \phi^2\right) \,.
\ee
With these expressions at hand we can obtain a general result for the expectation value.
Write the full behaviour of the scalar field near the boundary as
\be\label{eq:phi1}
\phi(r) = \left(\frac{r}{L} \right)^{d-\Delta} \phi_{(0)} +
\left(\frac{r}{L} \right)^{\Delta} \phi_{(1)} + \cdots \quad \text{as} \quad r \to 0 \,. 
\ee
We are assuming that there are no terms in the power series expansion in between the two shown.
This holds for several cases of interest, but not in general. If such terms do arise, the renormalisation of the action is more complicated than what we are doing here, see e.g. \cite{Skenderis:2002wp} for a systematic treatment.
Substituting the expansion (\ref{eq:phi1}) into the formulae we have given for the action and expectation value, one straightforwardly derives that
\be\label{eq:vev2}
\langle \ocal \rangle = \frac{2 \Delta-d}{L} \phi_{(1)} \,.
\ee
This is often summarised as saying that the `non-normalisable' falloff $\phi_{(0)}$ gives the source
whereas the `normalisable' falloff $\phi_{(1)}$ gives the expectation value.
The result (\ref{eq:vev2}) can
be applied to other fields such as components of the metric and Maxwell fields. It is necessary
to firstly make sure that the kinetic term is written in the same form as that of the scalar field in (\ref{eq:einsteinscalar}).

As an example we can (re-)compute the charge density
of the field theory placed at a finite chemical potential. The chemical potential $\mu = A_{t(0)}$ is a source for the charge density $J^t$. Defining $\phi = r A_t$
one can obtain an action for an effective scalar field from the Maxwell action. The conformal dimension of the charge density is $\Delta = d-1$. It follows from (\ref{eq:At}), (\ref{eq:vev2}) and comparing the overall normalisation of the actions (\ref{eq:einsteinmaxwell}) and (\ref{eq:einsteinscalar}) that
\be
\langle J^t \rangle = \frac{\mu (d-2) L^{d-3}}{g^2 r_+^{d-2}} \,.
\ee
Putting $d=3$ in this expression recovers our previous result (\ref{eq:rho}).

One can similarly obtain directly the energy density and pressure as the expectation values of
$\langle T^{tt} \rangle$ and $\langle T^{tx} \rangle$. In later sections we will use (\ref{eq:vev2})
to obtain expectation values for currents and scalar condensates.

\subsection{Dissipative dynamics close to equilibrium}

So far we have discussed time independent, homogeneous backgrounds. We have seen how
a solution to the classical bulk equations of motion determines the thermodynamic
variables of a dual field theory. In particular, given time independent sources (e.g. temperature,
chemical potential, relevant deformations of the action) the AdS/CFT dictionary
determines the response of the theory (e.g. energy density, charge density, vacuum
expectation values).

The next step is to consider small space and time dependent perturbations about
equilibrium. This is the domain of linear response theory and relates
directly to important experimental processes such as transport and spectroscopy.
The basic object we wish to compute is the retarded Green's function,
which is defined to linearly relate sources and corresponding
expectation values. In general there will be coupling between different operators,
so we can write (in frequency space)
\be\label{eq:greens}
\delta \langle \ocal_A \rangle(\w,k) = G^R_{\ocal_A \ocal_B}(\w,k) \delta \phi_{B(0)}(\w,k) \,.
\ee
We will review some basic aspects of the theory of retarded Green's functions in
a later section. In this section we will discuss how to compute
the Green's function appearing in (\ref{eq:greens}) via the AdS/CFT
correspondence.

In the geometry dual to the field theory at equilibrium there will be profiles
for the various bulk fields of interest, $\phi_A(r)$, with corresponding
boundary values $\phi_{A(0)}$. If we wish to perturb the boundary value,
then in order to satisfy the bulk equations of motion we will need to perturb
the entire bulk field:
\be\label{eq:perturbation}
\phi_A(r) \to \phi_A(r) + \delta \phi_A(r) e^{- i \w t + i k \cdot x} \,.
\ee
The equation of motion for $\delta \phi_A(r)$ is obtained by
substituting the perturbation (\ref{eq:perturbation}) into the bulk
equations of motion and linearising. We will give an example shortly.
As well as the equation we need boundary conditions.
Asymptotically we impose
\be
\delta \phi_A(r) = r^{d-\Delta} \delta \phi_{A(0)} + \cdots \quad \text{as} \quad r \to 0 \,.
\ee
We must then turn to the boundary condition in the interior, which is more involved.

We will consider circumstances in which the interior boundary condition is imposed
at a regular future horizon in the spacetime. This is always the case at finite temperature
and is often true at zero temperature also (e.g. pure AdS space has such a horizon in the
conformal frame we are using). A future horizon
is defined for us as a null surface beyond which events cannot causally propagate to the
asymptotically AdS region of the spacetime (the boundary region). For the types of spacetime we are considering,
the horizon is characterised by $g_{tt} \to 0$.
For instance, in the black hole metric
(\ref{eq:schwarzads}) the horizon is at $r=r_+$ while in AdS space the horizon
is at $r \to \infty$.  On a future horizon, regularity requires that modes are ingoing.
For a horizon at $r=r_+$ with a nonzero temperature this means\footnote{There is a linguistic
convention here: `ingoing' means into the horizon, not into the spacetime. To see that
ingoing is necessary for regularity, transform to Kruskal coordinates $\{\rho,\tau\}$
near the horizon: $\rho \pm \tau = e^{(8 \pi \log(r-r_+) \pm 2\pi T t)}$. The metric in Kruskal
coordinates is regular on the future ($\rho = \tau$) and past ($\rho=-\tau$) horizons.
On the future horizon an ingoing mode behaves as $(\rho+\tau)^{-i\w/(2\pi T)}$
and so is regular as $\rho \to \tau$ whereas an outgoing mode behaves as
$(\rho-\tau)^{i\w/(2 \pi T)}$ which oscillates with unbounded increasing frequency as the future horizon is approached.}
\be\label{eq:nonzeroT}
\delta \phi_A(r) = e^{- i 4 \pi \w/T \log(r-r_+)} \left( C_A + \cdots \right) \quad \text{as} \quad r \to r_+ \,.
\ee
In this expression $T$ is the temperature defined by the inverse of the length of the
Euclidean time circle needed to make the Euclidean geometry regular, as in
(\ref{eq:periodic}) above, and $C_A$ are constants. Clearly (\ref{eq:nonzeroT}) will break
down at zero temperature. For a zero temperature horizon, ingoing solutions satisfy
\be\label{eq:zeroT}
\delta \phi_A(r) = e^{i \w L^2_{2}/(r-r_+)} \left( C_A + \cdots \right) \quad \text{as} \quad r \to r_+ \,.
\ee
In this expression $L_2$ is the radius of the $AdS_2$ near horizon region and $r$
is such that the $\{t,r\}$ part of the near horizon metric is
\be\label{eq:ads2}
ds^2_{\{t,r\} \,\text{N.H.}} = - \frac{(r-r_+)^2}{L_2^2} dt^2 + \frac{L_2^2}{(r-r_+)^2} dr^2  \,.
\ee
Note that in (\ref{eq:zeroT}) the coefficient $C_A$ may have $r$ dependence that is subleading as $r \to r_+$.

In both the finite and zero temperature cases,
the choice of ingoing boundary condition breaks time reversal symmetry.
This is how we can obtain the retarded rather than, say, the advanced propagator.
Time reversal symmetry breaking was forced onto us by regularity because we are setting the boundary conditions on the future rather than past horizon. See figure 5.
\begin{figure}[h]
\begin{center}
\includegraphics[height=5cm]{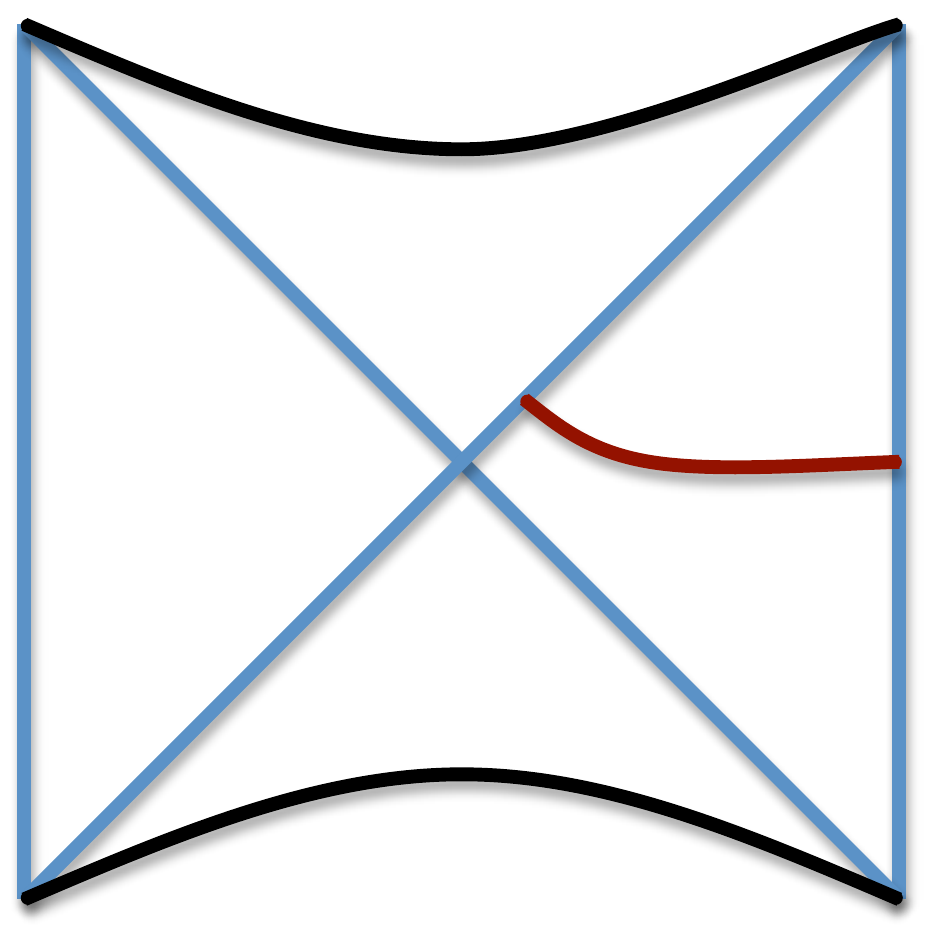}
\end{center}
\caption{Eternal black hole in AdS together with an initial spacelike slice for evolution in the right asymptotic region. Straight blue lines denote the asymptotically AdS boundaries and the future and past event horizons. Black lines are singularities. The red line denotes a possible time slice ending on the future event horizon.}
\end{figure}
This is a sensible thing to do because an initial spacelike slice ending on the past horizon does not provide a good Cauchy surface from which to evolve initial data; we would furthermore need to specify what is coming out of the `white hole'.
Furthermore, the presence of a future horizon in the spacetime is directly
connected to the possibility of dissipation. Energy flux across the horizon is lost
to an asymptotic observer. The connection between (future) horizons and dissipation
was at the heart of the `membrane paradigm' for black holes, see \cite{Damour:2008ji} for a review of
the original work.

Given a mode $\delta \phi_A$ satisfying the required boundary conditions and linearised
equations of motion, we immediately obtain from (\ref{eq:greens}) that
\be\label{eq:getG}
G^R_{\ocal_A \ocal_B} = \left. \frac{\delta \langle \ocal_A \rangle}{\delta \phi_{B(0)}} \right|_{\delta \phi = 0} = \left.  \lim_{r \to 0} \frac{\delta \Pi_A}{\delta \phi_{B(0)}} \right|_{\delta \phi = 0}  =  \frac{2 \Delta_A-d}{L} \frac{\delta \phi_{A(1)}}{\delta \phi_{B(0)}} \,.
\ee
Here the $\d$s denote functional derivatives, with $\phi_{B(0)}$ extended into the space via
$\phi_{B(0)}(r) = (r/L)^{\Delta_B-d} \phi_B(r)$. We have dropped the explicit $\w, k$ dependence.
The `momentum' $\Pi_A$ was defined in
(\ref{eq:momentum}), $\phi_{A(1)}$ was defined in (\ref{eq:phi1}) and we have
further used various results from the previous subsection in a hopefully obvious way.
The final term in (\ref{eq:getG}) is the most transparent: the retarded Green's function is given directly by how the coefficient of the `normalisable' falloff depends (linearly) on the `non-normalisable' falloff. The middle expression, however, is more robust and would apply also to modes which could not be written as an effective scalar field. Furthermore, by considering the quantity $\delta \Pi_A/\delta \phi_{B (0)}$ as a function of $r$ and moving away from the boundary (in particular, taking $r$ right up to the horizon) it is possible to make direct contact with results from the black hole membrane paradigm. This leads to easy proofs of universality results \cite{Iqbal:2008by}.

A prescription to compute retarded Green's functions in AdS/CFT was first introduced in
\cite{Son:2002sd}. This prescription was subsequently shown to emerge from an AdS/CFT implementation of the Schwinger-Keldysh double contour formalism for finite temperature quantum field theory in \cite{Herzog:2002pc} and more formally in \cite{Skenderis:2008dh}. These approaches viewed the Green's function as a two point function.
More recent works have viewed the Green's function as a ratio of a one point function (expectation value) and a source; this allows the Green's function to be computed directly from the basic relation (\ref{eq:deltaO}) and avoids subtleties with analytic continuation and contact terms. We have followed this approach here. The equivalence with the two point function prescription was recently emphasised in \cite{Iqbal:2008by} and we have adopted some notation from that paper.

\subsection{Example: How to compute electrical and thermal conductivities}
\label{sec:conductivity}

In this subsection we shall illustrate the above concepts by
describing electric and heat currents in a theory with a
gravity dual. We shall work in a 2+1 dimensional theory (with $z=1$) and
for simplicity stay at zero momentum: $k=0$.

We will consider the theory at a finite chemical potential, and hence
finite charge density. It is this fact which allows the heat (energy) and
electric currents to mix, as any equation relating these two currents needs
a proportionality factor carrying units of charge. Therefore, Ohm's law
must be generalised to
\be\label{eq:EandT}
\left(
\begin{array}{c}
\langle J_x \rangle \\
\langle Q_x \rangle
\end{array}
\right) = 
\left(
\begin{array}{cc}
\sigma & \alpha T \\
\alpha T & \bar \kappa T
\end{array}
\right)
\left(
\begin{array}{c}
E_x \\
-(\nabla_x T)/T
\end{array}
\right) \,,
\ee
where the heat current $Q_x = T_{tx} - \mu J_x$.
We will explain the difference between $Q_x$ and $T_{tx}$
shortly. There are three conductivities: electrical ($\sigma$),
thermal ($\bar \kappa$) and thermoelectric ($\alpha$).

The absence of a background magnetic field
(or any other source of time reversal symmetry breaking)
implies, as we prove below, that
$G^R_{J_i J_j} = G^R_{J_j J_i}$,
and similarly for the heat currents. Isotropy
implies that an off diagonal (Hall) conductivity would
be antisymmetric in spatial indices, e.g. $\sigma^H_{ij} \sim
\epsilon_{ij}$. Given that conductivities are proportional
to Green's functions, as we will also see shortly, it follows
that Hall conductivities vanish in time symmetric
configurations.  This is why we have taken all the currents
and sources to point in the $x$ direction in (\ref{eq:EandT}).
 
We now need to relate the sources $E_x$ and $\nabla_x T$
to background values for a vector potential $\delta A_{x(0)}$
and a metric fluctuation $\delta g_{tx(0)}$. The former case is
straightforward. At zero momentum,
in terms of the background gauge potential
\be\label{eq:Ej}
E_j = i \w \delta A_{j(0)} \,.
\ee
A short argument will now show that furthermore a thermal gradient leads to
\be\label{eq:gj}
i \w \delta g_{tj(0)} = - \frac{\nabla_j T}{T} \,, \qquad \text{and} \qquad i \w \delta A_{j(0)} = \mu \,\frac{\nabla_j T}{T} \,.
\ee
Recall that the period of Euclidean time is $1/T$.
In order to keep track of all the factors of the temperature,
let us rescale the time so that there is no $T$ dependence in the period:
$t = \bar t/T$. With the new dimensionless time coordinate,
the metric has $g_{\bar t \bar t (0)} = - \frac{1}{T^2}$. We are taking
the original metric to be Minkowski space. It follows
that a small constant thermal gradient, $T \to T + x \nabla_x T$, implies
\be\label{eq:perturbation2}
\delta g_{\bar t \bar t (0)} = - \frac{2 x \nabla_x T}{T^3} \,.
\ee
We can endow all quantities with a time dependence $e^{-i\bar \w \bar t}$.
Recall that diffeomorphisms acting on the background fields give the fluctuations
\be\label{eq:diffeo}
\delta g_{ab (0)} = \pa_a \xi_b + \pa_b \xi_a \,, \qquad \delta A_{a (0)} = A_{b (0)} \pa_a \xi^b +
\xi^b \pa_b A_{a(0)} \,.
\ee
The field theory is invariant
under background gauge transformations, this is precisely the content
of field theory Ward identities.
We can add such pure gauge modes to our perturbation (\ref{eq:perturbation2}).
Taking $\xi_{\bar t} = i x \nabla_x T/\bar \w T^3$ and
$\xi_{x} = 0$, one obtains that after the gauge transformation $\delta g_{\bar t \bar t (0)} = 0$,
$\delta g_{x \bar t (0)} = i \nabla_x T/ \bar \w  T^3$ and
$\delta A_{x (0)} = - i \mu \nabla_x T/ \bar \w  T^3$.
Scaling back to the original dimensionful time $t$, we obtain (\ref{eq:gj}).

Combining (\ref{eq:Ej}) and (\ref{eq:gj}) we can see that the the source term
in the action becomes
\bea
\delta S  & = &  \int d^{d-1}x dt \sqrt{-g_{(0)}} \left (T^{t x} \delta g_{tx (0)} + J^x A_{x (0)} \right) \nonumber \\
& = & \int d^{d-1}x dt \sqrt{-g_{(0)}}  \left ((T^{t x} - \mu J^x) \frac{- \nabla_x T}{i \w T} + J^x \frac{E_x}{i\w} \right) \,.
\eea
Thus we see that the current sourced by a thermal gradient is $Q_x = T_{tx} - \mu J_x$, as we
claimed above.
Substituting (\ref{eq:Ej}) and (\ref{eq:gj}) into (\ref{eq:EandT})
gives
\be\label{eq:EandTw}
\left(
\begin{array}{c}
\langle J_x \rangle \\
\langle Q_x \rangle
\end{array}
\right) = 
\left(
\begin{array}{cc}
\sigma & \alpha T \\
\alpha T & \bar \kappa T
\end{array}
\right)
\left(
\begin{array}{c}
 i \w (\delta A_{x(0)} + \mu \delta g_{tx(0)}) \\
 i \w \delta g_{tx(0)}
\end{array}
\right) \,,
\ee
This linear relation between a source and an expectation value
makes it clear that the conductivities are nothing other than the retarded Green's functions
\be
\sigma(\w) = \frac{- i G^R_{J_x J_x}(\w) }{\w} \,, \quad \a(\w) T = 
\frac{- i G^R_{Q_x J_x}(\w) }{\w} \,,\quad \bar \kappa(\w) T = \frac{- i G^R_{Q_x Q_x}(\w) }{\w} \,.
\ee

From our previous discussion we know that in order to compute the
response of the theory to these small background fields via AdS/CFT
we need to solve the equations of motion of perturbations
$\delta A_x$ and $\delta g_{tx}$ in the bulk. These perturbations do not
source any other fields (this simplification occurs because we have
set the momentum $k=0$). The bulk action we will
use is the Einstein-Maxwell action (\ref{eq:einsteinmaxwell}). The background
solution is given by the 4 dimensional Reissner-Nordstrom-AdS black hole,
discussed around (\ref{eq:RNads}). Linearising the Einstein-Maxwell equations of motion
(\ref{eq:EMeom}) about this background one obtains the following two
independent equations
\bea\label{eq:gp}
\d g_{tx}' + \frac{2}{r} \delta g_{tx} + \frac{4 L^2}{\g^2} A_t' \delta A_x & = &  0 \,, \\
(f \delta A_x')' + \frac{\w^2}{f} \delta A_x + \frac{r^2 A_t'}{L^2} \left(\d g_{tx}' + \frac{2}{r} \delta g_{tx} \right) & = & 0\,,
\eea
with $f, A_t$ and $\gamma^2$ given below (\ref{eq:RNads}) above.
Note in particular that $A_t' = - \mu/r_+$ is a constant. We can easily obtain a decoupled
equation for $\delta A_x$
\be\label{eq:Ax}
(f \delta A_x')' + \frac{\w^2}{f} \delta A_x - \frac{4 \mu^2 r^2}{\g^2 r_+^2} \delta A_x =  0\ \,.
\ee
It is straightforward to check that solutions to this equation behave near the
boundary as
\be\label{eq:a0a1}
\delta A_x = \delta A_{x(0)} + \frac{r}{L} \delta A_{x(1)} + \cdots \quad \text{as} \quad r \to 0 \,.
\ee
Because this is a linear equation of one variable, we know that $A_{x(1)}$ will depend linearly on $A_{x(0)}$. We will need to find the ($\w$ dependent) coefficient of proportionality numerically.
However, the fact that the equation for $\delta g_{tx}$ is first order means that we can obtain two of the three conductivities in (\ref{eq:EandT}) without solving any differential equations.

We are going to implement equation (\ref{eq:getG}) in order to compute the Green's functions.
Rather than mapping the equations onto a problem for scalars, let us compute the `momenta' (\ref{eq:momentum}) directly from the full action.
Using the Einstein-Maxwell action (\ref{eq:einsteinmaxwell}) together with the gravitational counterterms in (\ref{eq:fullaction}) one finds
\bea
\Pi_{g_{tx}} & = & \frac{\delta S}{\delta g_{tx (0)}} = - \rho \, \delta A_{x (0)} + \frac{2 L^2}{\k^2 r^3} (1 - f^{-1/2}) \d g_{tx (0)} \,, \\
\Pi_{A_x} & = & \frac{\delta S}{\delta A_{x (0)}} = \frac{f \d A_{x(0)}'}{g^2} - \rho \, \delta g_{tx (0)} \,.
\eea
A few comments are in order. Firstly, we have introduced the charge density $\rho$
as defined in (\ref{eq:rho}). Secondly, in computing the derivatives of the action with respect to
$g_{tx(0)}$ one should be careful to first eliminate second derivative terms in the action by integrating
by parts. The resulting first derivative terms on the boundary are precisely cancelled by the Gibbons-Hawking boundary term in (\ref{eq:fullaction}). We also used (\ref{eq:gp}) to simplify $\Pi_{g_{tx}}$.
Finally, as above, we defined $g_{tx(0)}$ at finite $r$ by $g_{tx(0)} = r^2/L^2 \, g_{tx}$ and $A_{x(0)} = A_x$. 

In the limit $r \to 0$ and using (\ref{eq:getG}) and (\ref{eq:a0a1}), in particular $\delta A_{x (0)}' \to \delta A_{x (1)}/L$, we find
\be\label{eq:prelim}
\left(
\begin{array}{c}
\langle J_x \rangle \\
\langle T_{tx} \rangle
\end{array}
\right) = 
\left(
\begin{array}{cc}
\frac{1}{g^2 L} \frac{\delta A_{x(1)}}{\delta A_{x(0)}} & -\rho \\
-\rho & -\epsilon
\end{array}
\right)
\left(
\begin{array}{c}
 \delta A_{x(0)} \\
 \delta g_{tx(0)}
\end{array}
\right) \,,
\ee
where we introduced the energy density $\epsilon = - L^2 (1+r_+^2\m^2/\g^2)/\k^2/r_+^3$. The energy density is obtained by the fact, mentioned in section \ref{sec:mu} above, that conformality implies $\epsilon = -2 \Omega/V_2$. 

Comparing (\ref{eq:EandTw}) and (\ref{eq:prelim}) gives the conductivities
\be\label{eq:conduct}
\sigma(\w) = \frac{-1}{g^2 L} \frac{i}{\w} \frac{\d A_{x(1)}}{\d A_{x(0)}}  \,; \; \;  T \a(\w) = \frac{i \rho}{\w} - \mu \sigma(\w)\,; \; \; T \bar \kappa(\w) = \frac{i (\epsilon+P-2\mu \rho)}{\w} + \mu^2 \sigma(\w)\,.
\ee
We have not explained one step here: in the numerator in $\bar \kappa(\w)$ we added a $P$. This originates from a contact term that must be present due to translation invariance \cite{Hartnoll:2007ip}.
All that is left is to solve the differential equation (\ref{eq:Ax}) in order to obtain the electrical conductivity in (\ref{eq:conduct}). For numerical stability it is convenient to explicitly remove the near-horizon oscillations from $A_x$ by defining
\be
A_x(r) = f(r)^{-i4 \pi \w/T} S(r) \,,
\ee
where the temperature $T$ was defined in (\ref{eq:RNT}) above. Substituting into (\ref{eq:Ax})
we obtain a differential equation for $S$. Ingoing boundary conditions now amount to the requirement that near the horizon: $S = 1 + \a_1 (r-r_+) + \a_2 (r-r_+)^2 + \cdots$. The overall normalisation is not important as the equation is linear. Indeed we see in (\ref{eq:conduct}) that the conductivity is a ratio of two coefficients in the near-boundary expansion, so the overall normalisation will drop out.
The coefficients $\a_i$ are easily found by looking for Taylor series expansions of (\ref{eq:Ax}) at the horizon. We wish to numerically integrate the equation (\ref{eq:Ax}) from the horizon to the boundary.
The Taylor expansion at the horizon is necessary because the horizon is a singular point of the differential equation, so we cannot set the initial data exactly at the horizon. Therefore we must set the initial conditions a little away from the horizon. The essential lines of Mathematica code computing the conductivity will look something like
\bea
{\tt
{soln[\w\_ ]}} & := & {\tt{NDSolve[\{\text{AxEqn}[\w]==0,S[1-\epsilon]== Ser[1-\epsilon,\w],}} \nonumber \\
& & {\tt{ S'[1-\epsilon]==SerPrime[1-\epsilon,\w]\},S,\{r,\eta,1-\epsilon\}]}} \nonumber \\
{\tt{\sigma [\w\_] }}& := & {\tt{-I/\w \, S'[\eta]/S[\eta] \, /. \, soln[\w ][[1]][[1]]  }
}\nonumber
\eea
Here $\epsilon$ is small number setting the initial distance from the horizon and $\eta$ is a small
number determining the distance from the boundary at which the conductivity (\ref{eq:conduct}) is evaluated. The functions {\tt Ser} and {\tt SerPrime} are the Taylor series expansion at the horizon and the derivative thereof, respectively. In performing numerics it is generally convenient to set $L=1$ and furthermore to scale the horizon to $r_+=1$. However, one then needs to undo this scaling to recover physical units.

\begin{figure}[h]
\begin{center}
\includegraphics[height=4.7cm]{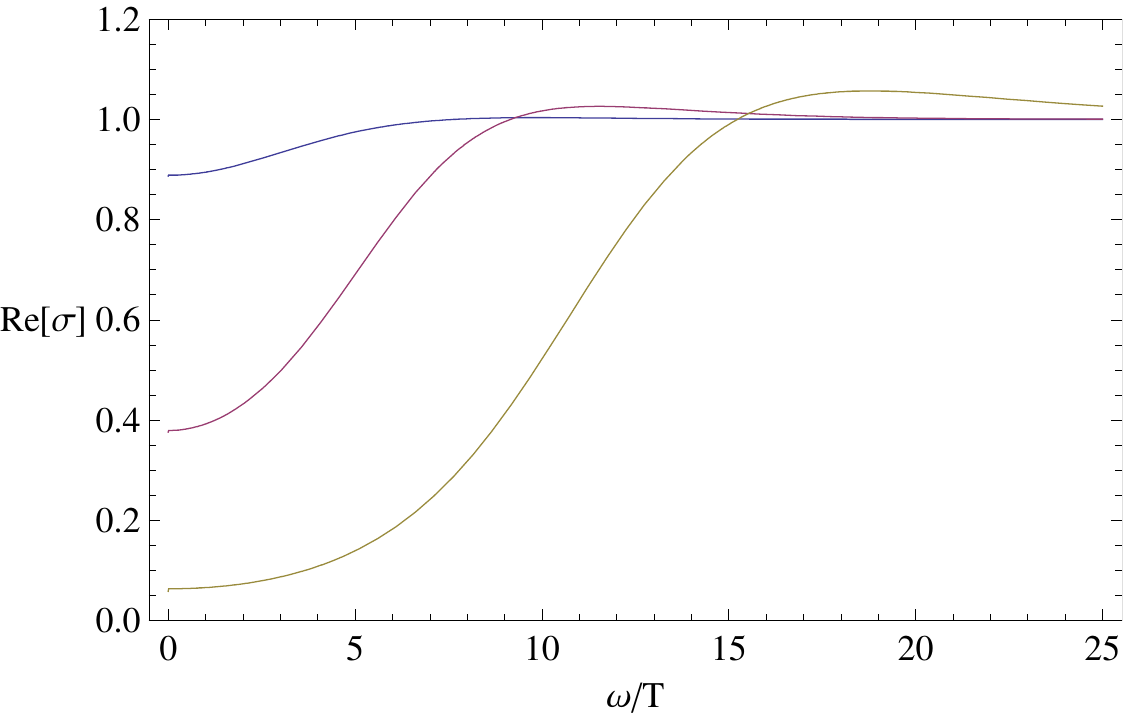}
\includegraphics[height=4.7cm]{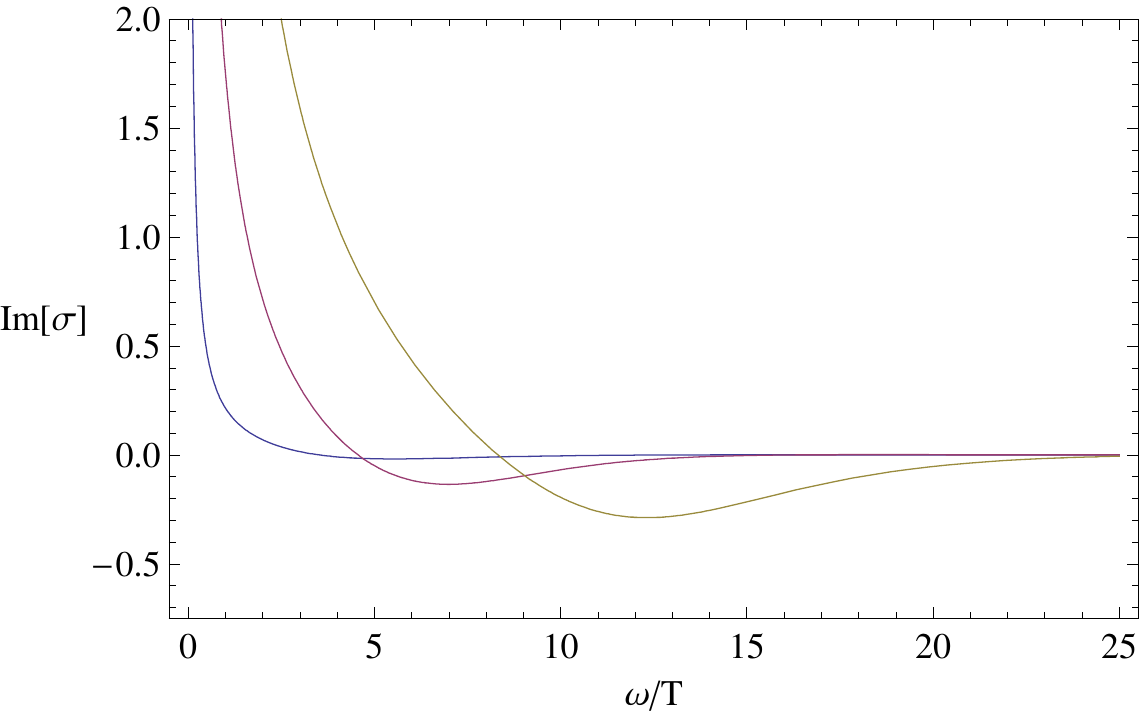}
\end{center}
\caption{The real (left) and imaginary (right) parts of the electrical conductivity computed via AdS/CFT as described in the text. The conductivity is shown as a function of frequency. Different curves correspond to different values of the chemical potential at fixed temperature. The gap becomes deeper at larger chemical potential. We have set $g=1$ in (\ref{eq:conduct}).}
\end{figure}

Numerical results for the real and imaginary parts of the electrical conductivity are shown in figure 6 above. These plots have not appeared elsewhere. We comment on the physical interpretation of these plots in the following subsection. Particularly suggestive is the depletion of the real part at frequencies below a scale set by the chemical potential.

\subsection{Comparison to experiments in graphene}
\label{sec:graphene}

It is amusing and instructive to compare our results for the conductivity in figure 6 to some
recent experimental data in graphene. Graphene is a natural material to compare to,
as at low energies it is described by a 2+1 dimensional relativistic theory with a chemical potential
determined by the gate voltage (see e.g. \cite{simons}).
It therefore has precisely the same kinematics as the AdS/CFT system we are studying.
Graphene has been subjected to intense study recently following the isolation of single layered
samples \cite{graphene}.

\begin{figure}[h]
\begin{center}
\includegraphics[height=5cm]{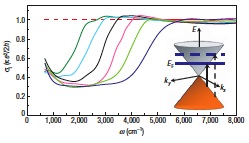} \\
\includegraphics[height=5.2cm]{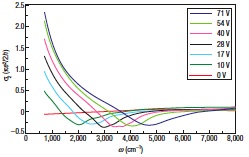}
\end{center}
\caption{Experimental plots of the real (top) and imaginary (bottom) parts of the electrical conductivity in graphene as a function of frequency. The different curves correspond to different values of the gate voltage. The inset in the upper plot shows an interband transition that is accessible at energies above $2 E_F$. Plots taken from \cite{graphene2}.}
\end{figure}

Figure 7 above shows experimental results for the real and imaginary parts of the conductivity
at different values of the chemical potential, taken from
\cite{graphene2}.

The similarity with the AdS/CFT plots of figure 6 is striking. Let us focus on the real part of the conductivity, the imaginary part can be determined from the real part through the Kramers-Kronig relations as we discuss below. There are three features in data. At large frequencies the conductivity tends to a constant; at low frequencies the (real part of the) conductivity is depleted below a scale set by the chemical potential; at very small frequencies the conductivity starts to rise again.
All of these features are straightforward to explain and bring out the similarities and differences with the AdS/CFT results.

The fact that the conductivity tends to a constant at large frequencies in both figures 6 and 7 is consistent with the fact that the conductivity is dimensionless in $2+1$ dimensions. This statement does not rely on relativistic invariance (i.e. one only needs set $\hbar=1$).

A depletion in the real part of conductivity below frequencies set by the chemical potential occurs in both figures 6 and 7. As we will see in the following section, the real part of the conductivity is the dissipative part of the conductivity and measures the presence of charged states as a function of energy. The drop in the real part of the conductivity therefore corresponds to a drop in the density of excitations at energies below the chemical potential. In graphene there is a simple explanation for this fact. The chemical potential sets the size of the Fermi surface. At zero momentum (we are computing the conductivity at zero momentum) the only available single particle excitations are when an electron jumps between different bands. This is illustrated in the inset of figure 7. In graphene, such an excitation has energy $2 E_F$, where $E_F$ is the Fermi energy and is proportional to the chemical potential $\mu$. Therefore, the dissipative conductivity will be Boltzman suppressed up to an energy scale set by $\mu$, as observed in figure 7. Given that the same structure is observed in figure 6, in AdS/CFT, one is lead to wonder if there may also be an effective Fermi surface in the strongly coupled theories studied via AdS/CFT.

The increase in the conductivity at very small frequencies appears as the main difference between figures 6 and 7. In fact, the results are closer than they might seem.
In the AdS/CFT results there is a delta function in the real part of the conductivity at $\w = 0$.
The delta function cannot be resolved numerically, but we know it is there because the imaginary
part of the conductivity has a pole as $\w \to 0$. The Kramers-Kronig relations (see (\ref{eq:KK}) below) imply that the real part must therefore have a delta function. The divergence of the conductivity
at low frequencies is directly related to conservation of momentum, as we shall discuss below. In graphene, momentum conservation is broken by the presence of impurities and the ionic lattice. These effects introduce a momentum relaxation timescale $\tau$ so that at low frequencies one has a Drude peak described by
\be\label{eq:drude}
\sigma(\w) = \sigma_0 + \frac{\rho^2}{\e+P} \frac{1}{1/\tau - i \w} \,.
\ee
This formula (from \cite{Hartnoll:2007ih}) will be discussed in more detail in a later section. Letting $\tau \to \infty$ we recover the pole (and hence delta function) of the translationally invariant AdS/CFT case. In short: the increase in the conductivity at low frequencies in figure 7 is a smoothed out version of the delta function in figure 6. We will discuss the addition of impurities to AdS/CFT computations in section  \ref{sec:impure} below.

\section{The physics of spectral functions}

In this section we will review the physical interpretation of retarded
Green's functions, and in particular the spectral density. Most of this
material can be found in more detail in, e.g. \cite{chaikin} chapter 7.6 or \cite{simons}
chapter 7. We go on to illustrate the concepts with examples from
AdS/CFT computations.

\subsection{Relation to two point functions and symmetry properties}
\label{sec:twopoint}

The retarded Green's function for two observables $\ocal_A$ and $\ocal_B$
is given in terms of the expectation value of their commutator as follows
\be\label{eq:onetotwo}
G^R_{\ocal_A \ocal_B}(\w,k) = - i \int d^{d-1}x dt e^{i \w t - i k \cdot x}
\theta(t) \langle [\ocal_A(t,x), \ocal_B(0,0)] \rangle \,,
\ee
where $\theta(t)$ is the Heaviside step function: nonzero and equal to one for
$t>0$. This representation of the Green's function is very useful to
establish several important properties. A proof of this result goes as follows.

We wish to compute an expectation value in the presence of a time dependent
perturbation to the Hamiltonian
\be
\delta H(t) = \int d^{d-1}x \phi_{B(0)}(t,x) \ocal_B(x) \,.
\ee
This is given by
\be
\langle \ocal_A \rangle(t,x) = \Tr \rho(t) \, \ocal_A(x) \,,
\ee
where $\rho(t)$ is the time dependent density matrix, satisfying
$i \pa_t \rho = [H_0 + \delta H,\rho]$. Passing to an interaction picture
(so that the time dependence due to $H_0$ is absorbed into the operators $\ocal_A$)
one obtains
\be
\langle \ocal_A \rangle(t,x) = \Tr \rho_0 \, U^{-1}(t) \ocal_A(t,x) U(t) \,, 
\ee
where $\rho_0 = e^{-H_0/T}$, say, and as usual $U$ is the time ordered exponential
\be
U(t) = T e^{- i \int^t \delta H(t') dt'} \,.
\ee
Expanding to first order in the perturbation of the Hamiltonian gives
\bea\label{eq:positionG}
\delta \langle \ocal_A \rangle(t,x) & = & - i \Tr \rho_0 \int^t dt' [\ocal_A(t,x), \delta H(t')] \nonumber \\
& = & - i \int^t d^{d-1}x' dt' \langle [\ocal_A(t,x), \ocal_B(t',x')] \rangle \phi_{B(0)}(t',x') \,.
\eea
Taking a Fourier transform of this result (using spacetime translation invariance) and comparing with the definition of the Green's function in (\ref{eq:greens}) leads to (\ref{eq:onetotwo}).

From (\ref{eq:onetotwo}) we can show that if the system is time reversal invariant and if the operators $\ocal_A$ satisfy $T \ocal_A(t,x) T^{-1} = \epsilon_A \ocal_A(-t,x)$, with $\epsilon_A = \pm 1$, then the Green's function has the following symmetry property
\be\label{eq:symmetric}
G^R_{\ocal_A \ocal_B}(\w,k) =  \epsilon_A \epsilon_B G^R_{\ocal_B \ocal_A}(\w,-k) \,.
\ee
This result follows from recalling that time reversal is an anti-unitary operator and therefore
\bea
\lefteqn{\langle [\ocal_A(t,x), \ocal_B(0,0)]\rangle  = \langle T [\ocal_A(t,x), \ocal_B(0,0)] T \rangle^* }\nonumber \\
& & =  \epsilon_A \epsilon_B \langle [\ocal_B(0,0), \ocal_A(-t,x)]\rangle = \epsilon_A \epsilon_B \langle [\ocal_B(t,-x), \ocal_A(0,0)]\rangle \,.
\eea
If time reversal is broken by, say, the presence of a background magnetic field\footnote{That a background magnetic field breaks time reversal follows from the Lorentz force: $m \ddot x = q \dot x \times B$.},
then the symmetry property (\ref{eq:symmetric}) continues to hold except that one of the Green's function
is evaluated in a time reversed background (i.e. $B \to - B$).

One consequence of this symmetry, noted above, is that the thermoelectric conductivity determining how
an electric field generates a heat flow (related to the Peltier effect) is the same as the thermoelectric conductivity determining how a heat gradient generates an electric current (the Seebeck effect).

\subsection{Causality and vacuum stability}
\label{sec:vacstability}

In (\ref{eq:positionG}) and (\ref{eq:onetotwo}) we see explicitly that the retarded Green's function is causal. Namely, the expectation value at time $t$ only depends on the source at times $t' < t$.
Consider taking the inverse Fourier transform of the Green's function
\be\label{eq:inverseF}
G^R_{\ocal_A \ocal_B}(t,k) = \int \frac{d\w}{2\pi} e^{-i\w t} G^R_{\ocal_A \ocal_B}(\w,k) \,. 
\ee
For $t<0$ we can evaluate this integral by closing the $\w$ contour in the upper half plane. Causality implies that we must obtain zero for the Green's function at $t<0$. Therefore
\be\label{eq:wanalytic}
G^R_{\ocal_A \ocal_B}(\w,k) \quad \text{is analytic in $\w$ for} \quad \text{Im}\, \w > 0 \,. 
\ee

If a computation of the Green's function leads to, say, a pole in the upper half
frequency plane it is a signal that something is wrong. It is possible to
be precise about the pathology. Suppose that
$G^R_{\ocal_A \ocal_B}(\w,k)$ has a single pole at some $\w = \w_\star$ in the upper
half plane. Then from (\ref{eq:inverseF}) we will have for $t<0$ that
\be
G^R_{\ocal_A \ocal_B}(t,k) \sim e^{- i \w_\star t} \sim e^{|\text{Im} w_\star| t} \,.
\ee
This is an exponentially growing mode indicating that the vacuum in which
the Green's function has been computed is unstable.

The analyticity property (\ref{eq:wanalytic}) leads to several physically useful results,
simply from contour integration. Firstly one has the Kramers-Kronig relation
between the real and imaginary parts of any function satisfying (\ref{eq:wanalytic})
and vanishing as $|\w| \to \infty$
\bea\label{eq:KK}
\text{Re}\,G^R(\w) & = & P \int_{-\infty}^\infty \frac{d\w'}{\pi} \frac{\text{Im}\,G^R(\w')}{\w'-\w} \,, \\
\text{Im}\,G^R(\w) & = & - P \int_{-\infty}^\infty \frac{d\w'}{\pi} \frac{\text{Re}\,G^R(\w')}{\w'-\w} \,,
\eea
where $P$ denotes principal value of the integral, as usual. These results follow
from
\be\label{eq:contour}
G^R(z)=\oint_\Gamma \frac{d\zeta}{2\pi i} \frac{G^R(\zeta)}{\zeta-z} \,,
\ee
with $\Gamma$ running along the real axis and closing in the upper half plane,
and then taking $z = \w + i 0$. One obtains (\ref{eq:KK}) if the contribution
from the semicircle at infinity in the upper half plane gives a vanishing contribution.
If $G^R(\w)$ does not vanish asymptotically, one should subtract the nonvanishing
behaviour and hence obtain the Kramers-Kronig relations for the subtracted function.

One can also obtain sum rules from (\ref{eq:contour}). The simplest one is
\be\label{eq:sum}
\chi \equiv \lim_{\w \to 0+i0} G^R_{\ocal_A \ocal_B}(\w,x)
= \int_{-\infty}^{\infty} \frac{d\w'}{\pi} \frac{\text{Im}\,G^R_{\ocal_A \ocal_B}(\w',x)}{\w'} \,.
\ee
This is called the thermodynamic sum rule because
$\chi = \pa \langle \ocal_A \rangle/\pa \phi_{B(0)}$ is the static, thermodynamic,
susceptibility. This is a real quantity. The sum rule thus has rather nontrivial physical
content: it relates an equilibrium thermodynamic quantity to an integral over all
frequencies of a dissipative process.
The integral in (\ref{eq:sum}) is well defined because
by acting on (\ref{eq:onetotwo}) by complex conjugation one finds
\be
\text{Im}\,G^R_{\ocal_A \ocal_B}(\w,k) = - \text{Im}\,G^R_{\ocal_A \ocal_B}(-\w,-k) \,.
\ee
It follows that if the Green's function is even under $k \to - k$, as is typically the case
(from (\ref{eq:symmetric}), it is necessarily the case if $\ocal_A = \ocal_B$),
then the imaginary part of the Green's function is an odd function of $\w$ and therefore vanishes
as $\w \to 0$. If the Green's function is odd under $k \to -k$ then the imaginary
part of the Green's function is a symmetric function of $\w$ and the sum rule (\ref{eq:sum})
does not contain useful information. 

\subsection{Spectral density and positivity of dissipation}

The spectral representation of the Green's function follows from
(\ref{eq:onetotwo}) by inserting a complete basis of energy eigenstates
in between the two operators. Assuming for simplicity that we are in
the canonical ensemble, so that the density matrix $\rho_0 = e^{- H_0/T}$,
one obtains
\be\label{eq:spectral}
G^R_{\ocal_A \ocal_B}(\w,k) = \sum_{mn} e^{-E_n/T}
\left(\frac{A_{nm} B_{mn}  \delta^{(d)}(k_{nm} -k)}{E_n-E_m + \w + i0} -
\frac{A_{mn} B_{nm}  \delta^{(d)}(k_{mn} -k)}{E_m-E_n + \w + i0}  \right)
\,,
\ee
where $E_m$ are energy eigenvalues, $H_0 |m\rangle = E_m | m \rangle$, $k_{nm} = k_n - k_m$
with $k_m$ momentum eigenvalues,
$A_{mn} = \langle m| \ocal_A(0,0) | n \rangle$ and $B_{mn} = \langle m| \ocal_B(0,0) | n \rangle$.
We are dropping factors of the partition function (i.e. $Z=1$). The $+i0$s are important,
they remind us that the poles are in the lower half frequency plane, and also to
correctly use the identity
\be\label{eq:Ppart}
\frac{1}{x \pm i0} = P \frac{1}{x} \mp i \pi \delta(x) \,.
\ee

The spectral representation allows us to show that $i \w$ times
the anti-hermitian part of the retarded Green's function satisfies a positivity
property, namely, given a vector $v_A$
\be\label{eq:positivity}
i \w v^*_A \left[G^R_{\ocal_A \ocal_B}(\w,x-x') - G^{R}_{\ocal_B \ocal_A}(\w,x-x')^* \right] v_B \geq 0 \,.
\ee
This property is obtained by manipulations showing that (\ref{eq:spectral})
implies
\bea
\lefteqn{\frac{i\w}{2} \left[G^R_{\ocal_A \ocal_B}(\w,x-x') - G^{R}_{\ocal_B \ocal_A}(\w,x-x')^*\right] } \nonumber \\
& & = 2 \pi \w \sinh\frac{\w}{2T} \sum_{mn} e^{-(E_n+E_m)/2T} \delta^{(d)}(k_{nm} -k)
\delta(E_n - E_m + \w) A_{nm} B_{mn} \,,
\eea
where one uses (\ref{eq:Ppart}). This property leads to an interpretation of the
meaning of the anti-hermitian part of the Green's function, as we will now see.

The time-varying external source $\phi_{(0)}$ does work on the system.
The rate of change of the total energy due to this source is
\be
\frac{dW}{dt} = \frac{d}{dt} \Tr \, \rho H = \Tr \, \rho \frac{\pa \delta H}{\pa t} =
\int d^{d-1}x \left(\langle \ocal_A \rangle + \delta \langle \ocal_A \rangle \right) \pa_t \phi_{A(0)} \,.
\ee
In the second equality we used the Schr\"odinger equation
$i \pa_t \rho = [H,\rho]$ and the fact that $[H,H]=0$.
For the third equality recall that we are in the Schr\"odinger picture
so that operators are time-independent.

The positivity of the anti-hermitian part of the Green's function
times $i\w$ has the
physical consequence that the time average over a cycle of the rate of work
done is positive.
Taking a mode of a particular frequency
\be
\phi_{(0)}(t,x) = \text{Re}\, \left( \phi_{(0)}(x) e^{-i\w t} \right) \,,
\ee
and using the definition of the Green's function in (\ref{eq:positionG})
and (\ref{eq:onetotwo}), then to leading order in the external source $\phi_{(0)}$ one obtains
\bea
\overline{\frac{dW}{dt}} & \equiv &  \frac{\w}{2\pi} \int_0^{2\pi/\w} dt \frac{dW}{dt} \nonumber \\
 & = & \w \int d^{d-1}x d^{d-1}x' \phi^*_{(0) A}(x) \frac{i}{2}
\left(G^R_{\ocal_A \ocal_B}(\w,x-x') - G^{R}_{\ocal_B \ocal_A}(\w,x-x')^* \right) \phi_{(0) B}(x) \nonumber
\\
& \geq & 0 \,.
\eea
That is, dissipation is captured precisely by the anti-hermitian part of the
Green's function times $i \w$. In the derivation of the above result one uses the
fact, evident from (\ref{eq:onetotwo}) and mentioned already above, that
$G^R_{\ocal_A \ocal_B}(\w,k)^* = G^R_{\ocal_A \ocal_B}(-\w,-k)$.

In the case that $\ocal_A = \ocal_B$ and there are no other operators involved,
then one can introduce the spectral function
\be
\chi_A(\w,k) = - \text{Im} \, G^R_{\ocal_A \ocal_A}(\w,k) \,.
\ee
The positive property (\ref{eq:positivity}) clearly reduces to
\be\label{eq:spectraldensity}
\w \, \chi_A(\w,k) \geq 0 \,.
\ee
The spectral function directly measures the degrees of freedom in the theory
that have an overlap with the operator $\ocal_A$.
Even if there is more than one operator, one can diagonalise the
Green's function and define a spectral function for each of the
decoupled eigenoperators. If the Green's function is even under time reversal,
$\epsilon_A = \epsilon_B$ in (\ref{eq:symmetric}), and also even under $k \to -k$,
then one can easily obtain from our above results that
\be
\w \, \chi_{AB}(\w,k) \equiv -\w \, \text{Im} \, G^R_{\ocal_A \ocal_B}(\w,k) \geq 0 \,, \qquad \text{[with evenness conditions]}
\ee
for each component separately. The statement we derived previously about the anti-hermitian
part of the Green's function is more general, however.

\subsection{Quantum critical dynamics with particle-hole symmetry}
\label{sec:crossover}

Particle-hole symmetry means that $\mu=0$. That is, there is no
net charge density carried by say, `electrons' rather than `holes'.
In this section we will discuss, largely following \cite{Herzog:2007ij},
some general features of charge dynamics in relativistic
quantum critical systems in d = 2+1 dimensions with $\mu=0$.
In such theories, the finite temperature
retarded Green's function for a current $J^\mu$ is constrained by rotational invariance
and charge conservation to take the form
\be\label{eq:JJ}
G^R_{J^\mu J^\nu}(\w,k) = \sqrt{-\w^2+k^2} \left(P_{\mu \nu}^T K^T(\w,k)
+ P_{\mu \nu}^L K^L(\w,k) \right) \,,
\ee
where $P_{\mu \nu}^L = \eta_{\mu \nu} - \frac{p_\mu p_\nu}{p^2} - P_{\mu \nu}^T$,
$P_{i j}^T = \d_{ij} - \frac{k_i k_j}{k^2}$ and $P_{t \mu}^T = 0$. In these expressions
the 3-momentum $p^\mu = (\w,k)$. Thus the charge dynamics is parametrised by
two functions of frequency and momentum: $K^T(\w,k)$ and $K^L(\w,k)$.
Scale invariance implies that these functions are functions of $\w/T$ and $k/T$,
there can be no independent temperature dependence.

The dynamic simplifies considerably in the limits when the temperature is large or
small compared to the frequency and momentum. We will focus on the
charge correlation function for concreteness: $G^R_{J^t J^t}(\w,k)$.
When $\w/T \ll 1$ one is in the hydrodynamic or `collision-dominated' regime. This regime includes
the DC ($\w=0$) limit that is often taken in experiments. In this low frequency (and
momentum) limit one expects a diffusive behaviour
\be
G^R_{J^t J^t}(\w,k) = \frac{D \chi k^2}{-i \w + D k^2} \,.
\ee
This form of the Green's function is completely fixed by hydrodynamics. See for instance
\cite{chaikin} or, for the brave, \cite{Forster}. There are two constant parameters, the diffusion constant
$D$ and the susceptibility $\chi$. The electrical conductivity is a constant
$\sigma(0) = \chi D$. Recall again that conductivity is dimensionless in 2+1 dimensions.

In the opposite limit, $\w/T \gg 1$ one speaks of `collisionless' or coherent transport. This is
effectively the zero temperature limit. At zero temperature, scale invariance, Lorentz
symmetry and charge conservation completely determine the Green's function up to a constant.
In particular
\be\label{eq:loT}
G^R_{J^t J^t}(\w,k) = \frac{K k^2}{\sqrt{-\w^2 + k^2}} \,.
\ee
The conductivity in this limit is again a constant $\sigma(\infty) = K$. For general theories, the two conductivities $\sigma(0)$ and $\sigma(\infty)$ are not related, and describe very different
physical processes. Thus the $T \to 0$ and $\w \to 0$ limits do not commute. As we mentioned, experiments at low frequencies and temperatures typically correspond to taking the $\w \to 0$ limit
first \cite{damle}.

The two limits are characterised by different dispersion relations for the dominant pole. At low frequencies $\w \sim k^2$ whereas at high frequencies $\w \sim k$. Clearly there should be a
crossover between these behaviours as a function of $\w/T$ and $k/T$. While such a crossover is
anticipated on purely kinematic grounds, as we have just shown, prior to an AdS/CFT computation there was no 2+1 system for which the crossover could be exhibited in a direct computation. Using the Einstein-Maxwell action (\ref{eq:einsteinmaxwell}) about the Schwarzschild background (\ref{eq:schwarzads})
one numerically computes the Green's function for the currents (\ref{eq:JJ}) in much the same way as we did in section \ref{sec:conductivity} above, but generalised to allow for a finite momentum. Figure 8, which is taken from \cite{Herzog:2007ij}, shows how the dispersion relation changes from quadratic to linear as function of $\w/T$.

\begin{figure}[h]
\begin{center}
\includegraphics[height=7cm]{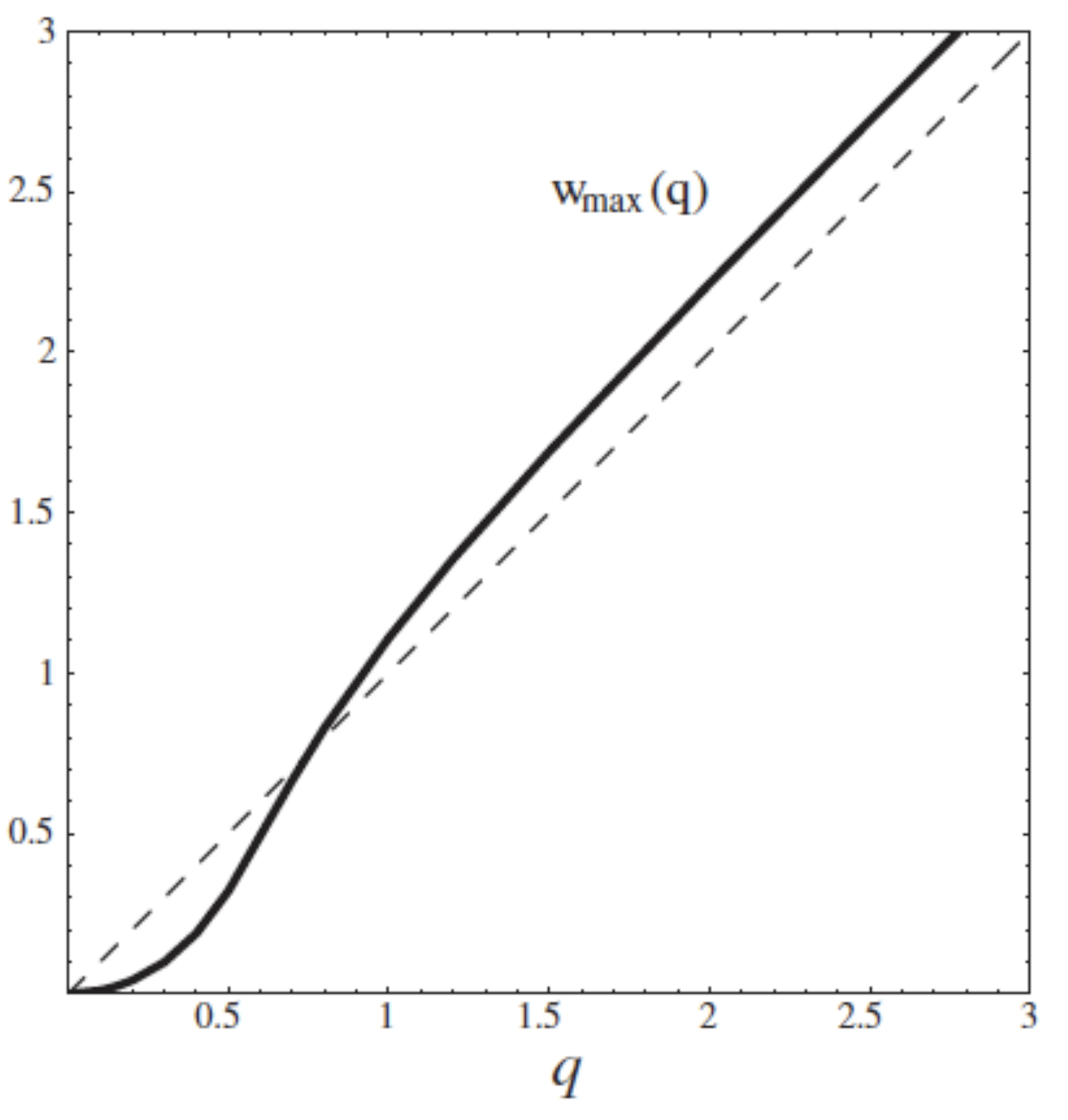}
\end{center}
\caption{Solid curve shows the location of the peak in the density-density spectral function at real frequencies $\w$ as a function of momentum $k$ ($q$ in the figure). The dispersion relation is quadratic at small momenta and linear at large momenta. Figure taken from  \cite{Herzog:2007ij}.}
\end{figure}

In performing the bulk gravitational computation one finds that $\sigma(0) = \sigma(\infty)$. In fact, at $k=0$ the conductivity is exactly constant and independent of $\w/T$. This non-generic fact
is interpreted in field theory as a self-duality under particle-vortex duality, and can be traced to the self-duality of the Maxwell equations in the bulk \cite{Witten:2003ya, Herzog:2007ij}. It is not surprising, perhaps, that the simplest theories in which the general feature of a coherent to hydrodynamic crossover can be shown explicitly also have special symmetries.

We close this section with two comments. Firstly, there is no coherent to hydrodynamic crossover in 1+1 dimensions. In 1+1 dimensions conformal invariance completely fixes the retarded Green's function of the charge to take a form analogous to (\ref{eq:loT}) at all temperatures. There is no hydrodynamic regime.

Finally, the formulae just given for the current and charge density Green's functions omit an important piece of physics. Due to fact that the zero momentum ($k=0$) modes are closely related to conserved charges, using hydrodynamics one finds that the Green's functions must decay like a power law at late times rather than exponentially, see e.g. \cite{Kovtun:2003vj}. Technically this arises because at $k=0$ the spatial derivative terms appearing in the hydrodynamical equations become less important than terms which are nonlinear in the hydrodynamical variables. These nonlinear terms give rise to the power law tails. In 2+1 dimensions, the Fourier transform of the power law tails gives a $\log|\w|$ divergence in response functions such as the conductivity as $\w \to 0$. That nonlinear terms in hydrodynamics are required to see this effect immediately explains why no such divergence was observed in our gravitational computations: nonlinearities involve one loop corrections to the fluctuation equations in the bulk and are therefore suppressed by inverse powers of $N$. Indeed, the suppression of power law tails at large $N$ was seen at weak coupling in \cite{Kovtun:2003vj}.
In a real life system, effects outside of a simple hydrodynamic approach, such as impurities or long range Coulomb interactions, will typically become important before the logarithmic divergence kicks in
\cite{markus}.

\subsection{Quantum critical transport with a net charge and magnetic field}
\label{sec:rhoandB}

It is of interest, for instance with a view to graphene \cite{markus} or to the high-$T_c$ cuprates \cite{Hartnoll:2007ih}, to extend the analysis of the previous section to include a net charge density and a background magnetic field. The structure of the Green's functions becomes significantly less constrained. Among the important new effects is that the charge current and the heat current can now mix, due to the presence of a background charge density. This leads to the generalised Ohm's law of equation (\ref{eq:EandT}) above. In a magnetic field we need to generalise the law further, to allow for off-diagonal elements in the conductivities (due to the background magnetic field)
\be\label{eq:EandT2}
\left(
\begin{array}{c}
\langle J_i \rangle \\
\langle Q_j \rangle
\end{array}
\right) = 
\left(
\begin{array}{cc}
\sigma_{ij} & \alpha_{ij} T \\
\alpha_{ij} T & \bar \kappa_{ij} T
\end{array}
\right)
\left(
\begin{array}{c}
E_j \\
-(\nabla_j T)/T
\end{array}
\right) \,.
\ee
As in section \ref{sec:conductivity}, we would like to study these conductivities as a function of frequency $\w$ at zero momentum, $k=0$. We shall work in $d=2+1$ dimensions for concreteness.

In section \ref{sec:conductivity} we found that thermoelectric and thermal conductivities could be computed in terms of the electrical conductivity alone. This is a general property of relativistic theories at a finite charge density (and/or magnetic field) and follows from Ward identities \cite{Hartnoll:2007ip, chris}. We will not give the full argument here, the essential point is the following. Recall from section \ref{sec:conductivity} that the conductivities were given in terms of the Green's functions $G^R_{J^i J^j}(\w)$, $G^R_{J^i Q^j}(\w)$ and $G^R_{Q^i Q^j}(\w)$. We saw in section \ref{sec:twopoint} that the retarded Green's functions are given by two point functions of the operators in question. One way of thinking about Ward identities is that they impose invariance of the partition function and correlation functions under gauge transformations acting on background gauge fields for all global symmetries of the field theory (see e.g. \cite{Policastro:2002tn} for uses of Ward identities in this manner). Gauge coordinate transformations (\ref{eq:diffeo}) act on background chemical potentials and magnetic fields, and this action in turn mixes the two point functions of the electric and thermal currents. In 2+1 dimensions, the end result is most cleanly expressed in terms of the following `complexified' conductivities
\be\label{eq:complexified}
\sigma_\pm = \sigma_{xy} \pm i \sigma_{xx} \,, \quad \alpha_\pm = \alpha_{xy} \pm i \alpha_{xx} \,, \quad \bar \kappa_\pm = \bar \kappa_{xy} \pm i \bar \kappa_{xx} \,.
\ee
The relationship between the conductivities is found to be \cite{Hartnoll:2007ip}
\bea
\pm \alpha_\pm T \w & = & (B \mp \mu \w) \sigma_\pm - \rho \,, \\
\pm \bar \kappa_\pm T \w & = & \left(\frac{B}{\w} \mp \mu \right) \alpha_\pm T \w - s T + m B \,.
\eea
We will therefore only present results for the electrical conductivity in the following. For expressions for all the transport coefficients as well as for applications to the Nernst effect (closely related to $\alpha_{xy}$), see \cite{bhaseen1, Hartnoll:2007ih, bhaseen2}.

In discussing the electrical response of the system, it is useful to distinguish two lengthscales. Firstly there is the hydrodynamic lengthscale $l_T = 1/T$. At scales $l \gg l_T$ one expects a description in terms of conserved quantities. In the presence of a background magnetic field $B$ there is another lengthscale, $l_B = 1/\sqrt{B}$. At scales $l \gg l_B$, momentum is not a locally conserved quantity, due to the Lorentz force from the background magnetic field. The limit $l \gg l_B, l_T$ was studied in 
\cite{Hansen:2008tq} using hydrodynamics, without conserved momentum. Other studies of hydrodynamics in a background magnetic field include
\cite{Buchbinder:2008dc, Buchbinder:2008nf, Buchbinder:2009mk, Caldarelli:2008ze}. Here we will start with the case in which the magnetic field is small so that we can study intermediate lengthscales $l_T \ll l \ll l_B$.
In this regime momentum is approximately conserved and leads to a rich (`magneto')hydrodynamics
\cite{Hartnoll:2007ih}, including a cyclotron resonance.

The conductivity $\sigma(\w)$ can be computed analytically in the regime $l_T \ll l \ll l_B$ using either hydrodynamic methods \cite{Hartnoll:2007ih} or the AdS/CFT correspondence \cite{Hartnoll:2007ip}.
The result is
\bea\label{eq:xx}
\sigma_{xx} & = & \sigma_Q \frac{\w (\w + i \gamma + i \w_c^2/\gamma)}{(\w+i \gamma)^2 - \w_c^2} \,, \\
\sigma_{xy} & = & - \frac{\rho}{B} \frac{-2 i \gamma \w + \gamma^2 + \w_c^2}{(\w+i \gamma)^2 - \w_c^2} \,. \label{eq:xy}
\eea
In these expressions
\be\label{eq:cpole}
\w_c = \frac{B \rho}{\epsilon + P} \,, \qquad \gamma = \frac{\sigma_Q B^2}{\epsilon + P} \,.
\ee
There is a resonance at the frequency $\w = \w_c$ with a width $\gamma$. The coefficient $\sigma_Q$ is the unique transport coefficient necessary to describe current dynamics in this regime. Hydrodynamics does not fix $\sigma_Q$, which depends on the microscopic theory. Using a bulk Einstein-Maxwell theory (\ref{eq:einsteinmaxwell}) one finds \cite{Hartnoll:2007ip}
\be
\sigma_Q = \frac{(sT)^2}{(\epsilon + P)^2} \frac{1}{g^2} \,.
\ee
An expression for the energy density $\epsilon$ was given in section \ref{sec:conductivity} while the pressure is in section \ref{sec:mu}. In a neutral background with no magnetic field $sT = \epsilon + P$ and hence $\sigma_Q = 1/g^2$.

There are various pieces of physics contained in the above expressions for $\sigma_{xx}$ and $\sigma_{xy}$. The most immediate is the cyclotron resonance at $\w_\star = \pm \w_c - i \gamma$ (consistently with the general theory of section \ref{sec:vacstability}, this pole is in the lower half plane). This is a collective cyclotron motion, it is not the `microscopic' cyclotron motion of individual electrons. Indeed the value of $\w_\star$ is an interesting prediction for future experiments on graphene \cite{subirmarkus}. The damping of the cyclotron motion can be thought of loosely as due to the fact that the positively and negatively charged modes of the system are undergoing cyclotron orbits in opposite directions and colliding.

It is instructive to take the strict DC limit ($\w=0$) of (\ref{eq:xx}) and (\ref{eq:xy}) in which one immediately obtains
\be\label{eq:xxB}
\sigma_{xx} = 0 \,, \qquad \sigma_{xy} = \frac{\rho}{B} \,.
\ee
This result was first obtained from AdS/CFT in \cite{Hartnoll:2007ai} and gives the Hall conductivity of the system. There is a simple physical interpretation of the fact that the Hall conductivity is proportional to the charge density. Under applied magnetic and electric fields, positively and negatively charged particles move in the same direction and therefore their currents cancel. In order to obtain a net Hall conductivity, there needs to be a surplus of either positive or negative charge.

Another instructive limit of (\ref{eq:xx}) and (\ref{eq:xy}) is to set $B=0$. One obtains
\be\label{eq:xxnoB}
\sigma_{xx} = \sigma_Q + \frac{\rho^2}{(\epsilon+P)} \frac{i}{\w} \,, \qquad \sigma_{xy} = 0 \,.
\ee
Here we recover the divergence of conductivity, discussed in section \ref{sec:graphene} above, as $\w \to 0$ if there is a nonzero charge density. Indeed (\ref{eq:xxnoB}) is the same as equation (\ref{eq:drude}) in the absence of impurity scattering. There is again a simple intuitive picture for this divergence. In a medium with no net charge, currents can relax while preserving momentum. Consider a head on collision between an electron and a hole such that they are both at rest afterwards. Before the collision there is a net current (the currents add) while afterwards there is none. Momentum is clearly conserved in this process but the current is eliminated. If there is a net (positive say) charge density, in contrast, then on average the positive charges cannot cancel all their momentum with negative charges.
A collision between two positive charges conserves both momentum and current. Thus conservation of momentum combined with a net charge prevents current from relaxing. Therefore, in the presence of a constant electric field, the positive charges will accelerate indefinitely, giving a divergent conductivity.
Consistently with this picture, we saw that in the presence of a magnetic field, which locally violates conservation of momentum at long timescales, the DC conductivity (\ref{eq:xxB}) does not diverge.
The finite conductivities obtained in the probe limit of AdS/CFT computations, e.g. \cite{Herzog:2002fn, Karch:2007pd}, arise because the coefficient of the $1/\w$ term in (\ref{eq:xxnoB}) is effectively taken to zero.

The hydrodynamic formulae discussed so far in this subsection can be obtained without any AdS/CFT input. An example of a microscopic question that cannot be addressed by hydrodynamics is: what is the fate of the cyclotron resonance at large magnetic fields? Within AdS/CFT this question can be answered by identifying the cyclotron resonance pole in a (numerically computed) Green's function. The dependence of the pole on the magnetic field is shown in figure 9 below together with the deviation from the hydrodynamic result. A qualitatively similar plot was obtained recently in a weakly coupled microscopic description of graphene \cite{subirmarkus2}.

\begin{figure}[h]
\begin{center}
\includegraphics[height=5cm]{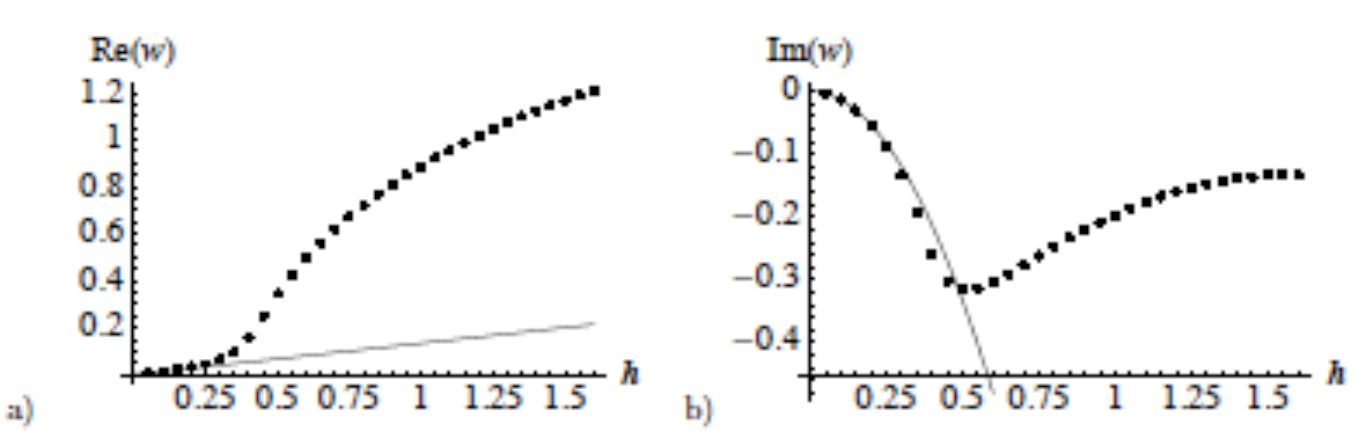}
\end{center}
\caption{The dots show the location of the cyclotron pole in the retarded Green's function in the complex frequency plane as a function of the magnetic field and at a fixed charge density. These points are obtained numerically using AdS/CFT. The curves show the hydrodynamic result (\ref{eq:cpole}). It is clear that the hydrodynamic expression is only correct at small magnetic fields. The x-axis is a dimensionless magnetic field $h = r_+^2 B$.  Figure taken from \cite{Hartnoll:2007ip}.}
\end{figure}

Finally, we note that electromagnetic duality in the bulk leads to an interesting relationship between the conductivity of the theory when the charge density and magnetic fields are swapped around \cite{Hartnoll:2007ip}. In particular under the action
\be
S: \quad B \to \rho \,, \quad \rho \to - B \,, \quad \sigma_Q \to \frac{1}{\sigma_Q} \,,
\ee
the complexified conductivity (\ref{eq:complexified}) built from the hydrodynamic expressions  (\ref{eq:xx}) and (\ref{eq:xy}) transforms as
\be
S: \quad \sigma_+(\w) \to \frac{-1}{\sigma_+(\w)} \,.
\ee
This can be thought of as the transformation of the conductivity under particle-vortex duality \cite{Hartnoll:2007ih}. This action can be somewhat trivially extended to a full $SL(2,\Z)$ acting in the standard way on the conductivity $\sigma_+$ \cite{Burgess:2000kj, Witten:2003ya, Hartnoll:2007ip}.

\subsection{A simple treatment of impurities}
\label{sec:impure}

In the previous sections we have seen that various quantities, such as the electrical conductivity at a finite charge density, diverged in the $\w \to 0$ limit due to translation invariance. In most physical condensed matter systems, translation invariance is broken by explicitly space-dependent potentials.
Two examples are the regular ionic lattices of solids and random impurities. Both of these effects are modeled by adding an explicitly space-dependent coupling $V(x)$ to the Hamiltonian
\be\label{eq:deltaH}
\delta H = \int d^{d-1}x V(x) \ocal(t,x) \,. 
\ee
Here $\ocal$ is some operator in the theory that couples to the impurities or lattice.

From the basic AdS/CFT formula (\ref{eq:deltaO}) we know how to implement (\ref{eq:deltaH}) in a gravitational dual. One should add a field $\phi$, dual to the operator $\ocal$, into the bulk and impose the boundary condition that $\phi(r,x,t) \to \phi_{(0)}(x) = V(x)$. Solving the bulk equations of motion subject to this boundary condition will give the geometry dual to the quantum critical theory in the presence of the potential $V(x)$.

The full problem appears daunting, as one will need to solve PDEs for the field $\phi(r,x)$ and the other bulk fields to which it couples. In this section we will treat impurities only and make the simplifying assumptions that the impurities are random, dilute and weak. Weakness means that we expand quantities to lowest nontrivial order in $V(x)$. Random and dilute means that we will then average over the impurity potential assuming no correlations
\be
\langle V(x) \rangle_\text{imp} = 0 \,, \qquad \langle V(x) V(y) \rangle_\text{imp} = \bar V^2 \delta^{(d-1)}(x-y) \,.
\ee
Note that the field theory path integral must be done with a fixed background $V(x)$, otherwise the averaging over impurity potentials will restore translation invariance.\footnote{The field theory path integral does not commute with the integral over $V(x)$ because the correlators are normalised by the partition function including $V(x)$, see e.g. \cite{simons}.}
The impurities are thus characterised by a single quantity: the strength of the impurity potential $\bar V$.
The scaling dimension of $\bar V$ is $(d+1)/2 - \Delta_\ocal$, where $\Delta_\ocal$ is the scaling dimension of $\ocal$. Impurities are a relevant perturbation of the dynamics if the scaling dimension of $\bar V$ is positive. This is called the Harris criterion.

Because the impurities break translation invariance, they should cause momentum to relax at late times. This will be reflected in the momentum density two point function. It is useful to parametrise the retarded Green's function for the momentum density in terms of the `memory function' $M(\w)$:
\be\label{eq:Mw}
G^R_{T^{tx} T^{tx}}(\w) = \frac{\chi_0 M(\w)}{\w + M(\w)} \,.
\ee
In a translationally invariant system, the zero momentum ($k=0$) retarded Green's function vanishes, $G^R_{T^{tx} T^{tx}}(\w) = 0$, at all nonzero frequencies.\footnote{This follows from the fact that for any operator $A$, the Heisenberg equations of motion imply that $\w^2 G^R_{AA}(\w) = - G^R_{[A,H] [A,H]}(\w) + G^R_{[A,H] [A,H]}(0)$. At zero spatial momentum, using the definition of the Green's function (\ref{eq:onetotwo}), the right hand side of this equation vanishes if $\int d^{d-1}x A(x)$ commutes with the Hamiltonian. That is, if $A$ is a conserved charge density, such as the momentum.} In the presence of impurities this will no longer be the case. Therefore we expect $M(\w) = \ocal(\bar V^2)$. In particular, the zero frequency limit of the memory function will give a pole in the Green's function very close to (just below) the real axis which sets the timescale for momentum relaxation
\be\label{eq:tauimp}
\lim_{\w \to 0} M(\w) = \frac{i}{\tau_\text{imp.}} \,.
\ee
The time dependence is computed by doing the inverse Fourier transform of the frequency space Green's function (\ref{eq:Mw}). At positive times we should close the contour in the lower half frequency plane. The pole at $\w = -i/\tau_\text{imp}$ is by assumption close to the real axis (the impurities are weak) and therefore dominates the late time behaviour
\be\label{eq:momentumrelax}
\langle T^{tx} \rangle(t) \sim G^R_{T^{tx} T^{tx}}(t) \sim e^{-t/\tau_\text{imp.}} \,.
\ee

The momentum relaxation (\ref{eq:momentumrelax}) can be directly incorporated into a hydrodynamic approach by adding a term to the hydrodynamic equations that imposes late time non-conservation of momentum
\be
\pa_\nu T^{\mu \nu} = F^{\mu \nu} J_\nu + \frac{1}{\tau_\text{imp.}} \left(\delta^{\mu}{}_\nu + u^u u_v \right) T^{\nu \gamma} u_\gamma \,,
\ee
where $u^\mu$ is the local fluid 3-velocity.
The effect of the last term in this equation upon the hydrodynamic expressions for the conductivity (\ref{eq:xx}) and (\ref{eq:xy}) is to let $\w \to \w + i/\tau_\text{imp}$ \cite{Hartnoll:2007ih}. This leads to, for instance, equation (\ref{eq:drude}) in which we see that the conductivity at $\w \to 0$ is finite in the presence of impurities, as we anticipated.

The overall scale of the momentum relaxation timescale $\tau_\text{imp.}$ is set by the phenomenological parameter $\bar V^2$. An important question, however, is whether $\tau_\text{imp.}$
has a strong dependence on the magnetic field and charge density. This requires a microscopic computation. It was shown in \cite{Hartnoll:2008hs} that to leading order in $\bar V$ the relaxation timescale, starting from (\ref{eq:tauimp}), can be given in terms of the dissipative part of the retarded Green's function of the operator $\ocal$ coupling to the impurity
\be\label{eq:ImOO}
\frac{1}{\tau_\text{imp.}} = \frac{\bar V^2}{2 \chi_0} \lim_{\w \to 0} \int \frac{d^{d-1}k}{(2\pi)^{d-1}} k^2 \frac{\text{Im}\, G^R_{\ocal \ocal}(\w,k)}{\w} \,.
\ee
This expression is general for any quantum field theory.
At the quantum critical points we are primarily interested in here,
conformal invariance requires that this expression has the scaling form
\be
\frac{1}{\tau_\text{imp.}} = \frac{\bar V^2}{T^{d- 2 \Delta_\ocal}} {\mathcal{F}}\left(\frac{\rho}{T^{d-1}}, \frac{B}{T^2} \right) \,.
\ee
The expression (\ref{eq:ImOO}) can be straightforwardly evaluated via AdS/CFT if we start with an action for the bulk field $\phi$ dual to the operator $\ocal$. The computation was performed in \cite{Hartnoll:2008hs} in 2+1 dimensions using a bulk scalar and a pseudoscalar field that were nonminimally coupled to the background field strength. The results are shown in figure 10 below.

\begin{figure}[h]
\begin{center}
\includegraphics[height=5.4cm]{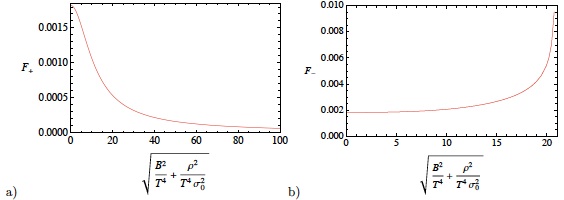}
\end{center}
\caption{The inverse relaxation timescale as a function of the magnetic field and charge density for two choices of the operator $\ocal$ arising via AdS/CFT in d=2+1 dimensions. The result only depends on the combination $B^2 + \rho^2/\sigma_0^2$ (with $\sigma_0 = 1/g^2)$ due to electromagnetic self-duality in the bulk. Figure taken from \cite{Hartnoll:2008hs}.}
\end{figure}

Both of these plots show that there is indeed a strong dependence on the magnetic field and charge density. In the right hand plot, the relaxation timescale goes to zero at a phase transition in which the operator $\ocal$ condenses. This transition is closely related to the Gubser-Mitra instability \cite{Gubser:2000ec}. The left hand plot may provide a prototype for the charge and magnetic field dependence of $\tau_\text{imp.}$ near stable quantum critical points. This dependence appears to match the scaling of the Nernst effect in an organic superconductor close to a Mott transition \cite{oxford, Hartnoll:2008hs, us}.

\section{Holographic superconductivity}

\subsection{What is a superconductor?}
\label{sec:generalSC}

So far we have described, from the viewpoint of a gravitational dual,
the equilibrium and linear response physics of quantum critical theories 
subjected to a finite temperature, external electromagnetic fields
and impurities. It is hoped that this framework will have applications to strongly
coupled condensed matter systems in the vicinity of quantum phase
transitions (and indeed more generally).
We noted in section \ref{sec:SC} that such systems include nonconventional
superconductors. However, the theories we have discussed so far have not
in themselves been superconducting.

Much of the richness of condensed matter physics concerns the dynamics
behind the onset of ordered phases at low temperatures. Ordered
phases appear as instabilities of the na\"ive vacuum towards the
formation of a symmetry breaking condensate. In this final section we want
to put our strongly coupled theories to work and see if their dynamics
can induce symmetry breaking phase transitions.

Superconductivity is among the most phenomenologically spectacular
of symmetry breakings. Two of the most important consequences,
an infinite DC ($\w=0$) conductivity and the expulsion of magnetic fields
(the Meissner effect), follow directly from electromagnetic gauge invariance.
We now review this fact,
see for instance the discussion in \cite{weinberg} or \cite{simons}.

It is useful to consider firstly the theory without dynamical photons.
We noted in section \ref{sec:mu} above that photons can correctly be treated
as an external electromagnetic field for many purposes in condensed matter
systems. We therefore have a global $U(1)$ symmetry which we assume to be
spontaneously broken. This breaking results in a massless Goldstone boson
$\theta$ which transforms under $U(1)$ by the shift $\theta \to \theta + \Lambda$.
Gauge invariance of the theory in an electromagnetic background $A$ means that
the free energy can be written as
\be\label{eq:SCfree}
F = \int d^dx \sqrt{g_{(0)}} {\mathcal{F}} \left[A - d \theta \right] \,,
\ee
for some function ${\mathcal{F}}$. Stability of the theory in the absence of Goldstone mode
excitations or background fields implies that ${\mathcal{F}}$ should have a minimum at
$A = d \theta$. The current generated by a small electromagnetic field is then
\be\label{eq:London}
J_i =  - \left. \frac{\delta F}{\delta A^i} \right|_{A = d \theta + \delta A}
= - {\mathcal{F}}''[0]  \delta A_i \,.
\ee
The minus sign arises because $F$ is the Euclidean action which differs from the Lorentzian
action by a minus sign.
For simplicity we are working in a gauge with $\delta A_t = 0$. Equation (\ref{eq:London}) is the (second) London
equation. The electric field in this gauge is just $\delta E_i = i \w \delta A_i$, in frequency space.
Therefore from (\ref{eq:London})
\be
J_i = \frac{i {\mathcal{F}}''[0]}{\w} \delta E_i \equiv \sigma(\w) \delta E_i \,.
\ee 
We see that the conductivity diverges as $\w \to 0$, as advertised. We can also note at this point
that stability of the theory (\ref{eq:SCfree}) requires ${\mathcal{F}}''[0]$ to be positive.

The Meissner effect follows from taking the curl of (\ref{eq:London}), whence
\be\label{eq:diamagnetic}
- {\mathcal{F}}''[0] \delta B_i = (\nabla \times J)_i \,.
\ee
We can now see that this current $J$ is diamagnetic, that is, it acts to expel the
external magnetic field $\delta B_i$ that we have applied. If we are considering
static fields, then (\ref{eq:diamagnetic}) can be combined with the Maxwell equation
$\nabla \times \delta B = \mu_0 J$ to obtain
\be\label{eq:meissner}
\left(\nabla_i^2 - \mu_0 {\mathcal{F}}''[0]  \right) \delta B_i = 0 \,.
\ee
This relation is called the first London equation and implies that the photon is massive
and hence exponentially suppressed inside a superconductor. The inverse of the mass,
$\mu_0 {\mathcal{F}}''[0]$, is the London penetration depth squared.
We should emphasise
that although a dynamical photon is necessary in order to see the expulsion of
magnetic fields, the essential physics underlying the Meissner effect is the generation
of diamagnetic currents (\ref{eq:diamagnetic}), which are computed within a theory
without dynamical photons in a background magnetic field.

There is a further important fact to be learnt from (\ref{eq:SCfree}). The momentum
conjugate to the Goldstone boson $\theta$ is the charge density
\be
\pi_\theta = - \frac{\delta F}{\delta \pa_t \theta} = \frac{\delta F}{\delta A_t} = J^t = \rho \,.
\ee
It follows that we have the commutator
\be\label{eq:commutator}
[\rho(x), \theta(y)] = i \delta(x-y) \,.
\ee
This shows that states in which the `phase' $\theta$ has a definite value
are maximally distinct from states with a definite charge. Of course a
generic state can have an expectation value for both $\theta$ and $\rho$
simultaneously. The relation (\ref{eq:commutator}) is useful to
emphasise that a finite charge density $\rho$ does not break the $U(1)$ symmetry.
After all, $\rho$ is a neutral operator which commutes with the charge operator (itself!).
In order to break the $U(1)$ symmetry, we need an expectation value for the phase
$\theta$.

Beyond the generalities above, one needs a microscopic theory to actually
determine whether or not a symmetry breaking condensate forms in a given material.
As we mentioned in section (\ref{sec:SC}) above, most traditional theories describing
the onset of superconductivity, BCS theory being the canonical example,
introduce charged `quasiparticles' (dressed electrons) that are then `paired'
into bosonic operators by a `gluing' interaction which is mediated by another
quasiparticle such as phonons. The composite charged bosonic operator is then
shown to condense. While this picture has been extremely successful for
conventional superconductors \cite{Parks}, one can argue that various
experimental `anomalies' of the high-T$_c$ superconductors indicate that
these materials cannot be accommodated into the framework just outlined
\cite{Polchinski:1992ed}.

One motivation for building a holographic superconductor is to have a
microscopic (that is, first principles rather than effective) description
of the onset of superconductivity in which there are no quasiparticles whatsoever:
there are no `electrons' and no `glue'. Instead, there is a strongly coupled
theory in which a charged operator condenses below a critical temperature.
That is, holographic superconductors provide a (unique) computationally tractable
model for the onset of what one might call `superconductivity without electrons'.
Time will determine the extent to which such models are useful for real life
nonconventional superconductors such as the cuprates. In that regard, it is
promising that recent theories of the cuprate superconductors do involve
$s$-wave superconductivity in a strongly coupled theory with an emergent
gauge field \cite{sachdev3}.

The holographic superconductors we are about to describe have a
deceptive similarity to the Landau-Ginzburg description of superconductivity
\cite{Hartnoll:2008kx}. One should bear in mind that AdS/CFT is not a low energy effective
field theory, but rather a dual description of the microscopic theory.
The classical nature of the gravitational description arises because of the
large $N$ limit, not because one is working at long wavelengths or low
energies.

\subsection{Minimal ingredients for a holographic superconductor}

We have already seen that in order to discuss charge transport in
field theory one is lead to Einstein-Maxwell theory (\ref{eq:einsteinmaxwell})
in the bulk. The dynamics of the current operator $J^\mu$ is captured
by the classical dynamics of the bulk photon field $A_\mu$.
We have furthermore just recalled that superconductivity is due
to spontaneous breaking of the electromagnetic $U(1)$ symmetry.
Spontaneous symmetry breaking occurs if a charged operator acquires a
vacuum expectation value. The basic relation (\ref{eq:deltaO}) implies
that such charged operators will be dual to charged fields in the bulk.
Thus we need to augment the Einstein-Maxwell action by additional
charged fields.

In a strongly coupled theory one might expect the symmetry broken
phase to have condensates of many  operators. This would involve
considering many coupled charged fields in the bulk. A simplification we will
make from the outset is to consider the minimal case of only a single charged field
in the bulk. The next question is what type of charged field to consider.
The operators that condense do not a priori need to be scalars. If the condensate
carries angular momentum one can talk about $p$-wave ($\ell=1$) or $d$-wave
($\ell=2$) superconductors, as opposed to $s$-wave superconductors with $\ell=0$.
We will focus here on the simplest case of $s$-wave superconductors in which
the charged operator is a scalar. For AdS/CFT work on $p$-wave superconductors
see for instance \cite{Gubser:2008zu, Gubser:2008wv, Roberts:2008ns, Ammon:2008fc, Basu:2008bh,
Herzog:2009ci}.

The upshot of the previous two paragraphs is that we need to consider
Einstein-Maxwell theory together with a charged (complex) scalar field. A minimal
Lagrangian (in $d+1$ dimensions) for such a system is
\be\label{eq:scaction}
\Lag = \frac{1}{2 \kappa^2} \left(R + \frac{d (d-1)}{L^2}
\right)  - \frac{1}{4 g^2} F^2 - |\nabla \phi - i q A \phi |^2 - m^2 |\phi|^2 - V(|\phi|) \,.
\ee
We will immediately specialise to the case $V(|\phi|) = 0$, again for simplicity.
We also specialise for concreteness to the case of $d=3$ dimensions for the boundary
field theory. Several interesting nonconventional superconductors in nature, such as the cuprates,
are layered and hence effectively $d=2+1 = 3$ dimensional. Qualitatively similar
behaviour to what we shall find also occurs for $d=4$ dimensional holographic
superconductors \cite{Horowitz:2008bn}.

The first step is to specify the normal, i.e. non-superconducting, state of the theory.
This will be dual to a solution to the equations of motion following from (\ref{eq:scaction})
with $\phi=0$. The simplest background we might consider is simply the Schwarzschild-AdS
metric (\ref{eq:schwarzads}), corresponding to a scale invariant theory at finite temperature. However, this will not work. We noted previously that in a scale-invariant
theory all nonzero temperatures are equivalent. In particular, there cannot be a preferred critical temperature, $T_c$, at which something special happens. In order to have a critical temperature, another scale must be introduced. If we wish to avoid adding any new ingredients into our theory, the simplest way to introduce a scale is to work at a finite chemical potential $\mu$. By dimensional analysis this allows $T_c \propto \mu$.\footnote{We will prefer to work in terms of the chemical potential rather than the (equivalent) charge density $\rho$ because the chemical potential is directly in units of energy whereas to convert the charge density to units of energy squared one needs to employ a preferred velocity (the `speed of light') which introduces an extra ambiguity when comparing to experimental systems.} A chemical potential appears in describing, for instance, the cuprate superconductors, as a measure of the doping away from critical doping \cite{sachdev2, Hartnoll:2007ih}. A chemical potential also describes graphene held at a finite gate voltage, as we noted in section \ref{sec:graphene} above. We have argued above that the theory at a finite chemical potential is dual to the Reissner-Nordstrom-AdS black hole solution (\ref{eq:RNads}). This is the desired gravitational dual to the normal state.

Given our background, the second step is to ask whether this phase can be unstable to the formation of a charged condensate. A charged condensate will be a nonvanishing expectation value $\langle \ocal \rangle$ for the charged operator dual to the bulk scalar field $\phi$. In section \ref{sec:vevs} we
saw that an expectation value requires the scalar field $\phi$ to be nonzero in the bulk, with the expectation value itself given by (\ref{eq:vev2}). In order for the bulk scalar field to turn on continuously below a critical temperature, there should be an instability of the Reissner-Nordstrom black hole against perturbations by the scalar field. 

To search for instability at a critical temperature one perturbs the Reissner-Nordstrom background (\ref{eq:RNads}) by the scalar field $\phi = \phi(r) e^{-i\w t}$. The equation of motion for $\phi(r)$ is
\be\label{eq:phiradial}
- r^4 \left(\frac{f}{r^2} \phi' \right)'  - \frac{r^2}{f} \left(\w + q \mu \left(1- \frac{r}{r_+} \right)\right)^2 \phi + (L m)^2 \phi = 0 \,.
\ee
The spacetime is unstable if there is a normalisable solution to this equation (\ref{eq:phiradial}) with ingoing boundary conditions at the horizon such that $\w$ has a positive imaginary part. Such a mode is growing exponentially in time. More formally, and connecting with our discussion in section \ref{sec:vacstability} above, we can note from (\ref{eq:getG}) that this mode would lead to a (pathological) pole in the retarded Green's function for $\ocal$ in the upper half frequency plane, because at the particular frequency at which the mode is normalisable, the denominator of (\ref{eq:getG}) vanishes.

At this point we are only interested in determining the critical temperature at which at instability first appears. At the critical temperature, the rate of growth of the unstable mode will go to zero and therefore we can put $\w = 0$ in (\ref{eq:phiradial}). That is, at the critical temperature we expect to find a static normalisable mode. In the following section we will see that this mode exponentiates into a new branch of static `hairy black hole' solutions which describe the superconducting phase.
Allowing ourselves to rescale the radial coordinate, one can check that equation (\ref{eq:phiradial}) with $\w=0$ only depends on three dimensionless parameters
\be\label{eq:3vars}
\gamma q \,, \quad \Delta \quad \text{and} \quad \frac{\gamma T}{\mu} \,.
\ee
Recall that we defined the dimensionless number $\gamma$ in (\ref{eq:gamma}) above. It is a free parameter within our phenomenological approach to the AdS/CFT correspondence but would be fixed by an embedding into string theory \cite{Denef:2009tp}. With $d=3$,
\be
\gamma^2 = \frac{2 g^2 L^2}{\kappa^2} \,.
\ee
$\Delta$ is the scaling dimension (\ref{eq:2sols}) of the operator $\ocal$ which in $d=3$ becomes
\be
\Delta (\Delta - 3) = (mL)^2 \,.
\ee
The charge $q$ of the scalar field is an integer, $q \in \Z$, if the symmetry group is $U(1)$ rather than $\R$, as we shall assume.\footnote{Note that a condensate $\langle \ocal \rangle$ does not break the symmetry completely, but rather breaks $U(1) \to \Z_q$.} That $\gamma$ appears multiplying $q$
in (\ref{eq:3vars}) is consistent with the fact that it parameterises the relative importance of the bulk electromagnetic and gravitational forces.

One therefore scans through values of $\Delta$ and $\gamma q$, and for each value determines numerically whether equation (\ref{eq:phiradial}) admits a normalisable solution with $\w=0$ for some critical value of $\gamma T/\mu$. The result of this scan, from reference \cite{Denef:2009tp}, is shown in figure 11 below.

\begin{figure}[h]
\begin{center}
\includegraphics[height=9cm]{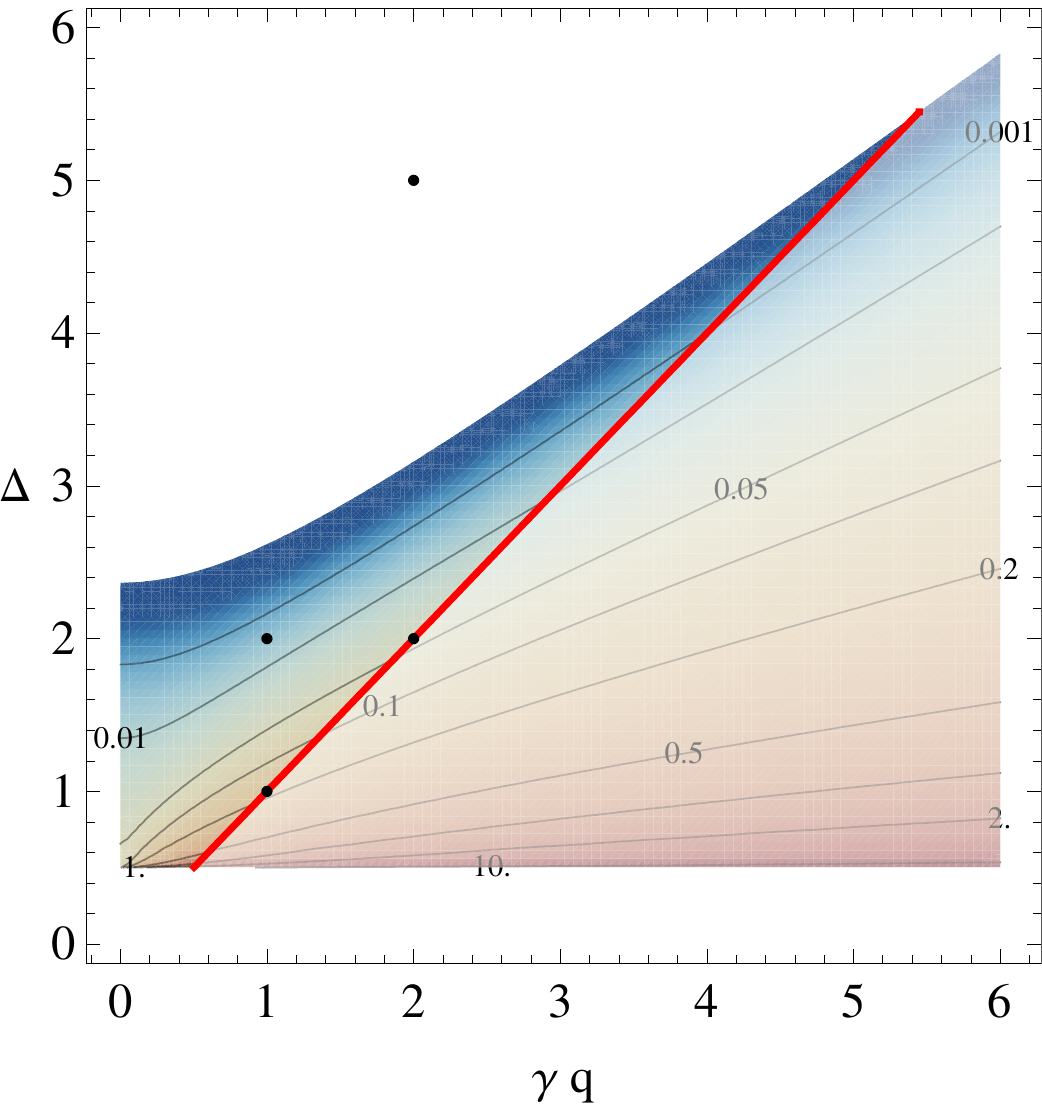}
\end{center}
\caption{The critical temperature $T_c$ as a function of charge $\gamma q$ and dimension $\Delta$. Contours are labeled by values of $\gamma T_c/\mu$. The BPS line $\Delta = \gamma q$ is shown in red. The top boundary (\ref{eq:criterion}) is a line of quantum critical points separating normal and superconducting phases at $T=0$. The bottom boundary of the plot is the unitarity bound $\Delta = 1/2$ at which $T_c$ diverges. Figure taken from \cite{Denef:2009tp}.}
\end{figure}

Figure 11 shows that there is a range of charges and masses for the bulk field $\phi$ in which the normal phase becomes unstable as the temperature is lowered. These are the holographic superconductors.
Note however that even neutral scalar fields ($q=0$) can become unstable. The criterion for instability can be read off from figure 11:
\be\label{eq:criterion}
q^2 \gamma^2 \geq 3 + 2 \Delta (\Delta - 3) \,.
\ee
This criterion can be understood analytically by combing the following two facts \cite{Gubser:2008px, Hartnoll:2008kx, Gubser:2008pf, Denef:2009tp}. Firstly, at zero temperature the background Reissner-Nordstrom black hole has an AdS$_2$ geometry (\ref{eq:ads2}) close to the horizon. The radius of this AdS$_2$ region is (from (\ref{eq:RNf}) and (\ref{eq:RNT})) $L_2^2 = L^2/6$. Secondly, the effective mass of the scalar field at the horizon is not $m^2$ but rather $m^2_+ = m^2 - \g^2 q^2/2L^2$. This can be seen by taking $r \to r_+$ in the equation of motion (\ref{eq:phiradial}) and again using (\ref{eq:RNf}) and (\ref{eq:RNT}).
Thus the background electric field has given an extra negative contribution to the mass squared of the scalar field, making it more likely to be unstable. This contribution comes from the gauge covariant derivative. In fact we can now ask precisely whether the scalar field is stable in the near horizon region. The criterion for stability in any AdS space is the Breitenlohner-Freedman bound. This bound requires that the mass squared of the field is sufficiently large that the corresponding scaling dimension $\Delta$ in (\ref{eq:2sols}) is real. For AdS$_2$ the bound becomes $(L_2 m_+)^2 \geq -1/4$. Using the relations just discussed we obtain that the field will be unstable if
\be
- \frac{1}{4} \geq (L_2 m_+)^2 = \frac{L^2}{6} \left(m^2 - \frac{\g^2 q^2}{2 L^2} \right) 
\ee
Rearranging this expression leads to the criterion (\ref{eq:criterion}).

To summarise, there are two distinct mechanisms causing a superconducting instability in the Reissner-Nordstrom background:
\begin{itemize}
\item At low temperatures an AdS$_2$ near-horizon throat appears in which an asymptotically stable negative mass squared scalar field can become unstable (because the Breitenlohner-Freedman bound is different for the near horizon AdS$_2$ and the asymptotic AdS$_4$).

\item In the presence of a background electric field, a charged scalar field acquires an effective negative mass squared which can drive the field unstable. This is closely related to superradiance instabilities and to pair production.
\end{itemize}
The combined effects of these two mechanisms is encapsulated in the criterion (\ref{eq:criterion}) and in figure 8. We did not put this physics in by hand, rather it emerged from the minimal kinematic ingredients we introduced. This is the bulk `microscopic' dynamics behind superconductivity. An interesting open question is to rephrase these processes in terms of more field-theoretic concepts.

\subsection{Superconducting phase}

\subsubsection{Condensate}

If we continue to cool the theory down below the critical temperature $T_c$ at which the bulk scalar field becomes unstable, we must switch to a different spacetime background. As the low temperature phase has a condensate for the operator $\ocal$, the bulk scalar field $\phi$ will be nonvanishing. This leads to the following ansatz, describing a charged `hairy' black hole
\be\label{eq:hairy}
ds^2 = \frac{L^2}{r^2} \left(- f(r) e^{-\chi(r)} dt^2 + \frac{dr^2}{f(r)} + dx^i dx^i \right) \,,
\ee
together with
\be
A = A_t(r) dt \,, \qquad \phi = \phi(r) \,.
\ee
The hairy black holes are then found by plugging this ansatz into the Einstein-Maxwell-scalar equations, following from the action (\ref{eq:scaction}), and solving numerically. We will not spell out the equations or the numerical method in detail, they may be found in \cite{Hartnoll:2008kx}. The earlier work \cite{Hartnoll:2008vx} considered the probe limit, $\gamma q \to \infty$, in which one can set $\chi = 0$ in 
(\ref{eq:hairy}) and $f$ equal to its value for the Schwarzschild-AdS metric (\ref{eq:fsads}). In this limit one only has to solve a simpler set of equations for $\{A_t(r), \phi(r)\}$.

Given the solution $\phi(r)$ one can read off the expectation value $\langle \ocal \rangle$ using (\ref{eq:vev2}). The result for several values of $q$ and for a particular value of the mass squared is shown in figure 12, taken from \cite{Hartnoll:2008kx}.

\begin{figure}[h]
\begin{center}
\includegraphics[height=4.7cm]{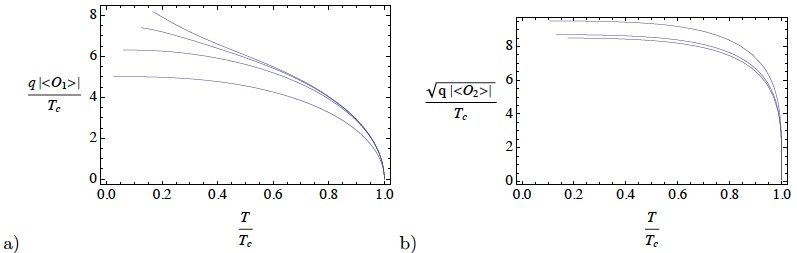}
\end{center}
\caption{The condensate as a function of the temperature for the case $\Delta=1$ (left) and $\Delta=2$ (right). In curve (a), from bottom to top, $\gamma q=1, 3, 6, 12$. In curve (b), from top to bottom, $\gamma q=3, 6, 12$. Figure taken from  \cite{Hartnoll:2008kx} ($q$ in the figure is $\gamma q$).}
\end{figure}

In figure 12 we see how the condensate turns on at $T=T_c$ and tends towards a finite value as $T \to 0$. The numerics become unreliable at very low temperatures and unfortunately do not let us determine, for instance, the fate of the increase of the condensate at low temperature in the left hand plot at large $q$. Some analytic results at low temperatures were obtained for a closely related system in \cite{Gubser:2008wz}. An important open question is to gain a better handle on the zero temperature limit of these hairy black holes.

The plots in figure 12 show the classic form for a second order phase transition and indeed one can check (numerically) that the condensate vanishes like $(T-T_c)^{1/2}$ and that across the transition the second derivative of the free energy is discontinuous.
We recalled in section \ref{sec:qcp} that second order transitions are forbidden in 2+1 dimensions at finite temperature, because the goldstone boson in the putative symmetry broken phase has large fluctuations which destroy the expectation value. We also noted in that section that taking a large $N$ limit, as we are doing, evades this result because fluctuations are suppressed in the large $N$ limit. At finite $N$, the phase transition we have just described will become a crossover. In 3+1 dimensions it would remain a genuine phase transition.

\subsubsection{Conductivity}
\label{sec:SCconductivity}

Using essentially the same procedure as in section \ref{sec:conductivity} we can compute the electrical conductivity in the hairy black hole background.\footnote{A full treatment of transport at finite momentum in the superconducting phase will have to allow for mixing of the conserved currents with excitations of the superfluid order parameter \cite{Herzog:2008he, Amado:2009ts}.} The result for the real (dissipative) part of the conductivity at low temperatures for several values of $q$ and for a particular value of the mass squared is shown in figure 13, again taken from \cite{Hartnoll:2008kx}. Computations for some different values of the mass squared can be found in \cite{Horowitz:2008bn}.

\begin{figure}[h]
\begin{center}
\includegraphics[height=4.9cm]{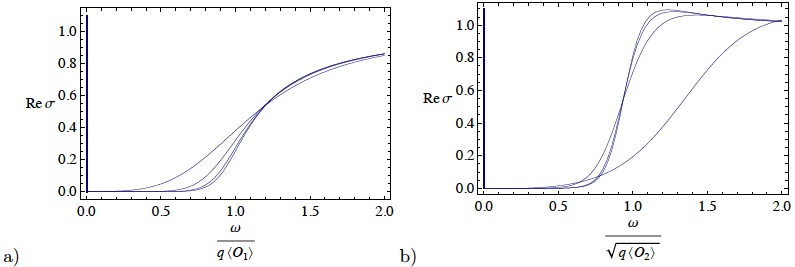}
\end{center}
\caption{The real (dissipative) part of the electrical conductivity at low temperature in the presence of a $\Delta = 1$ (left) and $\Delta = 2$ (right) condensate. The temperature taken was $T = 0.03 \gamma q \langle \ocal \rangle$ and $T = 0.03 \sqrt{\gamma q \langle \ocal \rangle}$, respectively, and the charges were $\gamma q = 1, 3, 6, 12$. The curves with steeper slope correspond to larger $\gamma q$. Figure taken from \cite{Hartnoll:2008kx} ($q \to \gamma q$).}
\end{figure}

There are three features of note in this plot: a delta function at the origin, $\w=0$, a gap at low frequencies $\w < \w_g$ and the fact that the conductivity tends to the normal state value at large frequencies. We will address the first two of these observations in more detail.

The delta function at the origin was anticipated on general grounds in (\ref{sec:generalSC}). Numerically one detects this delta function by observing a pole in the imaginary part of the conductivity and appealing to the Kramers-Kronig relations (\ref{eq:KK}). We noted in sections (\ref{sec:conductivity}) and (\ref{sec:rhoandB}) above that the DC ($\w=0$) conductivity also diverged in the normal state, due to translation invariance. In the absence of impurities it is difficult to distinguish these two (distinct) physical effects. However, it was shown in \cite{Hartnoll:2008kx} that the first derivative of the coefficient of the delta function is discontinuous across the phase transition. Furthermore, in the probe limit $\gamma q \to \infty$, we noted in (\ref{sec:rhoandB}) that the delta function in the normal state disappears, because momentum can be dissipated into the neutral metric (`gluon') degrees of freedom, whereas it was shown in \cite{Hartnoll:2008vx} that the superconducting delta function persists in this limit.

The coefficient of the delta function can be determined numerically \cite{Hartnoll:2008kx, Horowitz:2008bn} and is one definition of the superfluid density. From our general discussion in (\ref{sec:generalSC}) we see that it will also give us the magnetic penetration depth.

The absence of electric current dissipation for frequencies $\w < \w_g$ is indicative of a gap in the spectrum of charged excitations. Figure 13 seems to indicate that the gap is becoming exact at low temperatures, at least for sufficiently large charge $\gamma q$. We also found a gap at low temperatures in the normal state conductivity in section \ref{sec:conductivity} and suggested that that gap might indicate the presence of a Fermi surface and inter-band excitations. The superconducting gap in contrast appears to be tied to the presence of a condensate, with
\be\label{eq:wg}
\w_g \approx (q \langle \ocal \rangle)^{1/\Delta}_{(T=0)} \,.
\ee
This relation is exact in the probe limit \cite{Hartnoll:2008vx}, at least for the values of $\Delta$ studied
in \cite{Hartnoll:2008vx}.
A gap in the conductivity is typical of superconducting systems although, unlike the infinite DC conductivity, it does not follow from symmetry breaking alone. For instance, an important prediction of weakly coupled BCS theory is that $\w_g/T_c \approx 3.5$, which is indeed roughly observed in many conventional superconductors \cite{Parks}. The relation (\ref{eq:wg}) allows us
to identify the vertical axis in figure 9 with the gap and therefore obtain values for $\w_g/T_c$ in holographic superconductors. We can see that a range of values is possible, although in the probe limit the value $\w_g/T_c \approx 8$ appears to be fairly robust  \cite{Hartnoll:2008vx, Horowitz:2008bn}. It is amusing that this value is close to that reported in some measurements of the high-T$_c$ cuprates \cite{yazdani}. It is worth emphasising that $\w_g/T_c$ is a dimensioness order one quantity (i.e. it does not have a scaling with $N$) and is therefore a natural quantity to compare with experiment. It would be interesting to determine whether there are bounds on the values that this ratio can take in holographic superconductors.

In weakly coupled superconductors, the conductivity gap $\w_g = 2 E_g$, where $E_g$ is the energy gap in the charged spectrum. This is because the Cooper pairs have a negligible binding energy (in fact, they are not bound) and so the energy at which conduction becomes dissipative is the energy needed to produce a pair of electrons from the condensate, which is simply twice the energy of a single electron. Thus $\w_g=2 E_g$ is essentially the optical theorem, as illustrated in figure 14.

\begin{figure}[h]
\begin{center}
$\begin{array}{c}
\text{Re} \, \sigma(\w) \quad \sim \quad \text{Im} \;  \\
{} \\
{} \\
{}
\end{array}$
\includegraphics[height=2cm]{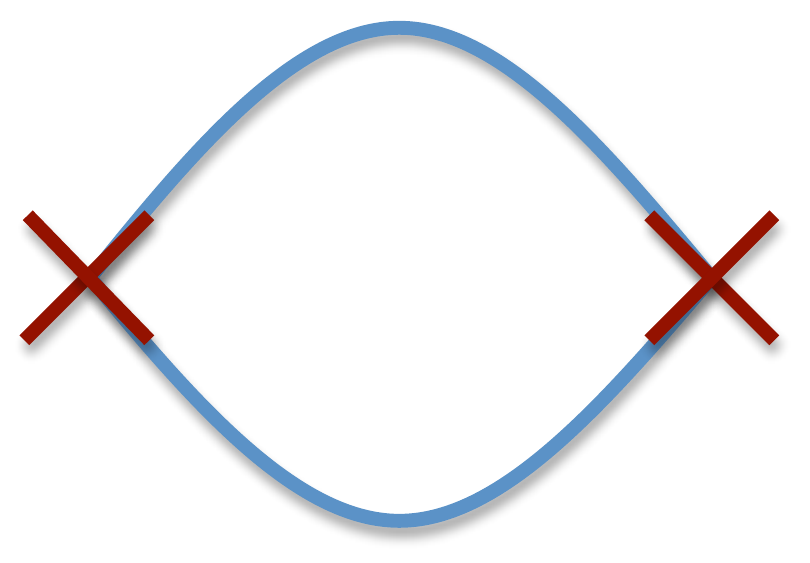}
$\begin{array}{c}
\; =\quad \sum_\text{On shell} \Bigg|  \\
{} \\
{} \\
{}
\end{array}$
\includegraphics[height=2cm]{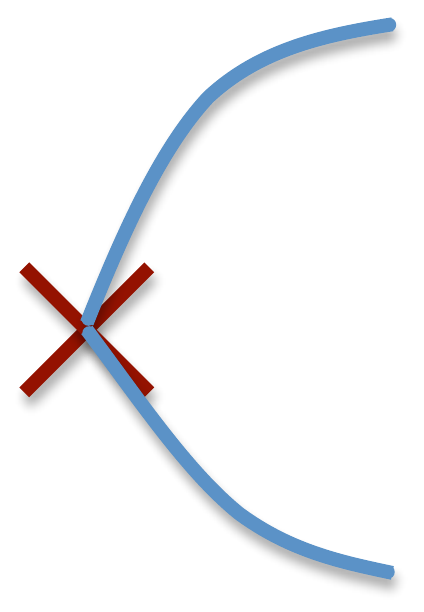}
$\begin{array}{c}
\Bigg|^2 \quad \sim \quad \theta(\w - 2 E_g) \,. \\
{} \\
{} \\
{}
\end{array}$
\vspace{-2cm}

\end{center}
\caption{At weak coupling the lowest order contribution to the conductivity involves two electrons appearing from the (nontrivial, with condensate) vacuum. For this process to contribute to the imaginary part of the Green's function, the virtual electrons must be on shell. Thus the energy in the background field ($\w$) must at least equal twice the energy need to produce an electron ($E_g$).}
\end{figure}

At strong coupling however there is no need for this relation to hold. The energy gap in the charged spectrum can be detected directly from looking at the conductivity at finite but small temperature and at very small frequencies. One will find the Boltzmann suppression
\be\label{eq:Eg}
\sigma(\w \to 0) = e^{- E_g/T} \,.
\ee
The gap $E_g$ can be computed for holographic superconductors \cite{Hartnoll:2008kx, Horowitz:2008bn} and indeed is not equal to $\w_g/2$ in general. An exception are the cases $\Delta = 2$ and $\Delta=1$ in the probe limit, in which indeed $\w_g=2 E_g$ \cite{Hartnoll:2008vx}. It is not yet clear whether this is a coincidence or indicative of an underlying weakly coupled `pairing' mechanism in this case.

There is in fact a slight puzzle in the observation of the gaps (\ref{eq:wg}) and (\ref{eq:Eg}), namely that holographic superconductors are not gapped! There is a Goldstone boson arising from the breaking of a global $U(1)$ symmetry. The Goldstone boson can run in loops in figure 14 and produce a nonvanishing imaginary conductivity all the way down to zero frequency. In weakly coupled BCS theory, this effect is higher order in a weak coupling expansion and therefore negligible. However, in strongly coupled holographic superconductors one would have anticipated a power law tail in the dissipative conductivity down to zero frequency. The puzzle is made sharper by the fact that such tails are observed in isotropic $p$-wave holographic superconductors \cite{Roberts:2008ns} which are in some ways similar to the $s$-wave models we are discussing (this shows that it is not purely a large $N$ effect). The existence of an exact gap may depend on the potential $V(|\phi|)$ in (\ref{eq:scaction}). However, a clean argument for when one expects the Goldstone boson to be important, and when not, is lacking at present. For instance, perhaps the Goldstone boson is responsible for the fact that the free energy appears to decrease like a power law at low temperatures in holographic superconductors \cite{Hartnoll:2008kx}, rather than exhibiting the exponential suppression expected for gapped systems.

\subsubsection{Magnetic fields}

There are two reasons why application of a magnetic field $B$ suppresses superconductivity.
The first is that
the Meissner effect (\ref{eq:meissner}) causes magnetic fields to be expelled from the superconducting region. This costs free energy of order $B^2$ integrated over the volume from which the magnetic field is expelled. For a sufficiently large magnetic field, this energy cost is greater than the free energy gained by being in the superconducting state $\Delta F = F_\text{norm.}(B) - F_\text{SC.}(0)$ and so the system will revert to the normal state above some critical field $B_c$. Such a transition will generically be first order and if it occurs the superconductor is called type I.

Different physics is observed if the magnetic field is not fully expelled from the superconducting phase. Instead, part of the flux forms vortices within the superconducting condensate. Vortices start to form at some $B_{c1}$ and then at a larger magnetic field $B_{c2}$ become sufficiently dense to destroy superconductivity. Both of these transitions are continuous and one now speaks of a type II superconductor.

While we can compute $B_c$ by comparing the free energy difference between the normal and superconducting states to $B^2$, the first order transition cannot occur dynamically in holographic superconductors as the photon is not dynamical (effectively, the coefficient of $B^2$ has been sent to infinity). This is called the extreme type II limit. The extreme type II limit is appropriate for 2+1 dimensional superconductors embedded into 3+1 dimensions in any case \cite{Hartnoll:2008kx}: while the free energy gain from superconductivity scales like the area $L^2$ of the sample, the magnetic field needs to be expelled from a volume of size $L^3$. Therefore in the large volume limit it is never favourable to expel the magnetic field.

Given that holographic superconductors are (extreme) type II, we should be able to compute $B_{c2}$, the magnetic field above which there is no superconductivity. The superconducting state in the presence of a penetrating magnetic field is difficult to study, as the bulk Einstein-Maxwell-scalar equations become PDES \cite{Hartnoll:2008kx}. It would be very interesting to find solutions in this phase, corresponding to black holes with magnetic charge forming a flux lattice at the horizon. What we can do more easily is to approach $B_{c2}$ from above. To do this we consider the normal phase in a background magnetic field and look for the onset of superconductivity.

Similarly to the case with no magnetic field, one takes the dyonic black hole background (\ref{eq:fmuB}) and (\ref{eq:Adyonic}) and perturbs by the scalar field $\phi$. Because the equations of motion for a charged scalar depend explicitly on the gauge potential (\ref{eq:Adyonic}), and this potential now depends explicitly on the coordinate $x$ we have to separate variables more carefully: $\phi = \phi(r) e^{- i \w t + i k y} X_\ell(x)$. Upon separating variables in the $\phi$ equations of motion, it is straightforward to check (for details see \cite{Albash:2008eh, Hartnoll:2008kx}) that the $X_\ell(x)$ are given by a Gaussian multiplied by Hermite polynomials, with eigenvalues $2 q B (\ell + \half)$, with $\ell \in \Z^+ \cup \{0\}$, and that the spectrum does not depend on $k$ (this is the degeneracy of the Landau levels). The equation for $\phi(r)$ becomes
\be\label{eq:phiradialB}
- r^4 \left(\frac{f}{r^2} \phi' \right)'  - \frac{r^2}{f} \left(\w + q \mu \left(1- \frac{r}{r_+} \right)\right)^2 \phi + \left( 2 q B (\ell + \half) r^2 + (L m)^2 \right) \phi = 0 \,.
\ee
To find the critical $B_{c2}$ one now simply puts $\w=0$ in this equation, because we are interested in the threshold unstable mode, and then scans for the magnetic field $B$, with $\ell=0$, such that there is a normalisable solution. The result is shown in figure 15, taken from  \cite{Hartnoll:2008kx}, for a couple of values of $\Delta$ and several values of the charge $q$.

\begin{figure}[h]
\begin{center}
\includegraphics[height=5.3cm]{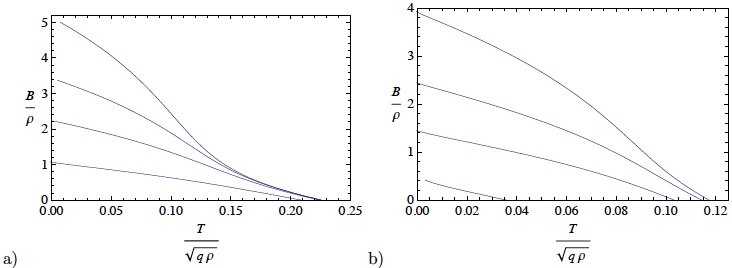}
\end{center}
\caption{(a) $\Delta = 1$ and, from right to left, $\gamma q = 12, 6, 3, 1$; (b) $\Delta = 2$
and from right to left, $\gamma q = 12, 6, 3, 1$. In the lower left region there is a superconducting
condensate. The critical magnetic field $B_{c2}$ is the limit of the curve as $T \to 0$. For $B > B_{c2}$ there is no superconductivity. Figure taken from \cite{Hartnoll:2008kx} ($q \to \gamma q$).}
\end{figure}

Although any one of the unstable modes is localised into a strip in the $x$ direction, with exponential suppression of the condensate beyond the width $(q B)^{-1/2}$, the center of the strip is in fact at $x = k/qB$. The degeneracy with respect to the momentum $k$ means that the condensate will form everywhere at once. This point was not appreciated in \cite{Albash:2008eh, Hartnoll:2008kx}.

\section{Potential and limitations of the holographic approach}

The application of holographic methods to condensed matter
phenomena is in its infancy. It is useful to assess what appear
to be the strengths and weaknesses of the program.

There are various formidable obstacles to making direct theoretical or experimental contact with `real world' systems. On the theoretical side, an important issue is that the type of theories that have weakly curved geometrical duals are likely to have significant differences with the field theories that typically arise in condensed matter physics. Although there is not as yet a precise characterisation of which field theories admit weakly coupled (i.e. classical) gravitational dual descriptions, the only properly understood cases are all supersymmetric large $N$ gauge theories. While emergent $U(1)$ fields are by now well understood and relevant for condensed matter systems, see e.g. \cite{subirrecent} for some examples along the lines of section \ref{sec:zA} above, even emergent $SU(2)$s are rather exotic and speculative \cite{lee}.

On the experimental side one problematic point is that many of the most precise experimental probes (such as tunneling microscopy) directly measure the electron densities. From a field theory point of view this corresponds to measuring the expectation values of `bare' or `UV' operators rather than operators that are natural in an effective field theory. In the holographic approach we only have access to the low energy observables (i.e. low energy compared to the lattice spacing etc.) and so we cannot compute these quantities.

The microscopic differences between (currently available) real experimental systems and theories with gravitational duals suggest that in the immediate future it is unlikely that values for experimental quantities obtained holographically could realistically aspire to more than being useful benchmarks.\footnote{Alternatively, the string landscape might suggest approaching these values statistically \cite{Denef:2009tp}.} Also important in this regard is the necessity of taking the large $N$ limit in holographic computations. For instance, it is now clear that while the value of the ratio of shear viscosity to entropy density, $\eta/s = 1/4 \pi$, is universal in classical gravity \cite{Kovtun:2004de, Iqbal:2008by}, there are controlled $1/N$ corrections to this result that can be both positive and negative and which for realistic values of $N$ give significant changes to the numerical value of the ratio \cite{Kats:2007mq, Brigante:2007nu, Buchel:2008vz}.

The strongest case for the usefulness of AdS/CFT for condensed matter physics rests on two pillars which we consider in turn. The first is that while theories with holographic duals may have specific exotic features, they also have features that are expected to be generic of strongly coupled (for instance, but not necessarily, quantum critical) theories. Insofar as theories with gravitational duals are computationally tractable examples of generic strongly coupled field theories, then we can use them to both test our generic expectations and guide us in refining these expectations. We can recall four examples of this approach considered in these lectures.
All four examples are special cases of the fact that finite temperature real time transport is much easier to compute via AdS/CFT than in almost any other microscopic theory: as we saw most explicitly in section \ref{sec:conductivity} it reduces to solving ODEs:

Firstly, we saw in section \ref{sec:crossover} that AdS/CFT provided the first explicitly computable example of the anticipated hydrodynamic to collisionless crossover in spectral densities of a CFT. Secondly, although we did not give details here, the magnetohydrodynamic results of section \ref{sec:rhoandB} were simultaneously derived using general hydrodynamic methods and AdS/CFT techniques. Given that the hydrodynamics is somewhat subtle in this case, it was extremely useful to have a computationally tractable microscopic model at hand. Thirdly, also in section \ref{sec:rhoandB}, we obtained results for the cyclotron resonance beyond the hydrodynamic regime. Fourthly, in section \ref{sec:impure} we computed the dependence of a momentum relaxation timescale due to weak impurities on the magnetic field and charge density.

A second promising aspect of the holographic approach is that it provides (unique in 2+1 and higher dimensions?) explicit examples of theories without a quasiparticle description in which computations are nonetheless feasible. This will hopefully force a certain conceptual re-evaluation which should ultimately clarify which commonly assumed properties of states of matter fail when there is not a weakly coupled description. For instance, it is highly unusual from a weakly coupled perspective that a theory with charged bosons can be stable against condensation at zero temperature in the presence of a chemical potential. Yet in figure 11 above we see that AdS/CFT provides strongly coupled theories in which there is a nontrivial criterion (\ref{eq:criterion}) for the stability of the vacuum in the presence of a chemical potential. A perhaps more obvious example already discussed, cf. figure 14, is that we can have theories for the onset of superconductivity in which the mass gap and gap in the conductivity are not related: $\w_g \neq 2 E_g$.

To rephrase the previous paragraph: theories with gravitational duals are well defined exotic theories against which the arsenal of condensed matter concepts can be tested. This must be done without recourse to a weak coupling language: one can ask questions about charges and currents and order parameters, but not about `electrons' or `phonons'. Two clear challenges for the immediate future are firstly to properly understand the dynamics of the ground state of holographic theories at finite chemical potential (is there a fermi surface?\footnote{Recent works in this direction include \cite{Rozali:2007rx, Shieh:2008nf, Karch:2009zz, Kulaxizi:2008kv, Kulaxizi:2008jx, Lee:2008xf, Liu:2009dm, Cubrovic:2009ye}. The last two of these shows strong evidence
of Fermi-surface related behaviour.} what is the correct strong coupling characterisation of a fermi surface? why can the bosons be stable?) and secondly to understand the dynamics behind the emergence of superconductivity from such a state. This second question will require an answer to the first. In the best of all possible worlds, the answers to these questions may shed light on nonconventional superconductors.

\section*{Acknowledgements}

Much of the material in these lectures I have learnt with or from my collaborators and colleagues.
I'd like to mention in particular many helpful discussions with Frederik Denef, Chris Herzog, Gary Horowitz, Pavel Kovtun, Prem Kumar, John McGreevy, Markus M\"uller and Subir Sachdev.

\end{document}